\documentclass[11pt,a4paper]{article}
\usepackage{epsfig}
\usepackage[T1]{fontenc}    
\usepackage{graphics}
\usepackage{graphicx}
\usepackage{pstricks,pst-coil,pst-fill,pst-plot}
\usepackage[fleqn]{amsmath}    
\usepackage{amssymb}    
\usepackage{amsfonts}   
\usepackage{verbatim}   
\usepackage{mathrsfs}   
\usepackage{dsfont}
\usepackage{euscript}
\usepackage{yfonts}
\usepackage{txfonts}
\usepackage{marvosym}
\usepackage{vmargin}        

\setmarginsrb{1.8cm}{2cm}{1.8cm}{2cm}{1cm}{1cm}{1cm}{1.6cm}
 \makeatletter
 \@addtoreset{equation}{section}
 \makeatother

\let\tend=\rightarrow

\providecommand{\bysame}{\leavevmode\hbox to3em{\hrulefill}\thinspace}

\newtheorem{theorem}{Theorem}[section]
\newtheorem{prop}{Proposition}[section]

\newtheorem{defin}{Definition}[section]

\newtheorem{lemme}{Lemma}[section]

\def\Proof{\medskip\noindent {\it Proof --- \ }}

\def\qed{\hfill\rule{2mm}{2mm}}

\newcommand\beq{\begin{equation}}
\newcommand\enq{\end{equation}}
\newcommand\bem{\begin{multline}}
\newcommand\enm{\end{multline}}

\def\beqa{\begin{eqnarray}}
\def\eeqa{\end{eqnarray}}
\def\ba{\begin{array}}
\def\ea{\end{array}}

\newcommand{\f}[2]{{\ensuremath{%
    \mathchoice%
    {\dfrac{#1}{#2}}
    {\dfrac{#1}{#2}}
    {\frac{#1}{#2}}
    {\frac{#1}{#2}}
}}}
\newcommand{\tf}[2]{\ensuremath{#1/#2}}
\newcommand{\pa}[1]{\ensuremath{\left(#1\right)}}

\newcommand{\pac}[1]{\ensuremath{\left[#1\right]}}
\newcommand{\paf}[2]{\ensuremath{\left(\f{#1}{#2}\right)}}

\def\a{\alpha}

\def\ga{\gamma}
\def\Ga{\Gamma}
\def\de{\delta}

\def\eps{\epsilon}
\def\veps{\varepsilon}
\def\la{\lambda}
\def\La{\Lambda}

\def\sg{\sigma}

\def\Ups{\Upsilon}

\def\Om{\Omega}

\def\vp{\varphi}

\newcommand{\mc}[1]{\ensuremath{\mathcal{#1}}}
\newcommand{\mf}[1]{\ensuremath{\mathfrak{#1}}}
\newcommand{\msc}[1]{\ensuremath{\mathscr{#1}}}

\newcommand{\bs}[1]{\ensuremath{\boldsymbol{#1}}}

\newcommand{\ov}[1]{\ensuremath{\overline{#1}}}
\newcommand{\wt}[1]{\ensuremath{\widetilde{#1}}}
\newcommand{\wh}[1]{\ensuremath{\widehat{#1}}}

\newcommand{\Int}[2]{\ensuremath{\int\limits_{#1}^{#2}}}
\newcommand{\Oint}[2]{\ensuremath{\oint\limits_{#1}^{#2}}}

\newcommand{\sul}[2]{\ensuremath{\sum\limits_{#1}^{#2}}}
\newcommand{\pl}[2]{\ensuremath{\prod\limits_{#1}^{#2}}}

\newcommand{\R}{\ensuremath{\mathbb{R}}}
\newcommand{\Cx}{\ensuremath{\mathbb{C}}}

\newcommand{\Dp}[1]{\ensuremath{\partial_{#1}}}

\newcommand{\limit}[2]{\ensuremath{\underset{#1 \tend #2}{\longrightarrow} }}

\newcommand{\ex}[1]{\ensuremath{\e{e}^{#1}}}

\def\id{\operatorname{id}}

\newcommand{\norm}[1]{\ensuremath{ || #1 || }}

\newcommand{\dd}{\mathrm{d}}
\newcommand{\e}[1]{\ensuremath{\mathrm{#1}}}

\newcommand{\intff}[2]{\ensuremath{\left [ \, #1 \,; #2 \, \right ] }}

\newcommand{\intoo}[2]{\ensuremath{\left ] \, #1 \,; #2 \, \right [ }}

\newcommand{\intn}[2]{\ensuremath{[\![ \, #1 \,;\, #2 \,]\!]}}

\begin{document}

\begin{flushright}

\end{flushright}
\par \vskip .1in \noindent

\vspace{14pt}

\begin{center}
\begin{LARGE}
{\bf Unitarity of the SoV transform for the Toda chain.}
\end{LARGE}

\vspace{30pt}

\begin{large}

{\bf K.~K.~Kozlowski}\footnote[1]{Universit\'{e} de Bourgogne, Institut de Math\'{e}matiques de Bourgogne, UMR 5584 du CNRS, France,
karol.kozlowski@u-bourgogne.fr}. 
\par

\end{large}

\vspace{40pt}

\centerline{\bf Abstract} \vspace{1cm}
\parbox{12cm}{\small
The quantum separation of variables method consists in mapping the original Hilbert space where a spectral problem is formulated
onto one where the spectral problem takes a simpler "separated" form. 
In order to realise such a program, one should construct the map explicitly and then show that it is unitary. 
 In the present paper, we develop a technique which allows one to prove the unitarity of this map 
in the case of the quantum Toda chain. Our proof solely builds on objects and 
relations naturally arising in the framework of the so-called quantum inverse scattering method. Hence, 
with minor modifications, it should be readily transposable to other quantum integrable models
solvable by the quantum separation of variables method. 
As such, it provides an important alternative to the proof of the map's unitarity 
based on the group theoretical interpretation of the quantum Toda chain, which is absent for 
more complex quantum integrable models.}

\end{center}

\vspace{40pt}

\section*{Introduction}

The quantum inverse scattering method is a powerful tool for 
solving a vast class of so-called quantum integrable models. 
The idea of the method consists in tailoring a particular quadratic algebra, the Yang-Baxter algebra, 
associated with the model of interest,
this in such a way that the original Hamiltonian belongs to a specific, one-parameter $\la$, 
commutative subalgebra $\{ \bs{\tau}(\la) \}_{\la \in\R}$ thereof. The algebraic relations associated with the 
Yang--Baxter algebra are most conveniently expressed on the level of the so-called monodromy matrix 
$T(\la)$ which is a matrix on some auxiliary space whose entries are operators on the model's Hilbert space. 
The family  $\{ \bs{\tau}(\la) \}_{\la \in\R}$ provides one with 	a set of commuting self-adjoint Hamiltonians, 
for instance through an expansion of the map $\la \mapsto \bs{\tau}(\la)$ around some point $\la_0$. 
Thus, the problem of obtaining the spectrum of the original Hamiltonian is mapped into a  multi-variable and multi-parameter
spectral problem associated with the family of commuting self-adjoint operators $\{ \bs{\tau}(\la) \}_{ \la \in \R } $. 
The resolution of this spectral problem can be carried out in two ways. The first one, referred to as the 
algebraic Bethe Ansatz has been developed in 1979 by Faddeev, Sklyanin and Takhtadjan
\cite{FaddeevSklyaninTakhtajanSineGordonFieldModel} and takes its roots in the 1931 seminal paper of Bethe 
\cite{BetheSolutionToXXX} where the so-called coordinate Bethe Ansatz solution of the XXX spin 1/2 Heisenberg has been proposed. 
The second one has been developed in 1985 by Sklyanin \cite{SklyaninSoVFirstIntroTodaChain} and can be thought of as the
quantum version of the classical separation of variables method. 
Although both methods take their roots in the representation theory of quantum groups, 
the algebraic Bethe Ansatz and the quantum separation of variables are, from a technical point of view, 
quite different. They can also be thought of as complimentary since, apart from very exceptional 
cases\footnote{such as certain sectors of the $\mf{sl}(2,\Cx)$ or $\mf{sl}(2,\R)$ XXX spin chain \cite{DerkachovKorchemskyManashovXXXSoVandQopNewConstEigenfctsBOp,DerkachovKorchemskyManashovXXXreelSoVandQopABAcomparaison}
or the so-called $sl(2)$-Gaudin model \cite{SklyaninSoVForSL2Gausin} } only one of the methods
is applicable for solving the model. This paper deals with certain technical aspects arising in the 
implementation of the quantum separation of variables method. 
The idea behind the latter method consists in mapping bijectively the multi-variable, multi-parameter
spectral problem associated with $\{ \bs{\tau}(\la) \}_{\la \in\R}$ onto an auxiliary multi-parameter 
spectral problem in \textit{one} variable which takes the form of a scalar $\bs{\tau}-\bs{Q}$ equation
\cite{BaxterPartitionfunction8Vertex-FreeEnergy,GaudinPasquierQOpConstructionForTodaChain,SklyaninSoVFirstIntroTodaChain}.
To achieve such a setting, one builds a unitary map $\msc{U}$  whose purpose is to send the original Hilbert space $\mf{h} $
of the model onto another Hilbert space $ \mf{h}_{\e{sep}} $
where the separation of variables, in the above sense, occurs. 
Thus, the implementation of the method involves solving three tasks. One should first find a convenient
explicit representation for the map $\msc{U}$, second one should prove its unitarity and, third, 
one should establish the equivalence of the original spectral problem on $\mf{h}$ with the scalar $\bs{\tau}-\bs{Q}$
equation on $\mf{h}_{\e{sep}}$. This paper deals with the quantum inverse scattering method approach to the 
resolution of the second problem. We shall now be slightly more explicit about the model
of interest and the quantum separation of variables method. This will allow us to formulate
the main achievement of the paper. 

\subsubsection*{The quantum Toda chain}

The very ideas of the quantum separation of variables take, in fact, their roots in the work of Gutzwiller on the 
quantum Toda
chain \cite{GutzwillerResolutionTodaChainSmallNPaper1,GutzwillerResolutionTodaChainSmallNPaper2}
which refers to a quantum mechanical $N+1$-body Hamiltonian 
\beq
\bs{ H }_{\kappa} \; = \; \sul{a=1}{N+1} \f{p_{a}^2}{2} \; + \;\kappa \ex{ x_{N+1} - x_1} 
\; + \;  \sul{a=1}{N} \ex{x_a -x_{a+1} } \qquad \e{acting}  \; \e{on} \quad 
\mf{h} \; = \; \bigotimes_{n=1}^{N+1} \mf{h}_n \; \simeq \; L^2\big(\R^{N+1}, \dd^{N+1}x \big) \;, 
\label{ecriture Hamiltonien Toda}
\enq
where $\mf{h}_n \simeq L^2\big(\R, \dd x \big)$ are "local" quantum spaces attached to the $n^{\e{th}}$-"particle". Furthermore, in 
\eqref{ecriture Hamiltonien Toda}, $p_n$ and $x_n$ are pairs of conjugated variables satisfying 
the canonical commutation relations $\pac{x_k,p_{\ell}} = i\hbar\de_{ k \ell}$. 
In the following, we shall choose the realization $p_{n} = -i \hbar \Dp{x_{n} }$. 
Also, the index $n$ present in the operators refers to the quantum space $\mf{h}_n$ where these operators act non-trivially. 
When $ \kappa =1$, one deals with the so-called closed Toda chain whereas, at $\kappa=0$, the model 
is referred to as the open Toda chain.

In the early '80's, Gutzwiller \cite{GutzwillerResolutionTodaChainSmallNPaper1,GutzwillerResolutionTodaChainSmallNPaper2} 
has been able to characterize the  
spectrum of $\bs{H}_{\kappa=1}$ in the case of a small number N+1 of particles, namely for $N=0,\dots, 3$. 
He expressed the map realizing the quantum separation of variables for the $N+1$ particle chain at 
$N=0,\dots, 3$ in the form of an integral transform. 
His main observation was that the non-trivial part of the integral transform's kernel 
was given by the generalized eigenfunctions of the $N$-particle open Toda chain $\bs{H}_{\kappa=0}$. However, the real deep 
connection which allowed for a systematic development of the method is definitely to be attributed to Sklyanin. 
In \cite{SklyaninSoVFirstIntroTodaChain}, by using an analogy with the classical separation of variables,
Sklyanin gave a quantum inverse scattering method-based interpretation of the aforementioned integral transform.
In the case of a quantum integrable model with a six-vertex $R$-matrix -such as the quantum Toda chain-, 
the transform corresponds precisely to the map that intertwines the $T_{12}(\la)$ operator entry of the model's 
monodromy matrix $T(\la)$ with a multiplication operator. 
In other words, the kernel of the integral transform is given by the eigenfunctions, 
understood in the generalized sense, of $T_{12}(\la)$. 
This observation, along with the set of algebraic relations stemming from the Yang-Baxter algebra 
satisfied by the entries $T_{ab}(\la)$ of $T(\la)$, allowed 
Sklyanin to construct the so-called quantum separation of variables representation on the space 
$\mf{h}_{\e{sep}} \, = \, L^2\big( \R^{N+1}, \dd \nu \big)$. 
In fact, due to the translation invariance of the close quantum Toda chain, 
the measure $\dd \nu$ factorizes $\dd\nu = \dd \veps \otimes \dd \mu$ into a "trivial" one-dimensional Lebesgue measure $\dd \veps$ that 
takes into account the spectrum $\veps$ of the momentum operator and a non-trivial part $\dd \mu$ which is absolutely continuous in 
respect to $\dd^Ny$. The aforementioned map allows one to represent functions
$\Phi \in \mf{h}$ as $\Phi(\bs{x}_{N+1})= \msc{U}\big[ \wh{\Phi} \big](\bs{x}_{N+1})$ where, for sufficiently 
well-behaved functions $\wh{\Phi} $,  
\beq
\msc{U}\big[ \wh{\Phi} \big](\bs{x}_{N+1}) \; = \; 
\Int{\R^{N+1} }{}  \vp_{\bs{y}_{N}} (\bs{x}_{N})  \cdot  \ex{\f{i}{\hbar}(\veps-\ov{\bs{y}}_N)x_{N+1}}     
\cdot \wh{ \Phi }(\bs{y}_{N}; \veps) \cdot  \dd \veps \otimes \f{ \dd\mu(\bs{y}_{N}) }{ \sqrt{N!} }  \; \; . \quad \e{Here} \quad
\ov{\bs{y}}_N=\sul{a=1}{N} y_a 
\label{definition transfo integrale U super cal}
\enq
and $\vp_{\bs{y}_{N}} (\bs{x}_{N})$ is an integral kernel which will be of central interest to our study. 
Also, above, we have adopted convenient for 
further discussions vectors notation, namely the subscript $r$ in $\bs{x}_r$ indicates that it is a $r$-dimensional 
vector, \textit{ie} $\bs{x}_r=(x_1,\dots,x_r)$. Note that due to the factorization of the measure $\dd \nu$, the map $\msc{U}$ factorizes
$\msc{U} \; = \; \mc{U}_N \circ \mc{F}$, in which $\mc{F}$ corresponds to taking a Fourier transform in $\veps$ followed by the 
action of a multiplication operator whereas $\mc{U}_N$ constitutes the non-trivial part of $\msc{U}$.
For $F \in L^{1}_{\e{sym}}\big(\R^N , \dd \mu(\bs{y}_N)\big)$, it is given by 
\beq
\mc{U}_N[F] (\bs{x}_N) \; = \;\f{1}{ \sqrt{N!} }  \Int{ \R^N }{}  \vp_{\bs{y}_N}(\bs{x}_N) \cdot F(\bs{y}_N)  \cdot \dd \mu(\bs{y}_N)  \;. 
\label{definition transfo U_N} 
\enq
 The subscript $ {}_{\e{sym}}$ occurring in $L^{1}_{\e{sym}}$ indicates that the 
function $F$ is a symmetric function of its variables.
In the following, we will refer to the integral transform $\mc{U}_N$ as the separation of variables (SoV) transform. 
The main advantage of the SoV transform is that it provides one with a very simple form for the 
eigenfunctions of the family of transfer matrices $\bs{\tau}(\la)$. When focusing on a sector with a fixed momentum 
$\veps$, any eigenfunction $\mc{V}_{t}$
of $\bs{\tau}(\la)$ associated with the eigenvalue $t(\la)$ admits a factorized representation
in the space $\mf{h}_{\e{sep}}$, in the sense that 
\beq
\wh{\mc{V}_{t}}(\bs{y}_N;\veps) \; = \; \pl{a=1}{N} q_{t}(y_a) \; .
\label{definition forme fct propre espace separe}
\enq
The function of \textit{one} variable $q_{t}(y)$ appearing above is entire and solves the scalar form of the  so-called 
$\bs{\tau}-\bs{Q}$ equation
\beq
t(\la) \cdot  q_{t}(\la) \; = \; (i)^{N+1} q_{t}(\la+i\hbar) \; + \; (-i)^{N+1} q_{t}(\la-i\hbar) 
\label{ecriture eqn TQ}
\enq
under the below condition on the asymptotic behaviour of the solutions \cite{GaudinPasquierQOpConstructionForTodaChain}
\beq
q_{t}(\la) = \e{O}\Big(  \ex{- \f{N \pi}{2\hbar} |\Re(\la) |  } |\la |^{\f{N}{2\hbar}(2|\Im(\la)|- \hbar) }  \Big)
\;\;\e{uniformly} \;\; \e{in}  \;\; |\Im(\la)| \;\leq \; \f{\hbar}{2} \;,  
\label{ecriture eqn TQ conditions sur cptm asymptotique}
\enq
with $t(\la)$ being a monoic polynomial of degree $N+1$. We do stress that 
the $\bs{\tau}-Q$ equation is a \textit{joint} equation for the coefficients of the polynomial $t( \la )$ and the solution 
$q_{t}$. The asymptotic behaviour \textit{and} the regularity conditions  on $q_{t}$ can be met simultaneously only for well
tuned polynomials $t(\la)$ corresponding to eigenvalues of $\bs{\tau}(\la)$; this effect gives rise
to so-called quantization conditions for the Toda chain.  

\vspace{2mm}

To phrase things more precisely, within the framework of the quantum separation of variables, 
the resolution of the spectral problem for the quantum Toda chain amounts to
\begin{itemize}
\item[i)] building and characterizing the  kernel  $\vp_{\bs{y}_N}(\bs{x}_N)$ of the SoV transform ;
\item[ii)] establishing the unitarity of $\mc{U}_N \; : \; L^2\big(\R^N, \dd^N x \big) \;  \tend  \; 
L^2_{\e{sym}}\big(\R^N, \dd \mu(\bs{y}_N) \big) $; 
\item[iii)] characterizing all of the solutions to \eqref{ecriture eqn TQ} and \eqref{ecriture eqn TQ conditions sur cptm asymptotique} 
and proving the equivalence of this spectral problem to the original one formulated on $\mf{h}$.
\end{itemize}
  
\noindent Point $iii)$ has been first argued by Sklyanin \cite{SklyaninSoVFirstIntroTodaChain}
and the correspondence proved by An \cite{AnCompletenessEigenfunctionsTodaPeriodic}.

\subsubsection*{The $GL(N,\R)$-Whittaker function interpretation of $\vp_{\bs{y}_N}(\bs{x}_N)$}  
  
The resolution of point $i)$ takes its roots in the work of Kostant 
\cite{KostantIdentificationOfEigenfunctionsOpenTodaAndWhittakerVectors}. 
The author of \cite{KostantIdentificationOfEigenfunctionsOpenTodaAndWhittakerVectors} 
found a way to quantize the integrals of motion for the open  classical  Toda chain 
hence showing the existence of an abelian ring of operators containing the quantum open Toda chain Hamiltonian 
\eqref{ecriture Hamiltonien Toda}.
Furthermore, he was able \cite{KostantIdentificationOfEigenfunctionsOpenTodaAndWhittakerVectorsMoreDeep} to identify the system of joint generalized eigenfunctions to this ring as Whittaker functions for 
$GL(N,\R)$. Kostant's approach has been continued and extended so as to include other Toda Hamiltonians
such as the closed one, $\bs{H}_{\kappa=1}$, by Goodman and Wallach in \cite{GoodmannWallachQuantumTodaI,GoodmannWallachQuantumTodaII,GoodmannWallachQuantumTodaIII}. 
In particular they provided an explicit construction of a set of generators for the aforecited ring of
operators in involution. Recall also that the systematic study of Whittaker functions has been initiated by 
Jacquet \cite{JacquetFctWhittakerPrGrpesChevalley} and that the theory has been further developed
by  Hashizume \cite{HashizumeSomeCharacterizationWhittakerFunctions} and 
Schiffmann \cite{SchiffmannIntegralRepsAndMoreWhittakerFcts}. At the time, the Whittaker function were constructed by purely group theoretical handlings, what allowed to represent them by means of the so-called Jacquet's multiple integral. 
In 1990, Stade \cite{StadeNewIntRepWhittakerFctOutOfJacquetRep} obtained 
another multiple integral representations for the $GL(N,\R)$ Whittaker functions. 
The authors of \cite{GerasimovKharchevMarshakovMorozovMironovOlshanetskyGaussDecmpBasedIntRepWhittFcts}
proposed yet another multiple integral representation for these function 
which was based on the so-called Gauss decomposition\footnote{Note that the construction of Jacquet's multiple integral representation is based on the so-called Iwasawa decomposition.} of group elements. 
However, for many technical reasons, all these representations, although explicit, were 
hard to deal with or extract from them the sought informations on the functions. 
Nonetheless, this state of the art was already enough in what concerned applications 
to the quantum Toda chain.

Indeed, recall that Gutzwiller
\cite{GutzwillerResolutionTodaChainSmallNPaper1,GutzwillerResolutionTodaChainSmallNPaper2}  
constructed the eigenfunctions for the closed $N+1$-particle quantum Toda chain, 
at small values of $N$,  be means of an integral transform whose kernel corresponds to the 
eigenfunctions of the open $N$-particle quantum Toda chain. 
Building on this idea and implicitly conjecturing that the ring of operators found by 
Kostant actually coincides with 
the quantum inverse scattering method issued integrals of motion for the open $N$-particle Toda chain,
 Kharchev and Lebedev \cite{KharchevLebedevIntRepEigenfctsPeriodicTodaFromWhittakerFctRep} 
wrote down a multiple  integral representation for the eigenfunctions for the closed periodic Toda chain $\bs{H}_{\kappa=1}$
in the form $\msc{U}[\wh{\mc{V}_t} ]$, \textit{cf} 
\eqref{definition transfo integrale U super cal} and \eqref{definition forme fct propre espace separe}. 
Their construction worked for any value of $N$. 
The main point of their conjecture is that it allowed them
to use Kostant's characterization of the eigenfunctions of the open Toda abelian ring of operators as Whittaker functions for $GL(N,\R)$
so as to identify the kernel $\vp_{\bs{y}_N}(\bs{x}_N)$ with such Whittaker functions. 
At the time, they used the so-called Gauss decomposition based multiple integral representations for 
these Whittaker functions \cite{GerasimovKharchevMarshakovMorozovMironovOlshanetskyGaussDecmpBasedIntRepWhittFcts}. 
Later in \cite{KharchevLebedevMellinBarnesIntRepForWhittakerGLN,KharchevLebedevIntRepEigenfctsPeriodicTodaFromRecConstrofEigenFctOfB},
the two authors managed to connect their approach with Sklyanin's quantum separation of variables \cite{SklyaninSoVFirstIntroTodaChain} 
approach to the quantum Toda chain.

More precisely, Sklyanin's method relies on the observation that the integral kernel $\vp_{\bs{y}_{N}} (\bs{x}_{N})$ of the SoV transform 
corresponds, up to some minor modifications, to the eigenfunction of the $T_{12}(\la)$ operator entry of the monodromy matrix
for an $N+1$-particle Toda chain:
\beq
T_{12}(\la) \cdot \vp_{\bs{y}_{N}} (\bs{x}_{N}) \; = \; \ex{-x_{N+1}}\pl{a=1}{N} \big(\la-y_a\big) \cdot \vp_{\bs{y}_{N}} (\bs{x}_{N})\;. 
\label{ecriture equation definition heuristique des fct propres B}
\enq
In \cite{SklyaninSoVGeneralOverviewAndConstrRecVectPofB}, Sklyanin proposed a inductive scheme 
based on the recursive construction of the monodromy matrix which allowed one to 
build the eigenfunctions $\vp_{\bs{y}_N}(\bs{x}_N)$ inductively. Kharchev and Lebedev
\cite{KharchevLebedevMellinBarnesIntRepForWhittakerGLN,KharchevLebedevIntRepEigenfctsPeriodicTodaFromRecConstrofEigenFctOfB}
managed to implement this scheme on the example of the open Toda chain, hence obtaining
a new multiple integral representation for the $GL(N,\R)$ Whittaker functions
which they called Mellin-Barnes representation. 
Finally, in \cite{GerasimovKharchevLebedevRepThandQISM}, Gerasimov, Kharchev and Lebedev established a clear connection 
between the group theoretical and the quantum inverse scattering method-based approaches to the open 
Toda chain. In particular, that paper proved the previously used conjecture relative to 
the concurrency between  Kostant's ring of operators on the one hand and the quantum inverse scattering issued 
conserved charges on the other.  

\vspace{2mm} 

There exists one more multiple integral based representation for the generalized eigenfunctions 
of the open Toda chain due to Givental \cite{GiventalGaussGivIntRepObtainedForEFOfOpenToda}. 
The group theoretic interpretation of this type of multiple integral representation has been given in 
\cite{GerasimovKharchevLebedevOblezinGaussGiventalIntRepFromQISMAndQOp}. Since the 
corresponding proof built on a specific type of Gauss decomposition for the group elements of $GL(N,\R)$,
this multiple integral representation bears the name Gauss--Givental. 
Furthermore, paper \cite{GerasimovKharchevLebedevOblezinGaussGiventalIntRepFromQISMAndQOp} also contained 
some comments relative a connection between the Gauss--Givental representation and the model's
$\bs{Q}$-operator constructed earlier by Gaudin and Pasquier 
\cite{GaudinPasquierQOpConstructionForTodaChain}.
In fact, this connection, within the setting of another quantum separation of variables solvable model, 
the so-called non-compact XXX magnet, has been established, on a much deeper level of understanding, a few years earlier 
by  Derkachov, Korchemsky and Manashov \cite{DerkachovKorchemskyManashovXXXSoVandQopNewConstEigenfctsBOp}. 
These authors observed that one can build the
eigenfunctions of the $\la \mapsto T_{12}(\la)$ operators arising in the context of the non-compact XXX magnet 
out of the building bricks for the integral kernel of the model's $\bs{Q}$-operator. This allowed
Derkachov, Korchemsky and Manashov to propose a "pyramidal" representation for the quantum SoV's map kernel
for the non-compact XXX magnet. Later, in
\cite{SilantyevScalarProductFormulaTodaChain},  Silantyev applied the DKM  method so as to re-derive the 
 Gauss-Givental representation for the $GL(N,\R)$ Whittaker functions $\vp_{\bs{y}_N}(\bs{x}_N)$. 

\subsubsection*{The main result of the paper}

Until now, we have not yet discussed the completeness and orthonormalilty of the system of generalized eigenfunctions 
of the open Toda chain $\{ T_{12}(\la) \}_{\la \in \R}$ abelian ring of operators. These properties, in fact, 
boil down to proving  the unitarity of $\mc{U}_N$, \textit{viz} the point $ii)$ mentioned earlier. 
The completeness and orthonormalilty have, in fact, been already established within the framework of the group theoretical 
based approach to the model. Semenov-Tian-Shansky proved 
\cite{SemenovTianShanskyQuantOpenTodaLatticesProofOrthogonalityFormulaForWhittVectrs} the orthonormalilty 
of the system $\{ \bs{x}_N \mapsto  \vp_{\bs{y}_N}(\bs{x}_N) \}_{ \bs{y}_N \in \R^N }$. The latter, written formally, takes the form 
\beq
\Int{ \R^{N} }{} \Big( \vp_{\bs{y}_{N}^{\prime}} (\bs{x}_{N}) \Big)^{*}   \cdot 
 \vp_{\bs{y}_{N}} (\bs{x}_{N}) \cdot \dd^{N} x  \; = \; 
 \big[  \mu(\bs{y}_N) \big]^{-1}  
\sul{ \sg \in \mf{S}_N }{}  \pl{a=1}{N} \de\big( y_a - y^{\prime}_{\sg(a)} \big) \;. 
\label{ecriture relation completude par rapport espace originel}
\enq
Also, the completeness
of the system $\{ \bs{y}_N \mapsto \vp_{\bs{y}_N}(\bs{x}_N) \}_{ \bs{x}_N \in \R^N }$, which written formally, takes the form 
\beq
\Int{ \R^{N} }{} \Big( \vp_{\bs{y}_{N}} (\bs{x}_{N}^{\prime}) \Big)^{*}  \vp_{\bs{y}_{N}} (\bs{x}_{N})\; 
\mu(\bs{y}_N) \cdot   \dd^{N} y  \; = \; 
\pl{a=1}{N} \de(x_a - x^{\prime}_{a}) \;, 
\label{ecriture relation completude par rapport espace SoV}
\enq
follows from the material that can be found in chapters 15.9.1-15.9.2 and 15.11 of Wallach's book \cite{WallachRealReductiveGroupsII}.

We do stress that the proofs \cite{SemenovTianShanskyQuantOpenTodaLatticesProofOrthogonalityFormulaForWhittVectrs,WallachRealReductiveGroupsII}
are technically involved, rather long, and completely disconnected from the QISM description of the model. 
It is, in fact, the last fact that is the most problematic from the point of view of implementing
the quantum separation of variables to more complex quantum integrable models. Even for a relatively 
simple model such as the lattice discretization of the Sinh-Gordon model \cite{BytskoTeschnerSinhGordonFunctionalBA}, 
point ii) remains an open question. It is the quest towards obtaining a simple and systematic approach to the resolution 
of analogues of problems outlined in point $ii)$ but for more complex models that led us to developments described in the present paper. 
Namely, we propose a new approach to proving the unitarity of the quantum separation of variables map,
that is to say the 

\vspace{2mm}
\noindent {\bf Theorem}
\textit{ The integral transform $\mc{U}_N$ defined by \eqref{definition transfo U_N} for functions 
$ F \in \msc{C}^{\infty}_{\e{c};\e{sym}}\big(\R^N \big)$ extends to  
a unitary map $\mc{U}_N \; : \; L^2_{\e{sym}}\big(\R^N, \dd \mu(\bs{y}_N) \big)  \;  \tend  \; L^2\big(\R^N, \dd^N x \big)  $
where }
\beq
\dd \mu(\bs{y}_N) \; = \; \mu(\bs{y}_N) \cdot \dd^N y 
\quad \e{with} \quad  \mu(\bs{y}_N)  \; = \; 
\f{1}{(2\pi \hbar)^{N}}  \pl{ k \not= p }{ N } \Ga^{-1}\Big( \f{y_k-y_p}{i\hbar}  \Big) \;, 
\label{definition mesure Sklyanin}
\enq
\textit{and the functions $\vp_{\bs{y}_N}(\bs{x}_N)$  are defined by their Mellin-Barnes multiple integral representation 
\eqref{definition fct Whittaker par Mellin-Barnes}}

The method we develop for proving this theorem
\begin{itemize}
\item is completely independent from the previous scheme of works 
that require a group theoretical interpretation of the model in the spirit of \cite{GoodmannWallachQuantumTodaI,
GoodmannWallachQuantumTodaII,GoodmannWallachQuantumTodaIII,KostantIdentificationOfEigenfunctionsOpenTodaAndWhittakerVectors};
\item solely relies on structures and objects naturally arising within the quantum inverse scattering method;
\item   is quite simple - on a formal level of rigour, it is almost immediate to implement- and relatively short.
\end{itemize}

Since our method naturally fits into the quantum inverse scattering method approach, its 
 main advantage consists in allowing one, in principle, to apply it for proving the unitarity of the quantum separation of variables transform in the case of many other quantum integrable models. 
We chose to develop our method on the example of the Toda chain due to that model's simplicity;
this setting allowed us to avoid the technical ponderousness that would arise in the course of the analysis of more complex models.

\vspace{3mm} 
In order to be slightly more specific about our approach, we remind that the $\bs{Q}$-operator based DKM approach \cite{DerkachovKorchemskyManashovXXXSoVandQopNewConstEigenfctsBOp} allows one to derive 
the Gauss-Givental multiple integral representation for $\vp_{\bs{y}_N}(\bs{x}_N)$. 
In fact, it also allows one to establish, on a formal level of rigour, 
the relation \eqref{ecriture relation completude par rapport espace originel}, this in 
fairly simple way. Such a formal proof of the orthogonality condition has been given by Silantev \cite{SilantyevScalarProductFormulaTodaChain}.
In the present paper we, first of all, bring various elements of rigour
to Sylantev's manipulations \cite{SilantyevScalarProductFormulaTodaChain} leading to a 
completely independent in respect to Semenov-Tian-Shansky's work and much simpler proof 
of the isometric nature of $\mc{U}_N$.
Further, we provide a proof of the completeness relation
\eqref{ecriture relation completude par rapport espace SoV} which, also,  is completely 
independent from any group theoretical handlings. 
In fact, the proof we propose is based on the existence of two natural quantum inverse scattering method issued 
multiple integral representations for $\vp_{\bs{y}_N}(\bs{x}_N)$: the Gauss-Givental one and the Mellin-Barnes one. 
Knowing that 
\eqref{ecriture relation completude par rapport espace originel} holds and that $\vp_{\bs{y}_N}(\bs{x}_N)$
satisfies a Mellin-Barnes multiple integral representation issued recurrence relation allows us to deduce
\eqref{ecriture relation completude par rapport espace SoV}. In this respect, our proof highlights a sort of beautiful duality
between the two types of multiple integral representations. To the best of the author's knowledge,
this way of proving the completeness is based on completely new ideas. Furthermore, 
we do stress again that, on the formal level of rigour, the steps for proving the completeness relation are
extremely easy. 

\vspace{2mm}

The paper is organized as follows. In section \ref{Section Rep Mul Int} we introduce the Mellin-Barnes and 
Gauss-Givental multiple integral representations for $\vp_{\bs{y}_N}(\bs{x}_N)$ and establish
several basic properties of the latter. Then, in section \ref{Section Isometric character of UN},
we provide a proof, strongly inspired by the formal handlings of \cite{SilantyevScalarProductFormulaTodaChain},
of the isometric nature of the $\mc{U}_N$ transform. Then, in section \ref{Section Isometric Nature UN adjoint},
we provide a proof of the isometric nature of the formal adjoint of $\mc{U}_N$. 
All the ideas behind this proof are brand new, at least to the best of the author's knowledge. 
Several results of technical nature are gathered in the appendices. In appendix 
\ref{Appendix Proof of explicit behavior function varphi} we build on the Mellin-Barnes integral representation 
for $\vp_{\bs{y}_N}(\bs{x}_N)$ so as to derive uniform in $\bs{y}_N \in \R^N $, $\bs{x}_N \tend \infty$ asymptoics
of this function. 
Finally, in appendix \ref{Appendix Section from MB 2 GG} we establish a direct connection between the 
Mellin-Barnes and Gauss-Givental multiple integral representations, hence proving that, indeed, they do 
define the very same function. The proof given in \ref{Appendix Section from MB 2 GG} builds on several ideas introduced in  \cite{GerasimovLebedevOblezinBAxtOpMixedRepsForTodaWhittakerAndMore}. 
Still, the main difference between our proof and the one of \cite{GerasimovLebedevOblezinBAxtOpMixedRepsForTodaWhittakerAndMore}
is that we provide new arguments that allow us to circumvent the use of relations 
\eqref{ecriture relation completude par rapport espace originel}-\eqref{ecriture relation completude par rapport espace SoV}
in the proof.




\section{The kernel of the SoV transform for the Toda chain}
\label{Section Rep Mul Int}

As it has been mentioned in the introduction, there exists two quantum inverse scattering method issued multiple integral representations 
for the integral kernel $\vp_{\bs{y}_N }(\bs{x}_N)$  of the transform $\mc{U}_N$. 

In this section, we shall review the structure of these two representations, present some short proofs of 
several known facts about these representations as well as prove certain, yet unestablished, properties thereof. 
This preliminary analysis will allow us to introduce all the concepts and tools that will
be necessary for establishing the unitarity of $\mc{U}_N$. 

\subsection{The Mellin-Barnes representation}

Let $(\bs{y}_{N},  \bs{x}_N) \in \R^{N}\times \R^N$.  The functions $\vp_{\bs{y}_{N}}\big( \bs{x}_{N} \big)$ 
occurring in \eqref{definition transfo U_N} are defined as the unique solution to the induction
\beq
\vp_{\bs{y}_{N+1}}(\bs{x}_{N+1}) \; = \; \Int{ (\R-i\a)^{N} }{  } \hspace{-3mm} 
\ex{ \f{i}{\hbar} (\ov{\bs{y}}_{N+1} - \ov{\bs{w}}_{N}) x_{N+1} } 
\vp_{\bs{w}_{N}}(\bs{x}_{N}) \varpi( \bs{w}_{N} \mid \bs{y}_{N+1} ) \cdot \f{ \dd^{N} w  }{ N! (2\pi \hbar)^{N} } \;,
\label{definition fct Whittaker par Mellin-Barnes}
\enq
in which $\a>0$ is a free parameter 
\beq
\varpi( \bs{w}_{N} \mid \bs{y}_{N+1} ) \; = \; 
 \pl{a=1}{N} \pl{b=1}{N+1} \Big\{ \hbar^{\f{i}{\hbar}(w_a-y_b) } \Ga\Big( \f{y_b-w_a}{i\hbar} \Big) \Big\} 
 \cdot \pl{a \not= b}{N}  \Ga^{-1}\Big( \f{w_b-w_a}{i\hbar} \Big)  \;, 
\enq
and the inductions is subject to the initiation condition
\beq
\vp_y(x)  \; = \; \ex{\f{i}{\hbar} xy} \;.
\enq

It is straightforward to convince oneself that $\vp_{\bs{y_N}}(\bs{x}_N)$ admits the explicit expression
\beq
\vp_{\bs{y}_N}(\bs{x}_N) \; = \; \ex{ \f{i}{\hbar} \ov{\bs{y}}_{N} x_{N} }   \pl{s=1}{N-1} \Int{ (\R-i\a_s)^{N-s} }{  } \hspace{-3mm} 
\f{ \dd^{N-s} w^{(s)}  }{ (N-s)! (2\pi \hbar)^{N-s} } 
\pl{s=1}{N-1}  \ex{ \f{i}{\hbar} \ov{\bs{w}}_{N-s}^{(s)} \big( x_{N-s} - x_{N-s+1} \big) } 
\pl{s=1}{N-1}\varpi( \bs{w}_{N-s}^{(s)} \mid \bs{w}_{N-s+1}^{(s-1)} )  \;,
\label{definition fonction vp apres resolution MellinBarnes}
\enq
where we do agree upon $0 < \a_1 < \dots < \a_{N-1}$
\beq
\bs{w}_{N}^{(0)} = \bs{y}_N \qquad \e{and} \; \e{recall} \; \e{that} \qquad 
\ov{\bs{x}}_k \; = \; \sul{a=1}{k} x_a \quad \e{for} \; k-\e{dimensional} \; \e{vectors} \; \;\; \bs{x}_k \in \R^k \;. 
\enq
The iterated integral converges strongly (exponentially fast), see \textit{eg}
\cite{GerasimovKharchevLebedevRepThandQISM}, and defines a smooth function
\beq
(\bs{y}_N,\bs{x}_N) \; \mapsto \;  \vp_{\bs{y}_N}(\bs{x}_N)  \; \in \; L^{\infty}(\R^N\times \R^N, \dd^N y \otimes \dd^N x) \;.  
\enq
This ensures that the integral transform with 
$\dd \mu(\bs{y}_N)$ as defined in \eqref{definition mesure Sklyanin} 
\beq
\mc{U}_N[F] (\bs{x}_N) \; = \;\f{1}{ \sqrt{N!} }  \Int{ \R^N }{}  \vp_{\bs{y}_N}(\bs{x}_N) \cdot F(\bs{y}_N)  \cdot \dd \mu(\bs{y}_N)  
\label{definition noyau $U$ sur un espace dense}
\enq
is well defined for any $F\in L^1\big( \R^N, \dd \mu(\bs{y}_N) \big)$.

In fact, as we shall establish in appendix \ref{Appendix Proof of explicit behavior function varphi},
the function $\vp_{\bs{y}_N}(\bs{x}_N)$ can be recast in terms of a combination of 
oscillatory and exponentially decreasing terms in $\bs{x}_N\in \R^N$, this uniformly in $\bs{y}_N \in \R^N $. 
Such a representation allows one to carry out integration by parts in \eqref{definition noyau $U$ sur un espace dense}
which readily lead to the below proposition whose proof is postponed to appendix 
\ref{Appendix Proof of explicit behavior function varphi}:

\begin{prop}
\label{Proposition charactere de transfo UN}

Given any $F \in \msc{C}^{\infty}_{\e{c}}(\R^N)$, the integral transform $\mc{U}_N[F]$ is well 
defined and belongs to the Schwartz class $\mc{S}(\R^N)$. In particular,  for such functions $F$,
one has that $\mc{U}_N[F] \in L^{2}\big( \R^N, \dd^N x \big)$.  

\end{prop}

\subsection{The DKM construction of the eigenfunctions of the open Toda chain}

We now discuss the Gauss--Givental multiple integral representation for the integral kernel $\vp_{ \bs{y}_N }(\bs{x}_N)$
which has been obtained in 
\cite{GerasimovKharchevLebedevOblezinGaussGiventalIntRepFromQISMAndQOp,GiventalGaussGivIntRepObtainedForEFOfOpenToda,
SilantyevScalarProductFormulaTodaChain}, this by means of various different reasonings.
For our purpose, we shall follow the presentation of 
\cite{SilantyevScalarProductFormulaTodaChain} which was an adaptation of the method developed by 
Derkachov, Korchemsky and Manashov 
\cite{DerkachovKorchemskyManashovXXXSoVandQopNewConstEigenfctsBOp}.
More precisely, these authors have observed on the example of the so-called $\mf{sl}\big( 2, \Cx \big)$ XXX chain,
that one can extract the elementary building blocks  for the kernel $\vp_{\bs{y}_N }(\bs{x}_N)$ out of the integral kernel
$Q_{\la}(\bs{x}_N,\bs{x}^{\prime}_N)$ of the model's $\bs{Q}$-operator. 
The integral kernel of the $\bs{Q}$-operator for the Toda chain
has been constructed by Gaudin and Pasquier \cite{GaudinPasquierQOpConstructionForTodaChain}.
It admits the below representation 
\beq
Q_{\la}(\bs{x}_N,\bs{x}^{\prime}_N)   \; =  \; 
\pl{n=1}{N} V_{\la;-}\big( x_n-x^{\prime}_n \big) V_{\la;+}\big( x_n-x^{\prime}_{n-1} \big) \;, 
\label{ecriture noyau integral Q}
\enq
where 
\beq
V_{\la ; \pm }(x) \; = \; \exp\Big\{ -\f{1}{\hbar} \ex{\pm x} \; + \; i \f{\la x }{2 \hbar }  \Big\} \;. 
\enq

By sending $x_N^{\prime}\tend +\infty$ in the expression \eqref{ecriture noyau integral Q} 
for $Q_{\la}\big( \bs{x}_N,\bs{x}^{\prime}_N \big)$ we get 
\beq
Q_{\la}\big( \bs{x}_N,\bs{x}^{\prime}_N \big)  \; =  \; \La_{\la}^{(N)} \big( \bs{x}_N \mid \bs{x}^{\prime}_{N-1} \big) 
\,  \ex{-i\f{\la}{\hbar} x^{\prime}_N } \, 
\exp\Big\{  - \f{ 1 }{ \hbar } \ex{x^{\prime}_N-x_N^{} }  \Big\}  \cdot \big( 1+\e{o}(1) \big) \;, 
\enq
with
\beq
\La_{\la}^{(N)} \big( \bs{x}_N \mid \bs{x}^{\prime}_{N-1} \big)  \; = \;  \ex{ \f{i\la}{2\hbar} (x_1+x_N) }
\pl{n=1}{N-1} V_{\la;-}\big( x_n-x^{\prime}_n \big) \pl{n=2}{N} V_{\la;+}\big( x_n-x^{\prime}_{n-1} \big)   \;. 
\enq
 We do stress here that 
$\bs{x}^{\prime}_{N-1}=(x_1^{\prime}, \dots, x_{N-1}^{\prime})$ is a $N-1$ dimensional vector 
obtained from $\bs{x}^{\prime}_{N}$ by dropping the last coordinate $x_N^{\prime}$. 
The above function defines an integral kernel for the mapping 
\beqa
\La^{(N)}_y \quad  : \quad L^{\infty}(\R^{N-1})  &\tend &   L^{\infty}(\R^{N})  \\
\phantom{ \La^{(N}_y \quad  : \quad}  f  & \mapsto & \Int{ \R^{N-1} }{  } 
\La^{(N)}_y(\bs{x}_N \mid \bs{z}_{N-1} ) \, f(\bs{z}_{N-1})  \; \pl{a=1}{N-1}\dd z_a \;,
\eeqa
and plays a crucial role in the construction of the so-called Gauss-Givental representation for the 
function $\vp_{\bs{y}_N}(\bs{x}_N)$. 

We likewise define the conjugated operator 
\beqa
\ov{\La}^{\pa{N}}_{y} \quad  : \quad  \mc{S}(\R^N)  &\tend &  L^{\infty}(\R^{N-1}) \\
\phantom{ \ov{\La}^{\, \pa{N}}_{y} \quad  : \quad}  f  & \mapsto & \Int{\R^N}{} 
 {\ov{\La} }^{(N)}_y \big( \bs{z}_{N-1} \mid \bs{x}_{N} \big)   f(\bs{x}_N)  \,  \pl{a=1}{N}\dd  x_a \;,
\eeqa
where $\mc{S}(\R^N)$ refers to Schwartz functions on $\R^N$ whereas the integral kernel reads
\beq
\ov{\La}_{\la}^{ (N) } \big( \bs{z}_{N-1} \mid \bs{x}_{N}  \big)  \;  = \; \ex{ - \f{i\la}{2\hbar} \pa{x_1+x_N} }
\pl{n=1}{N-1} V_{- \la; - }\big( x_n-z_n \big) \pl{n=2}{N} V_{- \la;+ }\big( x_n-z_{n-1} \big) \;.
\enq

The operator $\La_{\la}^{ (N) }$ also defines a right-handed integral transform which is well behaved on appropriate classes of Schwartz functions
in that it does preserve the regularity properties of this class. More precisely, one has that 
\begin{lemme}
\label{Proposition character Schwartz des transformations Lambda a droite et gauche}

Let 
$G(\bs{z}_{N-1}, y) \in \mc{S}(\R^{N-1}\times \R)$, the function of $N$-variables
\beq
\wh{G}_L(\bs{x}_{N-2}) \; = \; 
\Int{ \R^N }{}    G(\bs{z}_{N-1}, y)  \cdot \La^{(N-1)}_y\big(\bs{z}_{N-1} \mid \bs{ x }_{N-2} \big) \cdot \dd^{N-1}z \cdot \dd y  
 \label{definition transformation integrale Lambda action gauche}
\enq
belongs to $\mc{S}(\R^{N-2})$.

\end{lemme}
\Proof 
Since $G$ belongs to the Schwartz class, for any $k\in \mathbb{N}^*$,  there exists $C_k>0$ such that 
\beq
\big|  G(\bs{z}_{N-1}, y) \big| \; \leq \; \f{ C_k }{ \big( 1+|y| \big)^{k+1} } \cdot 
\pl{a=1}{N-1} \bigg( \f{ 1 }{ 1+|z_a| }  \bigg)^{k+1}  \; .  \;
\enq
Thus, agreeing upon  $\norm{\bs{x}_N}_{\infty} \,= \, \max_{ 1\leq a \leq N }|x_a| $, 
and using $ \big| \La_{\la}^{ (N-1) } \big( \bs{z}_{N-1} \mid \bs{x}_{N-2}  \big) \big| \; \leq \; 1 $, 
\bem
\big| \wh{G}_L\big( \bs{x}_{N-2} \big)  \big| \; \leq \; 
\Int{  \norm{\bs{z}_{N-1} }_{\infty} \leq \f{ \norm{\bs{x}_{N-2}}_{\infty}}{2} }{} \hspace{-5mm} 
\big|  G(\bs{z}_{N-1}, y) \big| \cdot 
\pl{n=1}{N-2} \exp\Big\{ - \f{1}{\hbar}\ex{x_n-z_n} \, - \,  \f{1}{\hbar}\ex{z_{n+1}-x_n} \Big\} \cdot \dd^{N-1} z\cdot \dd y \\
\; + \; \sul{p=1}{N-1}  \Int{ 2 |z_p| >  \norm{\bs{x}_{N-2}}_{\infty}  }{} \hspace{-3mm}  \f{ C_k }{ \big( 1+|y| \big)^{k+1} } \cdot 
\pl{a=1}{N-1} \bigg( \f{ 1 }{ 1+|z_a| }  \bigg)^{k+1} \cdot \dd^{N-1} z\cdot \dd y  
\; \leq \; \f{ C }{ \Big( 1+\norm{\bs{x}_{N-2}}_{\infty} \Big)^{k} }  \;. 
\end{multline}

The first line can be bounded due to the extremely quick decay of the integrand when $ \norm{\bs{x}_{N-2}}_{\infty} \tend + \infty$
and the $L^{1}(\R^{N-1}\times \R)$ nature of $G$
whereas the bound on the \textit{rhs} of the second line follows from a direct integration. \qed 

\vspace{2mm}

The action of the operator $\La_{y}^{(N)}$ to the right produces a slightly less regular behaviour. 
\begin{lemme}
 \label{Lemme bornage transfo La a droite}
For any $\Phi \in L^{\infty}\big( \R^{N-1}, \dd^{N-1}x \big)$, the function $\big( \La_y^{(N)} \cdot \Phi\big)(\bs{x}_N)$
satisfies to the bounds
\beq
\Big|  \big( \La_y^{(N)} \cdot \Phi\big)(\bs{x}_N) \Big| \; \leq \; C \cdot \norm{\Phi}_{ L^{\infty}\big( \R^{N-1}, \dd^{N-1}x \big) }
\cdot \pl{n=1}{N-1} \exp \bigg\{ - \f{1}{\hbar} \ex{ \f{ x_{n+1}-x_{n} }{2} }   \bigg\}
\enq
for some $\Phi$-independent constant $C$, uniformly in $\bs{x}_N$ belonging to the domain 
\beq
\Big\{ \bs{x}_N \in \R^N  \; : \; x_1<0 \; \e{or} \; x_N >0 \Big\} \;. 
\label{definition domaine pour bonnes bornes operateur Lambda N}
\enq
\end{lemme}
In particular, the bounds guarantee an exponentially fast decay at infinity across the domain 
\eqref{definition domaine pour bonnes bornes operateur Lambda N}. 

\Proof 
Direct bounds lead to 
\beq
\Big|  \big( \La_y^{(N)} \cdot \Phi\big)(\bs{x}_N) \Big| \; \leq \;  \norm{\Phi}_{ L^{\infty}\big( \R^{N-1}, \dd^{N-1}x \big) }
\pl{n=1}{N-1} \Int{ \R }{} \exp \bigg\{ -\f{2}{\hbar} \ex{  \f{x_{n+1}-x_n}{2} }  \cosh(\tau)  \bigg\} \cdot \dd \tau \;. 
\enq
Then, the bound for $a>0$
\beq
\Int{ \R }{} \ex{ - a \cosh(\tau) } \cdot \dd \tau \; \leq \; 
\ex{-a} \Int{ \R }{}  \ex{- \f{a}{2} \tau^2 } \cdot \dd \tau    \; \leq \; \ex{-a} \sqrt{ \f{ 2\pi }{ a } }
\enq
accompanied by straightforward estimates allows one to conclude. \qed

\vspace{3mm}

The two operators $\La^{(N)}_y$  and $\ov{\La}^{(N)}_y$ satisfy an important exchange relation

\begin{lemme}
\label{Lemme exhange operateur Lambda bar Lambda}
For any $\eps_1, \eps_N >0$ one has 
\bem
\Int{\R^N}{} \ov{\La}_{ y^{\prime} }^{(N)}\big( \bs{\tau}^{\prime}_{N-1}\mid \bs{x}_N \big) 
\cdot \ex{ \f{\eps_N x_N}{\hbar} } \cdot \ex{ -\f{\eps_1 x_1}{\hbar} }  \cdot 
 \La_{ y }^{(N)}\big(  \bs{x}_N  \mid \bs{\tau}_{N-1} \big) \cdot \dd^N x  
 \; =  \; 
 \Ga\Big( \f{ y - y^{\prime} - i\eps_N }{ i\hbar } \Big) \cdot  \Ga\Big(  \f{ y^{\prime} - y - i\eps_1 }{ i\hbar } \Big) \\
\times \bigg( \f{ \ex{\tau_1}+\ex{\tau_1^{\prime}} }{\hbar}  \bigg)^{ -\f{\eps_1}{\hbar}} 
\cdot \bigg( \f{ \ex{-\tau_{N-1}}+\ex{-\tau_{N-1}^{\prime}} }{\hbar}  \bigg)^{ -\f{\eps_N}{\hbar}} 
 \; \Big( \La_{ y }^{(N-1)} \cdot \ov{\La}_{ y^{\prime} }^{(N-1)}  \Big)\big( \bs{\tau}^{\prime}_{N-1}\mid \bs{\tau}_{N-1} \big) 
\;. 
\label{equation echange entre Lambdas}
\end{multline}
\end{lemme}
Note that the $\eps_1, \eps_N$ dependent factors in the integrand are there so as to ensure 
the convergence of the  integral.

\Proof
The proof goes through a direct calculation. Namely, denote by $\mc{I}_N\big( \bs{\tau}^{\prime}_{N-1}\mid \bs{\tau}_{N-1} \big) $
the \textit{lhs} of \eqref{equation echange entre Lambdas}. Then, 
\beq
\mc{I}_N\big( \bs{\tau}^{\prime}_{N-1}\mid \bs{\tau}_{N-1} \big)
\; =\;   I_1^{(\eps_1)}(\tau_1^{\prime},\tau_1) 
\cdot \mc{I}_{2,\dots, N-1}(\bs{\tau}_{N-1}^{\prime}, \bs{\tau}_{N-1}) 
\cdot I_N^{(\eps_N)}(\tau_{N-1}^{\prime},\tau_{N-1})  .
\enq
in which:
\bem
 I_1^{(\eps_1)}(\tau_1^{\prime},\tau_1)  \; = \; \ex{i \f{ y^{\prime} \tau_1^{\prime} }{2\hbar}  } 
\ex{-i \f{ y \tau_1 }{2\hbar}  } 
 \Int{ \R }{}  \ex{ \f{i x}{\hbar}(y-y^{\prime} + i\eps_1) } 
\exp \Big\{ -\f{1}{\hbar} \ex{-x}  \cdot \big( \ex{\tau_1}+\ex{\tau_1^{\prime}} \big) \Big\} \cdot  \dd x \\
 \; = \; \ex{-i \f{ y^{\prime} \tau_1 }{2\hbar}  }  \ex{i \f{ y \tau_1^{\prime} }{2\hbar}  } 
\cdot \Ga\Big( \f{ y^{\prime} - y - i\eps_1 }{ i\hbar } \Big) \cdot 
\bigg( \f{ \ex{\tau_1}+\ex{\tau_1^{\prime}} }{\hbar}  \bigg)^{ - \f{ \eps_1  }{ \hbar } } 
\cdot \bigg[ \f{2}{\hbar} \cosh\Big(\f{\tau_1-\tau_1^{\prime} }{2}\Big)  \bigg]^{ \f{i}{\hbar}(y-y^{\prime}) }\;,
\nonumber
\end{multline}
as follow from implementing the change of variables $x = - \ln \pa{ t\hbar}  + \ln(\ex{z_1} + \ex{z^{\prime}_1}) $. 
Very similarly, one obtains 
\bem
I_N^{(\eps_N)}(\tau_{N-1}^{\prime},\tau_{N-1}) \; = \; 
\ex{-i \f{ y^{\prime}_{N-1} \tau_{N-1} }{2\hbar}  }  \ex{i \f{ y \tau_{N-1}^{\prime} }{2\hbar}  } 
\cdot \Ga\Big( \f{ y - y^{\prime} - i\eps_N }{ i\hbar } \Big) \cdot 
\bigg( \f{ \ex{-\tau_{N-1}}+\ex{-\tau_{N-1}^{\prime}} }{\hbar}  \bigg)^{ - \f{ \eps_N  }{ \hbar } } 
\cdot \bigg[ \f{2}{\hbar} \cosh\Big(\f{\tau_{N-1}-\tau_{N-1}^{\prime} }{2}\Big)  \bigg]^{ \f{i}{\hbar}(y^{\prime}-y) } \;.
\noindent
\end{multline}

Finally, using the identity 
\bem
\Int{\R}{}  V_{\la;+}\big( x_{p+1}-y \big)  V_{\la;-}\big( x_p-y \big) 
V_{\mu;+}\big( y - x^{\prime}_p \big)  V_{\mu;-}\big( y - x^{\prime}_{p+1} \big) \cdot \dd y \\ 
= \Bigg( \f{\cosh\big[ (x_{p}-x_{p}^{\prime})/2 \big] }
		 { \cosh\big[(x_{p+1}^{}-x_{p+1}^{\prime})/2 \big] } \Bigg)^{\f{i}{\hbar} (\la-\mu) } 
%
%
\Int{\R}{}  V_{\mu;+}\big( x_{p+1}-y \big)  V_{\mu;-}\big( x_p-y \big) 
V_{\la;+}\big( y - x^{\prime}_p \big)  V_{\la;-}\big( y - x^{\prime}_{p+1} \big) \cdot  \dd y 		 \; . 
\label{formule echange produit quatre fonction V}
\end{multline}
which is obtained through the change of variables 
\beq
y = -y^{\prime} -  \ln \paf{ \ex{-x_{p}} + \ex{-x^{\prime}_{p}}  }{ \ex{x_{p+1}}+\ex{x^{\prime}_{p+1}} }  \; , 
\enq
one gets 
\bem
\mc{I}_{2,\dots, N-1}(\bs{\tau}_{N-1}^{\prime}, \bs{\tau}_{N-1}) \; = \; 
\pl{p=2}{N-1} \bigg\{  \Int{\R}{}  V_{-y;+}(\tau_p-x)  V_{-y;-}(\tau_{p-1}-x) 
V_{-y^{\prime};-}(x-\tau^{\prime}_p) V_{-y^{\prime};+}(x-\tau^{\prime}_{p-1}) \cdot \dd x \bigg\}  \\
\; = \; \paf{ \cosh[(\tau_{N-1}-\tau_{N-1}^{\prime})/2] }{ \cosh[(\tau_{1}-\tau_{1}^{\prime})/2] }^{\f{i}{\hbar}(y-y^{\prime})}  
 \ex{  \f{i y^{\prime}}{2\hbar} (\tau_1+\tau_{N-1})} 
  \ex{- \f{i y}{2\hbar}(\tau_1^{\prime} + \tau_{N-1}^{\prime}) }
  \Big( \La^{(N-1)}_y \cdot \ov{\La}^{(N-1)}_{y^{\prime}} \Big) ( \bs{\tau}^{\prime}_{N-1} , \bs{\tau}_{N-1} ) \;. 
\label{definition integrale I2-N-1 pour lambda bar lambda}
\end{multline}
It remains to put all the three results together.  \qed

\vspace{3mm} 

It is also readily checked that the operators $\La_y^{(N)}$ commute in the sense that 

\begin{prop}
The operators $\La_{\la}^{(N)}$ satisfy to the commutation relations 
$\La_{\la}^{(N)} \La_{\mu}^{(N-1)} = \La_{\mu}^{(N)} \La_{\la}^{(N-1)}  $ which, in coordinates, reads 
\beq
\Int{\R^{N-1}}{}  \La_{\la}^{(N)} (\bs{x}_N \mid \bs{z}_{N-1}) \cdot  
\La_{\mu}^{(N-1)} (\bs{z}_{N-1} \mid \bs{x}^{\prime}_{N-2}) \cdot  \pl{a=1}{N-1}\dd z_a  
\; =  \; 
\Int{\R^{N-1}}{} 
\La_{\mu}^{(N)} (\bs{x}_N \mid \bs{z}_{N-1}) \cdot  \La_{\la}^{(N-1)}(\bs{z}_{N-1} \mid \bs{x}^{\prime}_{N-2}) 
\cdot  \pl{a=1}{N-1}\dd z_a \;. 
\enq

\end{prop}

Just as for the previous proposition, the proof goes through a direct calculation. We leave the details to the 
interested reader.

A repetitive application of proposition \ref{Proposition relation de recurrence fcts vp}
shows that the integral kernel $\vp_{\bs{y}_N}(\bs{x}_N)$ admits the alternative representation 
given by a multiple "pyramidal" action of the $\La$ operators
\beq
\vp_{\bs{y}_N}(\bs{x}_N) \; = \; \big( \La^{(N)}_{y_1} \dots  \La^{(1)}_{y_N} \big)(\bs{x}_N)   \; ,  
\label{ecriture forme pyramidale fct vp}
\enq
what corresponds precisely to the type of representations that have been developed for the SoV transform's kernel in  \cite{DerkachovKorchemskyManashovXXXSoVandQopNewConstEigenfctsBOp,SilantyevScalarProductFormulaTodaChain}. 
The representation \eqref{ecriture forme pyramidale fct vp} reads, in coordinates,
\beq
\vp_{\bs{y}_N}(\bs{x}_N ) \; = \;  \Int{ \R^{ \f{N(N-1)}{2}} }{} 
\La^{(N)}_{y_1}(\bs{x}_N \mid \bs{z}^{(1)}_{N-1}) \cdot  \La^{(N-1)}_{y_2}(\bs{z}^{(1)}_{N-1} \mid \bs{z}^{(2)}_{N-2} ) \, \dots  \, 
\La^{(1)}_{y_N}(z_{1}^{(N-1)} \mid - )  \cdot \pl{k=1}{N-1} \dd \bs{z}_k^{(N-k)} \;.
\label{definition noyau integral non trivial transfo U}
\enq
There, the $-$ in the argument of $\La^{(1)}_{y_N}(z_{1}^{(N-1)} \mid - )$  stresses that there is no dependence on the second type 
variable in this case. Furthermore, in \eqref{definition noyau integral non trivial transfo U}, 
the integrations ought to be considered, a priori, starting from $z_1^{(N-1)}$
and then gradually going up to the "exterior" vector $\bs{z}_{N-1}^{(1)}$.




\section{ Isometric nature of the $\mc{U}_N$ transform}
\label{Section Isometric character of UN}

Following the approach of \cite{DerkachovKorchemskyManashovXXXSoVandQopNewConstEigenfctsBOp,SilantyevScalarProductFormulaTodaChain}, we 
prove that the SoV transform $\mc{U}_N$ raises to an isometric map and, as such, is invertible from the left, the inverse being given by 
its formal adjoint $\mc{U}^{\dagger}_N $, \textit{viz} $ \mc{U}_N \cdot \mc{U}^{\dagger}_N \, = \, 
\e{id}_{L^{2}_{\e{sym}}\big(\R^N, \dd\mu(\bs{y}_N) \big)}$.  
We do stress that, in the course of the proof, we tackle the 
subtle problem of the exchangeability of various symbols which has not been considered in the aforecited papers.  
The corresponding result reads 
\begin{theorem}
\label{Theorem completude en x de la transfo SoV}
The map $\mc{U}_N$ defined for $\big(L^1\cap L^2\big)_{\e{sym}}\big(\R^N, \dd\mu(\bs{y}_N) \big)$ functions by the integral
transform \eqref{definition noyau $U$ sur un espace dense} extends into an isometric linear map 
\beq
 \mc{U}_N \; : \;  L^{2}_{\e{sym}} \big( \R^N, \dd^N \mu(\bs{y}_N) \big) \;  \tend \;   L^{2}\big( \R^N, \dd^N x \big)  \;. 
\enq
In other words, one has the equality 
\beq
\norm{ \mc{U}_N[F] }_{ L^{2}\big( \R^N, \dd^N x\big) }   \; = \; 
\norm{ F }_{ L^{2}_{\e{sym}}\big( \R^N,\dd^N \mu(\bs{y}_N) \big) } \;. 
\enq

\end{theorem}

\noindent Written formally, the isometric character of $\mc{U}_N$ translates itself into the orthogonality relations
\eqref{ecriture relation completude par rapport espace originel}. 

\vspace{2mm}

The idea of the proof consists in using the pyramidal structure \eqref{ecriture forme pyramidale fct vp}
of the kernels $\vp_{\bs{y}_N}(  \bs{x}_N )$
and the algebraic relations satisfied by the kernels $\ov{\La}^{(N)}_{y^{\prime}}$ and $\La^{(N)}_{y}$
as stated in lemma \ref{Lemme exhange operateur Lambda bar Lambda}. The subtle point here is that the algebraic exchange 
relations, in the $\eps_1, \eps_N \tend 0^+$ limit, 
 are only satisfied in a weak sense. Thus, one has to recourse to various regularizing steps
so as to justify the formal argument. 
We remind that the formal method goes back to the works of 
Derkachov, Manashov and Korchemsky \cite{DerkachovKorchemskyManashovXXXSoVandQopNewConstEigenfctsBOp}
 and has been applied by Silantev \cite{SilantyevScalarProductFormulaTodaChain} to the case of the Toda chain.

\Proof

Take $\eps>0$ and let $\Psi, \Phi\in \msc{C}_{\e{c};\e{sym} }^{\infty} \big(\mc{D}_{\eps} \big)$ where 
\beq
\mc{D}_{\eps} \; = \; \Big\{ \bs{y}_N \in \R^N \; : \; \min_{ \substack{ a ,b \\ a \not= b}  } |y_a-y_b| \; \geq \; 7 \eps   \Big\} \;, 
\enq
and $\msc{C}_{\e{c};\e{sym} }^{\infty} \big(\mc{D}_{\eps} \big)$  refers to smooth compactly supported functions
$\Psi(\bs{y}_N)$ on $\mc{D}_{\eps}$ that are, furthermore, symmetric in $\bs{y}_N$. 
Proposition \ref{Proposition charactere de transfo UN} ensures that
 $\mc{U}_N[\Phi]$ and $\mc{U}_N[\Psi]$ are both well-defined and belong to $\mc{S}\big( \R^N \big)$. 
In particular,  the scalar product 
\beq
P\; = \; N! \cdot \Int{ \R^N }{} \Big( \mc{U}_N[\Psi](\bs{x}_N) \Big)^* \;  \mc{U}_N[\Phi](\bs{x}_N)  \cdot \dd^N x 
\enq
is well defined. Let $\rho_{\eps} \in \msc{C}^{\infty}_{\e{c}}\big( \R \big)$ be such that 
\beq
0 \, \leq \, \rho_{\eps} \, \leq \, 1 \qquad 
\e{supp} \big( \rho_{\eps} \big) \; \subset \;  \intoo{-2\eps}{2\eps} \qquad \e{and} \qquad 
{\rho_{\eps}}_{ \mid \intff{-\eps}{\eps} } \; = \; 1 \;.  
\enq
Then introduce the function
\beq
\mf{n}_{\eps}\big( \bs{y}_N , \bs{y}^{\prime}_N \big) \; = \; \sul{ \sg \in \mf{S}_N }{}  
\pl{ \substack{ a, b   =1 \\  a \not= b  } }{ N } \Big( 1 \, - \,  \rho_{\eps}\big( y_a - y^{\prime}_{\sg(b)} \big)  \Big) \;. 
\enq
It is readily seen that, for any $(\bs{y}_N, \bs{y}_N^{\prime} ) \in \mc{D}_{\eps} \times \mc{D}_{\eps}$,
one has $\mf{n}_{\eps}\big( \bs{y}_N , \bs{y}^{\prime}_N \big) \geq 1$. 
Hence, the functions 
\beq
\varpi_{\sg}\big( \bs{y}_N , \bs{y}^{\prime}_N \big) \; = \; 
\f{1}{ \mf{n}_{\eps}\big( \bs{y}_N , \bs{y}^{\prime}_N \big) } \cdot 
\pl{ \substack{ a, b   =1 \\  a \not= b  } }{ N } \Big( 1 \, - \,  \rho_{\eps}\big( y_a - y^{\prime}_{\sg(b)} \big)  \Big)\qquad 
\e{with} \quad \sul{\sg \in \mf{S}_N  }{} \varpi_{\sg}\big( \bs{y}_N , \bs{y}^{\prime}_N \big)  \; = \; 1
\enq
provide one with a partition of unity on $\big( \bs{y}_N , \bs{y}^{\prime}_N \big) \; \in \;  \mc{D}_{\eps} \times \mc{D}_{\eps}$. 
As a consequence, one can recast the scalar product as 
\beq
P \; = \; \sul{ \sg \in \mf{S}_N }{} P_{\sg}  \qquad \e{with} \qquad  P_{\sg}= \Int{\R^N}{} \mc{I}_{\sg}(\bs{x}_N) \cdot \dd^N x
\enq
 and
\beq
\mc{I}_{\sg}(\bs{x}_N) \; = \;   \;  \Int{\R^N\times \R^N  }{} 
\hspace{-2mm} \vp_{\bs{w}_N}^{*}(\bs{x}_N) 
\cdot \vp_{\bs{y}_N}(\bs{x}_N) \cdot \Psi \big( \bs{w}_N \big)  \cdot \Phi \big( \bs{y}_N \big)  
\cdot \varpi_{\id}\big( \bs{y}_N , \bs{w}_N \big)  \cdot 
 \dd\mu\big( \bs{y}_N \big) \cdot \dd\mu\big( \bs{w}_N \big)
\enq
in which we agree upon $ \bs{w}_N =\big( y^{\prime}_{\sg(1)}, \dots, y^{\prime}_{\sg(N)} \big)$. 
Note that each integral over $\bs{x}_N$ does converges: the functions $\mc{I}_{\sg}(\bs{x}_N)$ all belong to the Schwartz class
 as can be seen by repeating the arguments that led to proposition \ref{Proposition charactere de transfo UN}. 

\vspace{1mm} 

We henceforth focus on the analysis of $P_{\sg}$. For this purpose, we introduce 
\bem
\msc{U}_{N-1}\Big[  \Psi \cdot \varpi_{\e{id}}\cdot  \Phi \Big]\big(y_1, w_N\mid \bs{\tau}_{N-1}, \bs{\tau}_{N-1}^{\prime} \big) \\
\; = \; 
\Int{ \R^{N-1}  }{} \pl{a=1}{N-1}\dd w_a  \hspace{-1mm} \Int{ \R^{N-1}  }{} \hspace{-1mm} \pl{a=2}{N}\dd y_a  
\, \cdot  \, \vp_{\bs{y}_N^{(2)}} (\bs{\tau}_{N-1} ) \cdot \vp_{\bs{w}_{N-1}}^{*} (\bs{\tau}_{N-1}^{\prime} ) \cdot  \Phi(\bs{y}_N)
\Psi(\bs{w}_N) \varpi_{\e{id}}\big( \bs{y}_N,  \bs{w}_N\big) \cdot \mu\big( \bs{y}_N \big) \cdot 
\mu\big( \bs{w}_{N} \big) \;,
\end{multline}
where we agree upon 
\beq
\bs{y}_N^{(k)} \; = \; \big( y_k, \dots ,y_N \big)  \; \in \; \R^{N+1-k}  \;. 
\label{definition variables yN ecoutes} 
\enq
By repeating the arguments that led to propositon \ref{Proposition charactere de transfo UN}, one gets 
that, uniformly in $y_1, w_N$ belonging to any fixed compact in $\R$, 
$\msc{U}_{N-1}\big[  \Psi\varpi_{\e{id}} \Phi \big]\big(y_1, w_N\mid \tau_{N-1}, \tau_{N-1}^{\prime} \big) $ is of Schwartz
class in $ \big( \tau_{N-1}, \tau_{N-1}^{\prime} \big)$. 
A straightforward application of Fubbini's theorem then yields
\beq
\mc{I}_{\sg}(\bs{x}_N) \; = \;   \;  \Int{\R^N\times \R^N  }{}
f_{0}\big( \bs{x}_N,\bs{\tau}_{N-1},\bs{\tau}_{N-1}^{\prime},y_1,w_N  \big)  \cdot 
\dd^{N-1}\tau \cdot \dd^{N-1}\tau^{\prime} \cdot \dd y_1 \cdot \dd w_N  
\enq
where 
\beq
f_{\nu}\big( \bs{x}_N,\bs{\tau}_{N-1},\bs{\tau}_{N-1}^{\prime},y_1,w_N  \big) \; = \; 
 \ex{\f{\nu}{\hbar}(x_N-x_1)  } \cdot \ov{\La}_{w_N}^{(N)}\big( \bs{\tau}_{N-1}^{\prime} \mid \bs{x}_N \big) 
\cdot \La_{y_1}^{(N)}\big(  \bs{x}_N \mid  \bs{\tau}_{N-1} \big)
\cdot \msc{U}_{N-1}\Big[  \Psi\varpi_{\e{id}} \Phi \Big]\big(y_1, w_N\mid \tau_{N-1}, \tau_{N-1}^{\prime} \big) \;. 
\nonumber
\enq

Since $\msc{U}_{N-1}\big[  \Psi \varpi_{\e{id}}  \Phi \big]$ is smooth and compactly supported in the two variables 
$(y_1, w_N)$ and bounded in the variables $\big(\bs{\tau}_{N-1}, \bs{\tau}_{N-1}^{\prime} \big)$, it follows
from lemma \ref{Lemme bornage transfo La a droite}
that the function 
\beq
\bs{x}_N \;  \mapsto \; \mc{I}_{\sg}(\bs{x}_N)
\enq
is bounded by a function decreasing faster than an exponential on the domain
\beq
\msc{D} \; = \; \Big\{ \bs{x}_N \in \R^N  \; : \; x_1 \leq 0 \; \e{or} \; x_N  \geq 0 \Big\}
\enq
Furthermore, since $ \mc{I}_{\sg} \in \mc{S}\big( \R^N \big)$, one gets that
 on $\R^N \setminus \msc{D} $,  given $\nu \geq 0$, 
\beq
\Big| \ex{\f{\nu}{\hbar}(x_N-x_1)  }  \mc{I}_{\sg}(\bs{x}_N)  \Big| \; \leq \; C \cdot 
\pl{a=1}{N} \bigg( \f{ 1 }{ 1+ |x_a| }  \bigg)^2 \; .  
\label{ecriture bornes exponentielles Isgima}
\enq
By the sur-exponential bounds on $\msc{D}$, \eqref{ecriture bornes exponentielles Isgima} holds, in fact, 
on the whole of $\R^N$. As a consequence, by dominated convergence,  
\beq
P_{\sg} \; = \; \lim_{\nu\tend 0^+} \Int{\R^N }{} \dd^N x 
\cdot \bigg\{ \Int{}{} 
f_{\nu}\big( \bs{x}_N,\bs{\tau}_{N-1},\bs{\tau}_{N-1}^{\prime},y_1,w_N  \big)  \cdot 
\dd^{N-1}\tau \cdot \dd^{N-1}\tau^{\prime} \cdot \dd y_1 \cdot \dd w_N  \bigg\} \;. 
\label{ecriture P sigma comme limit nu vers 0 avant echange ordre integral}
\enq

The $\nu$-regularization allows one to be in position so as to apply Fubbini's theorem
and take the $\bs{x}_N$-integration first. Indeed, 
it is readily seen that $| f_{\nu } | \leq g_{\nu}$ where, for some constant $C>0$, 
\bem
g_{\nu}
%
 \;=  \; C \cdot 
\f{  \bs{1}_K(y_1)\cdot  \bs{1}_K(w_N)  \cdot  \ex{ \f{ \nu }{ \hbar } (x_N -  x_1) } }
{ \pl{a=1}{N-1} \big( 1 + |\tau_a| \big)^3 \cdot\big( 1 + |\tau_a^{\prime}| \big)^3 } \cdot 
 \exp\bigg\{ -\f{ \ex{-x_1} }{ \hbar} (\ex{\tau_1}+\ex{\tau_1^{\prime}} ) 
 -\f{ \ex{x_N} }{ \hbar} (\ex{-\tau_{N-1} } \, + \,  \ex{-\tau_{N-1}^{\prime}} ) \bigg\}  \\
\times \pl{n=2}{N-1} \exp \bigg\{ - \f{2}{\hbar} \sqrt{ \big( \ex{-\tau_{n-1}}+ \ex{-\tau_{n-1}^{\prime} } \big) \cdot 
\big( \ex{\tau_{n}}+ \ex{\tau_{n}^{\prime} } \big) }  \cdot 
\cosh\big[ x_a -s_n(\bs{\tau}_{N-1}, \bs{\tau}^{\prime}_{N-1}  )  \big]  \bigg\}
\end{multline}
in which
\beq
s_n(\bs{\tau}_{N-1}, \bs{\tau}^{\prime}_{N-1}  ) \; = \; -\f{1}{2} 
\ln \bigg[ \f{ \ex{-\tau_{n-1}}+ \ex{-\tau_{n-1}^{\prime} } }{ \ex{\tau_{n}}+ \ex{\tau_{n}^{\prime} } } \bigg]
\enq
and $K$ denotes a compact such that $\e{supp}(\Psi) \cup \e{supp}(\Phi) \subset K^N$. 
The positive function $ g_{\nu} $  is readily seen to fullfill
\bem
\Int{ \R^N }{} g_{\nu}\big( \bs{x}_N,\bs{\tau}_{N-1},\bs{\tau}_{N-1}^{\prime},y_1,w_N  \big)  \cdot \dd^N x \; \leq \; 
 C \cdot 
\f{  \bs{1}_K(y_1)\cdot  \bs{1}_K(w_N)  }
{ \pl{a=1}{N-1} \Big\{  \big( 1 + |\tau_a| \big)^2 \cdot\big( 1 + |\tau_a^{\prime}| \big)^2 \Big\}  } 
 \cdot \Ga^2\big(\tf{ \nu }{\hbar}  \big)  \\
\times \f{ \pl{n=2}{N-1}\bigg\{ 
\exp \Big\{ - \f{2}{\hbar} \sqrt{ \big( \ex{-\tau_{n-1}}+ \ex{-\tau_{n-1}^{\prime} } \big) \cdot 
\big( \ex{ \tau_{n} } +  \ex{ \tau_{n}^{\prime} } \big) }   \,  \Big\}   
\cdot
\Big( 1+ \big| \ln \big( \ex{-\tau_{n-1}}+ \ex{-\tau_{n-1}^{\prime} } \big) \cdot 
\big( \ex{ \tau_{n} } +  \ex{ \tau_{n}^{\prime} } \big)  \big| \Big) \bigg\} }
{\Big( \big(\ex{\tau_1} + \ex{\tau_1^{\prime}} \big) \big(\ex{-\tau_{N-1}} + \ex{-\tau_{N-1}^{\prime}} \big) \Big)^{\f{\nu}{\hbar}}
\cdot  \pl{a=1}{N-1} \Big\{ \big( 1 + |\tau_a| \big) \cdot\big( 1 + |\tau_a^{\prime}| \big)  \Big\}  }
\label{equation bornage integrale g nu pour Fubbini}
\end{multline}
what follows from the bound 
\beq
\Int{ \R }{} \ex{-a \cosh(\tau) } \cdot \dd \tau  \; \leq \; C^{ \prime} \big( 1 \, + \,  | \ln(a) |  \big) \cdot \ex{-a} \qquad 
\e{for} \; \e{some} \quad C^{\prime} >0 \;. 
\enq
It is readily seen that the second line of \eqref{equation bornage integrale g nu pour Fubbini} is a bounded function of 
$\big( \bs{\tau}_{N-1},\bs{\tau}_{N-1}^{\prime} \big)\in \R^{N-1} \times \R^{N-1}$. 
Hence, 
\beq
\Int{\R^{2N} }{}  \dd^{N-1}\tau \cdot \dd^{N-1}\tau^{\prime} \cdot \dd y_1 \cdot \dd w_N \cdot 
\bigg\{ \Int{ \R^N }{} g_{\nu}\big( \bs{x}_N,\bs{\tau}_{N-1},\bs{\tau}_{N-1}^{\prime},y_1,w_N  \big)  \cdot \dd^N x
\bigg\} \; < \; + \infty \;. 
\enq
In virtue of Fubbini's theorem, $g_{\nu} \in L^{1}$ and thus $f_{\nu}$ as well. Therefore
one can change the orders of integration in \eqref{ecriture P sigma comme limit nu vers 0 avant echange ordre integral}
and compute the $\bs{x}_N$-integration first. The latter can then be taken by means of
 lemma \ref{Lemme exhange operateur Lambda bar Lambda},
thus leading to 
\bem
P_{\sg} \; = \; \lim_{\nu \tend 0^+} \Int{ \R^{2N} }{} 
 \Ga\Big( \f{y_1-w_N - i\nu}{i\hbar} \Big)\,  \Ga\Big( \f{w_N-y_1 - i\nu}{i\hbar} \Big) \cdot 
\f{ \Big( \La_{y_1}^{(N-1)}\cdot \ov{\La}_{w_N}^{(N-1)}\Big) \big( \bs{\tau}_{N-1}^{\prime} \mid \bs{\tau}_{N-1} \big) }
{ \Big( \big(\ex{\tau_1} + \ex{\tau_1^{\prime}} \big) \big(\ex{-\tau_{N-1}} + \ex{-\tau_{N-1}^{\prime}} \big) \Big)^{\f{\nu}{\hbar}} }
\cdot \hbar^{ \f{\nu}{\hbar} } \\
\times \msc{U}_{N-1}\Big[  \Psi \cdot \varpi_{\e{id}} \cdot \Phi \Big]\big(y_1, w_N\mid \tau_{N-1}, \tau_{N-1}^{\prime} \big)
\cdot \dd^{N-1}\tau \cdot \dd^{N-1}\tau^{\prime} \cdot \dd y_1 \cdot \dd w_N  \;. 
\end{multline}
One can take the limit under the integral sign by applying the dominated convergence theorem. 
Indeed, the function $\varpi_{\e{id}}$ ensures that the $\Ga$-functions are uniformly bounded
on the support of integrand whereas the remaining part of the integrand can be bounded
analogously to \eqref{equation bornage integrale g nu pour Fubbini}.

After a straightforward exchange of the order of integration, one is led to the representation 
\beq
P_{\sg} \; =  \;\Int{\R}{}  \dd w_N   \Int{ \R^{N-1} }{} \hspace{-1mm} \dd^{N-1}\tau \cdot  \Int{\R^{N-2}}{}\hspace{-2mm} 
\dd^{N-2}\xi  \, \cdot \, \ov{\La}_{w_N}^{(N-1)}  \big( \bs{\xi}_{N-2} \mid  \bs{\tau}_{N-1} \big) 
\, \cdot  \, 
\bigg\{  \Int{\R^{N-1}}{}  \dd\mu\big( \bs{y}_N^{(2)} \big) \cdot \vp_{\bs{y}_N^{(2)}}\big( \bs{\tau}_{N-1} \big) 
\cdot 
\Ups_{1}\big( w_N, \bs{y}_N^{(2)} \mid \bs{\xi}_{N-2} \big)  \bigg\} \;, 
\enq
where 
\bem
\Ups_{1}\big( w_N, \bs{y}_N^{(2)}\mid \bs{\xi}_{N-2} \big) \; = \; 
\Int{ \R }{}\hspace{-1mm} \f{ \dd y_1 }{ (2\pi \hbar) }\ga\big( y_1-w_N \big) \hspace{-2mm}
 \Int{\R^{N-1}}{} \hspace{-2mm} \dd^{N-1} \tau^{\prime}
\; \La_{y_1}^{(N)}\big( \bs{\tau}_{N-1}^{\prime} \mid \bs{\xi}_{N-2} \big) \\
\times  \Int{ \R^{N-1} }{}\hspace{-2mm} \dd^{N-1} w \cdot
\f{ \Phi(\bs{y}_N) \cdot  \varpi_{\e{id}}\big( \bs{y}_N, \bs{w}_{N} \big) \cdot \Psi\big( \bs{w}_N \big) }
{ \pl{a=2}{N}  \ga\big(y_a-y_{1}  \big) } 
\mu\big(\bs{w}_N\big)  \vp_{ \bs{w}_{N-1} }^{*} \big( \bs{\tau}^{\prime}_{N-1} \big)  \;. 
\end{multline}
Above, we have introduced 
\beq
\ga(x) \; = \; \Ga \Big( \f{x}{i\hbar} , - \f{x}{i\hbar} \Big)  \qquad \e{with} \qquad 
\Ga(x,y) \; = \;  \Ga(x) \cdot \Ga(y) \;. 
\enq
It follows from lemma \ref{Proposition character Schwartz des transformations Lambda a droite et gauche} that 
$\Ups_{1}\big( w_N, \bs{y}_N^{(2)}\mid \bs{\xi}_{N-2} \big)$ is smooth and compactly supported in respect 
to the first set of variables $w_N, \bs{y}_N^{(2)}$  and of Schwartz class in respect to the second 
argument $\bs{\xi}_{N-2}$, this uniformly in $w_N, \bs{y}_N^{(2)}$.

\vspace{2mm}

The remainder of the proof goes by induction. One defines a sequence of functions
\beq
\Ups_{k}\big( w_N, \bs{y}_N^{(k+1)}\mid \bs{\xi}_{N-k-1} \big) \; = \; 
\Int{ \R }{}\hspace{-1mm} \f{ \dd y_k }{ 2\pi \hbar } \ga\big( y_k-w_N \big) \hspace{-2mm}
 \Int{\R^{N-k}}{} \hspace{-2mm} 
\f{ \La_{y_k}^{(N-k)}\big( \bs{\tau}_{N-k}^{\prime} \mid \bs{\xi}_{N-k-1} \big)  \cdot 
\Ups_{k-1}\big( w_N, \bs{y}_N^{(k)}\mid \bs{\tau}_{N-k}^{\prime} \big)  }
{  \pl{a=k}{N}  \ga\big( y_a-y_{k+1} \big)    }  \cdot 
\dd^{N-k} \tau^{\prime} \;. 
\enq
Then, the induction hypothesis claims that $P_{\sg}$ can be recast as 
\bem
P_{\sg} \; = \; \Int{ \R }{}\dd w_N  \Int{ \R^{N-k} }{} \dd^{N-k} \tau  \Int{ \R^{N-k-1} }{}\dd^{N-k-1}\xi
\cdot \ov{\La}_{w_N}^{(N-k)}\big( \xi_{N-k-1} \mid \tau_{N-k}  \big)     \\
\times \Int{ \R^{N-k-1} }{} 
\vp_{  \bs{y}_N^{(k+1)} }\big( \bs{\tau}_{N-k}  \big)  \cdot \mu\big( \bs{y}_N^{(k+1)} \big) \cdot 
\Ups_{k}\big( w_N, \bs{y}_N^{(k+1)}\mid \bs{\xi}_{N-k-1} \big)
\pl{a=k+1}{N} \dd y_a \;.
\end{multline}

In which the functions $\Ups_{k}\big( w_N, \bs{y}_N^{(k+1)}\mid \bs{\xi}_{N-k-1} \big)$
are smooth and compactly supported in respect to the first set of variables $ w_N, \bs{y}_N^{(k+1)}$
and of Schwartz class in the second ones $\bs{\xi}_{N-k-1}$. 
The properties of the functions $\Ups_{k}$ are a consequence of 
lemma \ref{Proposition character Schwartz des transformations Lambda a droite et gauche}. 
The remaining contents of the induction can be established with the help of bounds quite similar to those used in the
"initialisation" part of the proof. The details are left to the reader. 

\vspace{2mm}
All in all, upon $k=N-1$ successive iterations, one gets 
\beq
P_{\sg} \; = \; \Int{}{}\dd w_N  \Int{ \R }{} \dd \tau  
\ov{\La}_{w_N}^{(N-k)}\big(- \mid \tau  \big)    \cdot \Int{ \R }{} 
\La_{y_N}^{(N-k)}\big( \tau \mid - \big) \cdot 
\Ups_{N-1}\big( w_N, y_N \mid - \big)  \cdot \f{ \dd y_N }{ 2\pi \hbar }  
\; = \; \Int{ \R }{}  \Ups_{N-1}\big( w_N, w_N \mid - \big)  \cdot \dd w_N   \;.
\nonumber
\enq
Then, by tracing backwards the chains of transformations, we are led to the representation 
\bem
P_{\sg} \; = \; \Int{ \R }{} \dd w_N \hspace{-2mm} \Int{ \R^{N-1} }{} \hspace{-2mm} \dd^{N-1}\tau \; \bigg\{ \Int{ \R^{N-1} }{} \hspace{-2mm} \dd^{N-1} w
\Int{ \R^{N-1} }{} \hspace{-2mm} \dd^{N-1} y  \cdot \pl{a=1}{N-1}\Big\{ \ga\big(y_a-w_N\big)  \Big\} 
\cdot  \varpi_{\e{id}}\Big( (\bs{y}_{N-1}, w_N), \bs{w}_N  \Big) \\
\times \Phi\big( (\bs{y}_{N-1}, w_N) \big) \cdot \Psi\big( \bs{w}_N \big) 
\cdot \vp^{*}_{\bs{w}_{N-1}}\big( \bs{\tau}_{N-1} \big) \cdot \vp_{\bs{y}_{N-1}}\big( \bs{\tau}_{N-1} \big) 
\cdot \mu\big( (\bs{y}_{N-1}, w_N) \big) \cdot \mu\big( \bs{w}_N \big) \bigg\} \;. 
\label{ecriture Psigma apres une etape de reduction}
\end{multline}
Having established \eqref{ecriture Psigma apres une etape de reduction}, it takes a straightforward induction to get 
\beq
P_{\sg} \; = \; \Int{ \R^N }{} \varpi_{\e{id}}\big( \bs{y}_{N}^{\prime}, \bs{y}_N^{\prime}  \big)
\cdot \Phi\big( \bs{y}_N^{\prime} \big) \cdot \Psi\big( \bs{y}_N^{\prime} \big) 
\cdot  \dd \mu\big( \bs{y}_N^{\prime} \big) \;. 
\label{ecriture P sg apres toute induction finie}
\enq
Above, we have recast the integrand in terms of the original variables $y_{\sg(a)}^{\prime} = w_a$. 
Note that in \eqref{ecriture P sg apres toute induction finie} one does end up with $\varpi_{\e{id}}$ instead of 
$\varpi_{\sg}$ since the first and second of its argument have \textit{both} been permuted. 

The integration in \eqref{ecriture P sg apres toute induction finie} runs, in fact, through $\mc{D}_{\eps}$.
Since, by definition, given any 
\beq
\bs{y}_{N}^{\prime} \in \mc{D}_{\eps} \qquad  \e{one} \; \e{has} \qquad  y_a^{\prime} - y_b^{\prime} \not\in 
\e{supp}\big(\rho_{\eps}\big) \; \quad \e{for} \; \e{any} \; \;  a \not= b, 
\enq
it follows that  $\varpi_{\e{id}}\big( \bs{y}_{N}^{\prime}, \bs{y}_N^{\prime}  \big) 
\; = \; \mf{n}_{\eps}^{-1}\big( \bs{y}_{N}^{\prime}, \bs{y}_N^{\prime}  \big)$.
Furthermore, going back to the definition of the function $\mf{n}_{\eps}$, it follows that 
the sole term that survives correspond to the identity permutation $\sg = \e{id}$, since, for any 
$\sg \in \mf{S}_N\setminus \{ \e{id} \}$, there exists $a \not= b $ such that $\sg(a)=b$, \textit{viz} 
$\rho_{\eps}(y_{\sg(a)}^{\prime}- y^{\prime}_b ) =1$. As a consequence, 
$\varpi_{\e{id}}\big( \bs{y}_{N}^{\prime}, \bs{y}_N^{\prime}  \big)= 1$ leading to 
\beq
N! \cdot \Int{\R^N}{}  \Big( \mc{U}_N\big[  \Psi^{*} \big](\bs{x}_N) \Big)^* \cdot  \mc{U}_N\big[  \Phi \big](\bs{x}_N) \cdot \dd^N x 
\; = \; P \; = \; N! \cdot \Int{ \mc{D}_{\eps} }{} 
\Phi\big( \bs{y}_N^{\prime} \big) \cdot \Psi\big( \bs{y}_N^{\prime} \big) 
\cdot  \dd \mu\big( \bs{y}_N^{\prime} \big) \;. 
\label{ecriture isometrie UN pour fcts nulle sur diag}
\enq

The result \eqref{ecriture isometrie UN pour fcts nulle sur diag} can be extended to functions 
$\Phi, \Psi \in \msc{C}^{\infty}_{\e{c}; \e{sym}}(\R^N)$. Indeed, let $\eta_{\eps}\in \msc{C}^{\infty}(\R^N)$ 
be such that 
\beq
0 \leq \eta_{\eps} \leq 1 \;  , \quad \e{supp}( \eta_{\eps} ) \; \subset \; \mc{D}_{\eps} \qquad \e{and} \qquad 
{ \eta_{\eps} }_{ \mid  \mc{D}_{2 \eps} } \; = \; 1  \;. 
\enq
Then, given a sequence  $\eps_M\limit{M}{+\infty}0$, the sequences of functions $\Psi_M  \; = \; \eta_{\eps_M} \cdot \Psi $, 
$\Phi_M  \; = \; \eta_{\eps_M} \cdot \Phi $ satisfy to the hypothesis previously used and are such that 
\beq
\Psi_M(\bs{y}_N)  \; \limit{M}{+\infty} \; \bs{1}_{\R^N \setminus \mc{D}_0} (\bs{y}_N) \cdot \Psi(\bs{y}_N) \; \; ,  
\qquad 
\Phi_M(\bs{y}_N)  \; \limit{M}{+\infty} \; \bs{1}_{\R^N \setminus \mc{D}_0} (\bs{y}_N) \cdot \Phi(\bs{y}_N)
\;. 
\enq
By using that $\R^N \setminus \mc{D}_0$ has zero
Lebesgue measure, it follows, 
\beq
\Int{\R^N }{} 
 \Phi_M\big( \bs{y}_N^{\prime} \big) \cdot \Psi_M\big( \bs{y}_N^{\prime} \big) 
\cdot  \dd \mu\big( \bs{y}_N^{\prime} \big)  \; \limit{M}{+\infty} \; 
\Int{ \R^N   }{} 
 \Phi\big( \bs{y}_N^{\prime} \big) \cdot \Psi\big( \bs{y}_N^{\prime} \big) 
\cdot  \dd \mu\big( \bs{y}_N^{\prime} \big) \;. 
\enq

This implies that the sequence $ \mc{U}_N\big[  \Psi^{*}_M \big] $,  resp. $ \mc{U}_N\big[  \Phi_M \big] $,  
is a Cauchy sequence in $L^{2} \big( \R^N,  \dd^N x \big)$, 
and thus converges to some function $\wt{\psi}\in L^{2} \big( \R^N,  \dd^N x \big)$, 
resp. $\wt{\phi}\in L^{2} \big( \R^N,  \dd^N x \big)$. However, since 
\beq
 \mc{U}_N\big[  \Psi^{*}_M \big] \big( \bs{x}_N \big) \; \limit{M}{+\infty} \; 
\mc{U}_N\big[  \Psi^{*} \big] \big( \bs{x}_N \big)   \qquad \e{resp}.  \quad 
 \mc{U}_N\big[  \Phi^{*}_M \big] \big( \bs{x}_N \big) \; \limit{M}{+\infty} \; 
\mc{U}_N\big[  \Phi^{*} \big] \big( \bs{x}_N \big) \;. 
\enq
uniformly in $\bs{x}_N \in \R^N $, it follows that $\wt{\psi} \; = \;  \mc{U}_N\big[  \Psi^{*} \big] $, resp. 
$\wt{\phi} \; = \;  \mc{U}_N\big[  \Phi \big] $. 
It remains to conclude by invoking the density of $\msc{C}_{\e{c}; \e{sym}}^{\infty} \big( \R^N \big) $
in $L^{2}_{\e{sym}} \Big( \R^N, \dd \mu\big( \bs{y}_N \big) \Big)$. \qed





\vspace{3mm}

\section{Isometric nature of the adjoint transform}
\label{Section Isometric Nature UN adjoint}

It is readily seen that the map $\ov{\mc{V}}_N$ defined on $L^1\big( \R^N, \dd^N x \big)$ through 
\beq
\ov{\mc{V}}_N[F](\bs{y}_N) \; = \; \f{ 1 }{ \sqrt{N!} } \Int{ \R^N }{} \Big( \vp_{\bs{y}_N}(\bs{x}_N) \Big)^* \cdot F(\bs{x}_N) \cdot \dd^{N} x \;
\enq
is a formal adjoint of $\mc{U}_N$. The purpose of the present section is to show that $\ov{\mc{V}}_N$
extends to an isometric operator $L^2\big(\R^N, \dd^N x \big) \; \tend \;  L^{2}_{\e{sym}}\big(\R^N, \dd^N\mu(\bs{y}_N) \big)$.
This will thus establish that $\mc{U}^{\dagger}_N = \ov{\mc{V}}_N$ and that $\mc{U}_N$ is a unitary map. 
In fact, since $\Big( \vp_{\bs{y}_N}(\bs{x}_N) \Big)^*  \; = \;  \vp_{-\bs{y}_N}(\bs{x}_N)  $, 
it is just as good to establish the isometricity of 
the operator $\mc{V}_N \; : \; L^2\big(\R^N, \dd^N x \big) \; \tend \;  L^{2}_{\e{sym}}\big(\R^N, \dd^N\mu(\bs{y}_N) \big)$
whose action on $\big(L^1 \cap L^2 \big)\big( \R^N, \dd^N x \big)$ is given by the integral transform
\beq
\mc{V}_N[F](\bs{y}_N) \; = \; \f{ 1 }{ \sqrt{N!} } \Int{ \R^N }{} \vp_{\bs{y}_N}(\bs{x}_N) F(\bs{x}_N) \cdot \dd^{N} x \;. 
\label{definition transformation VN}
\enq

We do stress that the method allowing us to prove the isometric nature of $\mc{V}_N$ 
has never been proposed, even on a formal level of rigour, previously. Thus the whole content of the
present section is completely new. Furthermore, from the point of view of formal manipulations,
our method is quite simple. Our proof builds on isometric properties of two auxiliary integral transforms $\msc{H}_N$ and $\msc{J}_N$. 
The isometriticity of the latter transforms is deduced from theorem \ref{Theorem completude en x de la transfo SoV}
adjointed to the Mellin-Barnes type recurrence relation satisfied by the functions $\vp_{\bs{y}_N}(\bs{x}_N)$.  
In this respect, the proof highlights a sort of duality between the isometric character of the 
operator $\mc{U}_N$ and $\mc{V}_N$. As soon as one isometry is proven,
the second one follows from it. 
 The steps of our proof do not rely on any property specifically associated with the Toda chain. Thus, it can quite probably be applied 
for proving the unitarity of SoV transforms arising in the context of other quantum integrable models solvable by the 
quantum separation of variables method.

\subsection{The $\msc{H}_N$ transform}

\begin{defin}
Let $\msc{H}_N$ be the operator defined, for $G \in \msc{C}_{\e{c};\e{sym}}^{\infty}(\R^N)$, as 
\beq
\msc{H}_N[G](\bs{w}_N) \; = \; \f{1}{  N! }  \lim_{\a \tend 0^+} \Int{ \R^N }{} 
\pl{a,b=1}{N} \bigg\{ \Ga\Big( \f{ y_b - w_a + i\a}{ i \hbar}  \Big) \bigg\}
\cdot \pl{a\not= b}{N} \bigg\{ \Ga^{-1}\Big( \f{ y_b - y_a}{ i \hbar}  \Big) \bigg\} \cdot   G(\bs{y}_N)
\f{ \dd^N y }{ (2\pi \hbar)^N   }  \;. 
\label{definition operateur HN comme integral}
\enq

\end{defin}

Prior to establishing the isometric character of the operator $\msc{H}_N$, we establish certain of its most basic properties. 

\begin{lemme}
\label{Lemme well definiteness and L2 character of H-transform}
Let $G \in \msc{C}_{\e{c}; \e{sym} }^{\infty}(\R^N)$. Then the map
\beq
\bs{w}_N \mapsto \pl{a<b}{N}(w_b-w_a) \cdot \msc{H}_N[G](\bs{w}_N) 
\enq
 is smooth and $\msc{H}_N[G] \in L^{2}_{\e{sym}}\big(\R^N, \dd \wt{\mu}(\bs{w}_N) \big)$ 
with the measure $\wt{\mu}$ being given by 
\beq
\dd \wt{\mu}(\bs{w}_N) \; = \;  \wt{\mu}(\bs{w}_N) \cdot \dd^N w \qquad \e{where} \qquad 
\wt{\mu}(\bs{w}_N) \; = \; \ex{\f{\pi}{\hbar} \ov{\bs{w}}_N } \cdot \mu(\bs{w}_N) \;. 
\label{definition mesure mu tilde}
\enq
\end{lemme}
\Proof 

We begin by showing that $\bs{w}_N \mapsto \pl{a<b}{N}(w_b-w_a) \cdot \msc{H}_N[G](\bs{w}_N)$ is smooth. 
For this purpose, we first need to define the skeleton $\Ga^{(k)}$ in $\Cx^k$:
\beq
 \Ga^{(k)}  \; = \;  \Dp{}\mc{D}_{y_1+i\a, \f{\a}{2} } \times \dots  \times \Dp{}\mc{D}_{y_k + i\a, \f{\a}{2}}  \;.  
\enq
Then, for pairwise distinct real numbers $\{w_a\}_1^N$ and $\{y_a\}_1^k$ one has that the contour integral 
%
%
\beq
\Oint{ \Ga^{(k)} }{} \pl{a=1}{k} \Big\{ \f{ 1 }{ z_a - y_a  - i \a } \Big\} 
\cdot \pl{ a > b }{ k } \big( z_a - z_b \big)  \cdot  \pl{a=1}{k} \pl{b=1}{N} \Big\{  \f{ 1 }{ z_a - w_b } \Big\}  
\cdot \f{ \dd^k z}{(2i\pi)^k}
\enq
can be estimated by taking the one dimensional residues in each variable either
in respect to the simple poles located inside of the respective contour or outside thereof. 
Choosing to compute the integral either by means of taking all residues inside or all outside of the 
contour leads to the identity
\beq
\pl{r > \ell }{k}\big( y_r - y_{\ell}  \big)  \cdot 
\pl{r=1}{k} \pl{\ell =1}{N} \bigg\{  \f{ 1 }{ y_r - w_{\ell}  + i \a } \bigg\} 
 \; = \; 
\sul{ \substack{ b_1\not= \dots \not= b_k \\ 1} }{ N }  
\pl{r=1}{k} \hspace{-2mm}  \pl{ \substack{ \ell=1 \\ \not= b_1, \dots, b_k } }{N}  \hspace{-2mm} 
\bigg\{  \f{ 1 }{ w_{b_r} - w_{\ell}  } \bigg\}  \;  \cdot \; 
 \pl{r<\ell}{k} \bigg\{  \f{ 1 }{ w_{b_r} - w_{b_{\ell} }  } \bigg\}  
 \;  \cdot \; \pl{s=1}{k} \bigg\{ \f{ 1 }{  y_s - w_{b_s} + i \a }  \bigg\} \;. 
\label{ecriture identite vers decomposition produits transfo Cauchy simples}
\enq

Implementing it at $k=N$ yields, upon invoking the symmetry of the integrand in $\bs{y}_N$, 
\beq
\pl{r < \ell }{N}\big( w_r - w_{\ell}  \big)  \cdot \msc{H}_N\big[ G \big] (\bs{w}_N) \; = \; 
%
%
\f{1}{N!} \lim_{\a \tend 0^+}  \Int{ \R^N }{ } \f{  N!  (i\hbar)^N G(\bs{y}_N)   }{ \pl{a=1}{N} (y_a - w_{a} + i \a) } 
 \cdot \pl{r < \ell }{N}\big( y_r - y_{\ell}  \big) 
 \cdot \f{ \pl{a,b=1}{N} \Ga\Big( \f{ y_b - w_a }{ i \hbar} +1 \Big)   }
 {  \pl{a\not= b}{N} \Ga\Big( \f{ y_b - y_a}{ i \hbar}  +1 \Big)  } 
\cdot \f{ \dd^N y }{ (2\pi \hbar)^N  } \;.  
\label{ecriture HN comme succession Transfos Cauchy}
\enq

Thus $\pl{r < \ell }{N}\big( w_r - w_{\ell}  \big)  \cdot \msc{H}_N\big[ G \big] (\bs{w}_N)$ 
can be recast as $+$ boundary values of $N$ one-dimensional 
Cauchy transforms. Since $G$ is smooth, it is readily seen by repeating the arguments holding for the one-dimensional 
case that 
\newline $\pl{a<b}{N}(w_b-w_a) \cdot \msc{H}_N\big[ G \big] (\bs{w}_N)$ is smooth.

\vspace{2mm}
We shall now establish the $L^{2}_{\e{sym}}\big(\R^N, \dd \wt{\mu}(\bs{w}_N) \big)$  character of $ \msc{H}_N[ G ] (\bs{w}_N) $.
This amounts to bounding the function when part of the variables goes to infinity. 
Since this function is manifestly symmetric in $\bs{w}_N$, it is enough to bound it on the domain 
 $ \{ \bs{w}_N \in \R^N \; : \; w_1 < \dots < w_N \}$. 
Let $R>0$ be such that $\e{supp}(G)$, the support of $G$, verifies $\e{supp}(G) \subset \intoo{-R}{R}^N $. We then introduce
\beq
\mc{D}_{k;\ell} \; = \; \Big\{ \bs{w}_N \in \R^N \; : \; w_1 < \dots <w_k < -R < w_{k+1} < \dots < w_{\ell} < R 
			< w_{\ell+1} <\dots < w_N   \Big\} \;. 
\enq
It is readily seen that it is enough to prove the $L^{2}$ character of $\msc{H}_N\big[ G \big] (\bs{w}_N)$ on each $\mc{D}_{k;\ell}$
with $0\leq k < \ell \leq N $.

\noindent $\pl{a<b}{N}(w_b-w_a) \cdot \msc{H}_N\big[ G \big] (\bs{w}_N)$ being smooth, it is bounded on $\mc{D}_{0,N}$ and, as such, 
$\msc{H}_N\big[ G \big]\in L^{2}\big(\mc{D}_{0,N}, \dd \wt{\mu}(\bs{w}_N) \big)$.

\noindent It thus remains to consider the case  $(k , \ell) \not= (0,N)$. Then, since 
\beq
\big( \msc{H}_N[ G ] (\bs{w}_N)  \big)^* \; = \; \msc{H}_N[ G^{(-)} ] (-\bs{w}_N) \qquad \e{with} \qquad 
G^{(-)}(\bs{y}_N) \; = \; G(-\bs{y}_N)
\enq
we may just as well assume that $\ell\leq N-1$. Having this in mind, we define
\beq
I_{k,\ell} \; = \; \left\{ \ba{c c } \intn{1}{k} \cup \intn{\ell+1}{N}  & k \geq 1  \vspace{1mm}  \\ 
									\intn{\ell+1}{N}  & k = 1        					\ea \right.
										\qquad \e{and}  \qquad
I_{k,\ell}^{\e{c}} \; = \; \intn{k+1}{\ell}    \;. 
\enq

Then, a direct application of \eqref{ecriture identite vers decomposition produits transfo Cauchy simples}
in respect to the variables $w_a$, $a \in I_{k,\ell}^{\e{c}}$  yields
\beq
\msc{H}_N\big[ G \big] (\bs{w}_N) \; = \;  \f{ |I_{k,\ell}^{\e{c}}| ! }{N!} 
\pl{ \substack{ a > b \\ a,b \in I_{k,\ell}^{\e{c}} } }{} \bigg\{ \f{1}{w_a - w_b }\bigg\}
\cdot \pl{a\in  I_{k,\ell} }{} \bigg\{ \Ga \Big( \f{ -w_a }{ i\hbar } \Big) \bigg\}^N   \lim_{\a \tend 0^+} \Int{ \R^N }{}  
 \f{  G_{ I_{k,\ell} }( \bs{y}_N, \bs{w}_N)  }{ \pl{a \in I_{k,\ell}^{\e{c}} }{} ( y_a - w_a + i\a) }  
\cdot  \Big(   \f{ i w_N }{\hbar}  \Big)^{ - \f{ i y_N}{\hbar} }  \cdot \f{ \dd^N y }{ (2\pi \hbar)^N } \;. 
\label{formule reecriture expression HN}
\enq
There, we have set, 
\bem
G_{ I_{k,\ell} }( \bs{y}_N, \bs{w}_N)  \; = \; (i\hbar)^{N | I_{k,\ell}^{\e{c}} |} \cdot  
\Big(   \f{ i w_N }{\hbar}  \Big)^{ i \f{y_N}{\hbar} }  \cdot 
 \pl{a \in I_{k,\ell}^{\e{c}} }{}\pl{b=1}{N} \Ga\Big( \f{ y_b - w_a }{ i \hbar} +1 \Big) 
\pl{a \in I_{k,\ell} }{}\pl{b=1}{N}\bigg\{  \Ga\Big( \f{ y_b - w_a }{ i \hbar} \Big) \cdot 
							\Ga^{-1}\Big( \f{-w_a}{i\hbar} \Big) \bigg\} \\
 \times (-1)^{N(m+1)} \cdot  \pl{ \substack{ a < b \\ a,b \in I_{k,\ell}^{\e{c}} } }{} \bigg\{ \f{1}{y_a - y_b }\bigg\} 
\pl{ a \in I_{k,\ell}^{\e{c}} }{}   \pl{ b \in I_{k,\ell} }{} \bigg\{ \f{ 1 }{ y_a - y_b  } \bigg\}
 \cdot  \pl{a\not= b}{N} \Ga^{-1}\Big( \f{ y_b - y_a}{ i \hbar}\Big) \cdot  G(\bs{y}_N) \;. 
\end{multline}

Further, we integrate by parts $1+|I_{k,\ell}^{\e{c}}|$  times in respect to $y_N$ the term
$\big(   \tf{ i w_N }{\hbar}  \big)^{ - \f{ i y_N}{\hbar} }$
in \eqref{formule reecriture expression HN} and also integrate by parts once  each of the 
singular factors in the variables $y_a$, with $a \in I_{k\ell}^{\e{c}}$, leading to 
\bem
\msc{H}_N\big[ G \big] (\bs{w}_N) \; = \; - \f{ |I_{k,\ell}^{\e{c}}|! }{ N! } 
\cdot \pl{ \substack{ a > b \\ a,b \in I_{k,\ell}^{c} } }{} \bigg\{ \f{1}{w_a - w_b }\bigg\}
\cdot \pl{a\in  I_{k,\ell} }{} \bigg\{ \Ga \Big( \f{ -w_a }{ i\hbar } \Big) \bigg\}^N   
 \; \cdot \; \bigg( \f{  i \hbar }{ \ln [ \tf{i w_N }{\hbar} ]   }  \bigg)^{1+ |I_{k,\ell}^{\e{c}}| }  \\
\times 
\Int{ \e{supp}(G)  }{}  \Big(   \f{ i w_N }{\hbar}  \Big)^{ -i \f{y_N}{\hbar} } 
\pl{a \in I_{k,\ell}^{\e{c}} }{} \bigg\{ \ln|y_a - w_a| + i \pi \bs{1}_{\R^+}(w_a - y_a)  \bigg\}
 \f{ \Dp{}^{1+  |I_{k,\ell}^{\e{c}}| } }{ \Dp{}y_N^{1 + |I_{k,\ell}^{\e{c}}| }  }  
\cdot \pl{a \in I_{k,\ell}^{\e{c}} }{}  \f{ \Dp{} }{ \Dp{}y_a }  
\cdot \Big\{ G_{ I_{k,\ell} }( \bs{y}_N, \bs{w}_N)  \Big\}  \cdot \f{ \dd^N y }{ (2\pi \hbar)^N } \;. 
\label{ecriture representation regularisee fonctionnelle HN}
\end{multline}

It follows from the uniform differentiability and uniformness in $\bs{y}_N \in \e{supp}(G)$
of the remainder in  the large $w_a$, $a \in I_{k,\ell}$, expansion 
\beq
\pl{ a \in I_{k , \ell}  }{} \pl{b=1}{N} 
\bigg\{ \Ga\Big( \f{ y_b - w_a }{ i \hbar}\Big)  \cdot \Ga^{-1}\Big( -\f{w_a}{i\hbar} \Big) \bigg\} 
\; = \;  \pl{a \in I_{k, \ell} }{} \bigg\{ \Big( -  \f{ w_a }{i\hbar}  \Big)^{ \f{\ov{\bs{y}}_N}{i\hbar} }  \bigg\}  \cdot 
\bigg\{  1 \; + \; \sul{ a \in I_{k, \ell} }{  } \e{O}\Big( \f{1}{w_a} \Big) \bigg\} 
\label{ecriture comportement infini produit fonctions gamma}
\enq
that there exists a constant $C>0$ such that 
\beq
 \Bigg| \f{ \Dp{}^{1+  |I_{k,\ell}^{\e{c}}| } }{ \Dp{}y_N^{1 + |I_{k,\ell}^{\e{c}}| }  }  
\pl{a \in I_{k,\ell}^{\e{c}} }{}  \f{ \Dp{} }{ \Dp{}y_a }  
\cdot \Big\{ G_{ I_{k,\ell} }( \bs{y}_N, \bs{w}_N)  \Big\}   \Bigg|   \; \leq \; 
 C  \Big( \ln(-w_1) + \ln(w_N) +1 \Big)^{ |I_{k,\ell}^{\e{c}}|  }  
\enq
uniformly in $\bs{w}_N \in \mc{D}_{k,\ell}$ and $\bs{y}_N \in \e{supp}(G)$. Using that the integration in 
\eqref{ecriture representation regularisee fonctionnelle HN} runs through a compact, 
it is readily inferred that there exists some constant $C^{\prime}$ such that 
\beq
 \wt{\mu}(\bs{w}_N) \cdot \big| \msc{H}_N\big[ G \big] (\bs{w}_N)  \big|^2   \; \leq  \; 
\f{ C^{\prime}  \cdot \big[ \ln(-w_1) + \ln(w_N) + 1 \big]^{ 2 |I_{k,\ell}^{\e{c}} |}  }
{ |w_1| \cdot \big( \ln w_N \big)^{ 2|I_{k,\ell}^{\e{c}} |}  \cdot  |w_N|  \cdot \big( \ln |w_N| \big)^2  } 
\cdot  \pl{ p =1 }{ |I_{k,\ell}|  } \ex{ - \f{\pi}{\hbar} \big[w_{a_p} ( |I_{k,\ell}| -2p ) + |I_{k,\ell}| \cdot |w_{a_p}|\big] }  \;. 
\label{ecriture bornage H_N transform}
\enq
The last product in \eqref{ecriture bornage H_N transform} utilises the parametrisation 
$I_{k,\ell} \; = \; \big\{ a_1, \dots, a_{ |I_{k,\ell}| }  \big\}$
with the additional assumption  $a_1 < \dots < a_{ |I_{k,\ell}| }$. 
The \textit{rhs} is clearly in  $L^1\big( \mc{D}_{k,\ell}, \dd^N w \big)$. \qed

\vspace{3mm}

We are now in position to prove the isometric character of $\msc{H}_N$ when restricted to 
$\msc{C}_{\e{c};\e{sym}}^{\infty}(\R^N)$. This represents the hardest result obtained 
in this paper. 

\begin{prop}
\label{Proposition charactere isometrique de la transfo H}

For any $G \in \msc{C}_{\e{c}; \e{sym}}^{\infty}(\R^N)$ one has 
\beq
\big| \big| \msc{H}_N\big[ G \big] \,   \big| \big|_{L^{2}_{\e{sym}}\big(\R^N, \dd \wt{\mu}(\bs{w}_N)\big)}
\; = \;  || \, G \, ||_{L^{2}_{\e{sym}}\big(\R^{N}, \, \dd \wt{\mu}(\bs{y}_{N}) \big)} \;, 
\label{ecriture character isometrique HN}
\enq
 where the operator $\msc{H}_N$ is as defined by \eqref{definition operateur HN comme integral}.

\end{prop}

It thus follows that $\msc{H}_N$ can naturally be extended into an isometric operator 
\beq
\msc{H}_N  \; :  \;  L^{2}_{\e{sym}}\Big(\R^{N}, \dd \wt{\mu}(\bs{w}_{N}) \Big) \quad   \longrightarrow \quad 
				L^{2}_{\e{sym}}\Big(\R^N, \dd \wt{\mu}(\bs{y}_N)\Big) \;. 
\enq

\Proof 

 Given any $F \in \msc{C}^{\infty}_{\e{c};\e{sym}}(\R^{N+1})$, we define 
\beq
\wt{F}(\bs{y}_{N+1}) \; = \; \hbar^{\f{i}{\hbar} N \ov{\bs{y}}_{N+1} } F(\bs{y}_{N+1})  \;. 
\enq
An application of the Mellin-Barnes recurrence \eqref{definition fct Whittaker par Mellin-Barnes} satisfied by the integral kernel $\vp_{\bs{y}_{N+1}}(\bs{x}_{N+1})$ of the transform $\mc{U}_{N+1}$, 
\textit{cf} \eqref{definition noyau $U$ sur un espace dense}, followed by an application of Fubbini's 
theorem\footnote{what is licit since the integral converges strongly in $\big( \bs{w}_N, \bs{y}_{N+1} \big)$
due to the compact support of $\wt{F}$ and the fast decay in $\bs{w}_N$ of \newline
$| \vp_{\bs{w}_N}(\bs{x}_N) \cdot  \varpi(\bs{w}_N\mid \bs{y}_{N+1}) |$, \textit{cf} \textit{eg} \cite{GerasimovKharchevLebedevRepThandQISM}.}, leads to the representation
\bem
\mc{U}_{N+1}\big[ \wt{F}  ](\bs{x}_{N+1}) \; =  \hspace{-3mm} \Int{ (\R-i\a)^N }{} \hspace{-3mm} 
\dd\mu(\bs{w}_N) \vp_{\bs{w}_N}(\bs{x}_N)\ex{-\f{i}{\hbar} \ov{\bs{w}}_N x_{N+1}}
\cdot \hspace{-2mm} \Int{ \R^{N+1} }{}  \ex{\f{i}{\hbar} \ov{\bs{y}}_{N+1}x_{N+1} } 
\f{ \varpi(\bs{w}_N\mid \bs{y}_{N+1}) }{ \sqrt{(N+1)!} \cdot \mu(\bs{w}_N) } 
\cdot  \wt{F}(\bs{y}_{N+1}) \cdot \f{\dd \mu(\bs{y}_{N+1}) }{ (2\pi \hbar)^N \cdot N! } \\
  \; = \; \f{   \mc{U}_N\Big[  \wt{\mc{S}}_N[F](*,x_{N+1}) \Big](\bs{x}_{N}) }{ (2\pi \hbar)^N \cdot N! \cdot \sqrt{N+1}}  \;,
\label{equation reecrtiure action UN+1 par action UN}
\end{multline}
where $*$ refers to the group of variables $\bs{w}_N$ in the function 
\beq
\wt{\mc{S}}_N[F]( \bs{w}_N ,x) \;= \; \ex{-\f{i}{\hbar} \ov{\bs{w}}_N x} \cdot \hbar^{\f{i}{\hbar}(N+1)\ov{\bs{w}}_N } 
\cdot \mc{S}_N[F](\bs{w}_N,x)
\label{definition transformation S tilde}
\enq
on which the transform $\mc{U}_N$ acts. The integral transform $\mc{S}_N$ introduced in \eqref{definition transformation S tilde} reads 
\beq
\mc{S}_N[F](\bs{w}_N,x) \; = \;   \lim_{\a \tend 0^+}
\Int{ \R^{N+1} }{} \hspace{-2mm} \ex{\f{i}{\hbar} \ov{\bs{y}}_{N+1}x  }
\pl{a=1}{N} \pl{b=1}{N+1} \Ga\Big( \f{ y_b - w_a + i\a}{ i \hbar}  \Big)
\cdot \pl{a\not= b}{N+1} \Ga^{-1}\Big( \f{ y_b - y_a}{ i \hbar}  \Big) 
\cdot F(\bs{y}_{N+1})  \cdot \f{ \dd^{N+1} y }{ 2 \pi \hbar } \;. 
\label{definition operateur integral S}
\enq

\noindent In fact, in order to establish the second line of \eqref{equation reecrtiure action UN+1 par action UN}, 
one should check that 
\begin{itemize}
\item  $\mc{S}_N[F](\bs{w}_N-i\a \bs{e}_N,x)\in L^{1}\big(\R^N , \dd\mu(\bs{w}_{N}) \big)$
with $\bs{e}_N = (1,\dots, 1) \in \R^N$ so that the action of $\mc{U}_N$ on this function is well defined;
\item that $\big| \mc{S}_N[F](\bs{w}_N-i\a \bs{e}_N,x) \big|$ can be bounded by an $L^{1}\big( \R^N , \dd\mu(\bs{w}_{N}) \big) $ function,
independently of $\a$ small enough. 
\end{itemize}
These properties allow then to move the $\a \tend 0^+$ limit past the action of $\mc{U}_N$. 
These two properties can be readily inferred by following the reasoning outlined in the proof of lemma  
\ref{Lemme well definiteness and L2 character of H-transform}, so we do not reproduce
the arguments once again. In fact, this very reasoning - with the sole difference being that one should integrate  by parts and in respect 
to $y_{N+1}$ the oscillating exponent $\exp[i \tf{ \ov{ \bs{y} }_{ N+1 }  x }{ \hbar }]$ so as to generate an explicit 
algebraic decay in $x$- allows one to establish that 
$S_N[F] \in L^{2}_{\e{sym}\times -}\big( \R^N \times \R , \dd\mu(\bs{w}_{N}) \otimes \dd x \big)$, and hence $\wt{S}_N[F]$ as well. 
Note that the subscript $ {\e{sym}\times -} $ refers 
to functions that are symmetric in respect to the first $N$ variables, in accordance with the Carthesian 
product decomposition of $\R^N\times\R$.  

The isometric character of $\mc{U}_{N+1}$ and $\mc{U}_N$ leads to the relation
\beq
\mc{N}[F] \; \equiv \; 
 \big| \big| \mc{S}_N[F] \, \big| \big|_{ L^2_{\e{sym}\times -}\big( \R^N \times \R ,\,  \dd \mu(\bs{w}_{N}) \otimes \dd x \big) }
 \; = \; \sqrt{N+1} \cdot N! (2\pi \hbar)^N\cdot  | | \, F \, ||_{ L^2_{\e{sym}}\big( \R^{N+1} , \, \dd \mu(\bs{y}_{N+1})\big) } \;. 
\label{ecriture des deux representations pour la norme de F}
\enq

We now build on the above two possible representations for $\mc{N}[F]$ so as to deduce
the relation \eqref{ecriture character isometrique HN}, this by using a specific choice for the 
function $F$. 
More precisely, from now on, we shall take $F=F_K  \in \msc{C}^{\infty}_{\e{c}; \e{sym} }(\R^{N+1})$
given by 
\beq
F_{K}(\bs{y}_{N+1}) \; = \; \e{Sym} \bigg\{  G(\bs{y}_N) \, u( \ov{\bs{y}}_{N+1} - K ) \, 
\ex{\f{\pi}{\hbar}\ov{\bs{y}}_{N} (1+\tf{N}{2})} \,  \Ga^{N}\Big( \f{ \ov{\bs{y}}_{N+1} }{i\hbar} + 1 \Big)
\cdot 
\Big( - \f{ \ov{\bs{y}}_{N+1} }{i\hbar} - 1 \Big)^{-N}   \cdot 
\Big( \f{ \ov{\bs{y}}_{N+1}+i}{\hbar}  \Big)^{ \f{ N }{i\hbar} \ov{\bs{y}}_{N} }  \bigg\} \;,
\label{Definition fonction F necessaire pr reduction integrale}
\enq
where $G$ and $u$ are such that 
\beq
G \in \msc{C}^{\infty}_{\e{c}; \e{sym} }(\R^N) \quad \e{and} \quad u \in \msc{C}^{\infty}_{\e{c}}(\R)
\qquad \e{with} \quad \Int{\R}{} |u(y)|^2 \cdot \f{\dd y }{2 \pi  \hbar} =1 \;. 
\enq
Finally, in \eqref{Definition fonction F necessaire pr reduction integrale}, 
$\e{Sym}$ stands for the symmetrization operator in respect to the variables $\bs{y}_{N+1}$, \textit{viz}. 
for any function $J$ of the variables $\bs{y}_{N+1}$, one has 
\beq
\e{Sym}\big\{ J \big\} \big(\bs{y}_{N+1}\big)  \; \equiv \; \f{1}{(N+1)!} \sul{ \sg \in \mf{S}_{N+1} }{}   J\big(\bs{y}_{N+1}^{\sg} \big) 
\qquad \e{where} \qquad \bs{y}_{N+1}^{\sg} \; = \; (y_{\sg(1)}, \dots, y_{\sg(N+1)} ) \;. 
\enq

We shall compute the $K \tend + \infty$ limit of $\mc{N}[F_K]$ in two ways by using 
\eqref{ecriture des deux representations pour la norme de F}. Ultimately, this will yield us
the sought isometricity of $\msc{H}_N$.

\subsection*{The $K\tend +\infty$ limit of $| |  F_{K}  ||_{ L^2_{\e{sym}}\big( \R^{N+1} , \dd \mu(\bs{y}_{N+1})\big) }$ }

Since $G$ is symmetric in $\bs{y}_N$, it is readily seen that the sum over the permutation group  in 
\eqref{Definition fonction F necessaire pr reduction integrale} can be recast as 
\beq
F_K(\bs{y}_{N+1}) \; = \; \f{  u(\ov{\bs{y}}_{N+1} - K )  }{N+1} \Ga^{N}\Big( \f{ \ov{\bs{y}}_{N+1} }{i\hbar} + 1 \Big)
\cdot \Big( - \f{ \ov{\bs{y}}_{N+1} }{i\hbar} - 1 \Big)^{-N}
 \sul{k=1}{N+1} G(\bs{y}_N^{(k)} ) \,  \, \ex{\f{\pi}{2\hbar}\ov{\bs{y}}_{N}^{(k)} (2+N)} \,     \cdot 
\Big( \f{ \ov{\bs{y}}_{N+1}+i}{\hbar}  \Big)^{ \f{ N }{i\hbar} \ov{\bs{y}}_{N}^{(k)} }
\enq
in which, from now on, $\bs{y}^{(k)}_N$ stands for the $N$- dimensional vector\footnote{ We urge the 
reader not to confuse this notation with the one introduced in \eqref{definition variables yN ecoutes} } 
\beq
\bs{y}_N^{(k)} \; = \; \big( y_1,\dots, y_{k-1} , y_{k+1},\dots , y_{N+1} \big) \;. 
\enq

Assume that is $K$ large enough. Then, the $k^{\e{th}}$ term in the above sum
does not vanish only if $\bs{y}_N^{(k)} \in \e{supp}(G)$. Thus, since both $\e{supp}(G)$ and $\e{supp}(u)$
are compact, the function $G(\bs{y}_N^{(k)} ) u(\ov{\bs{y}}_{N+1} - K )  $ will be non zero
solely for $y_k$ belonging to some $K$-independent compact centered at $K$. 
As a consequence, for $K$ large, the only non-zero contributions to 
the product $F^*_K F_K $ will stem from the diagonal terms, \textit{viz}
\beq
\big| F_K(\bs{y}_{N+1}) \big|^2 \; = \; \Big| 
\f{ u( \ov{\bs{y}}_{N+1} - K )  }{N+1} \Ga^{N}\Big( \f{ \ov{\bs{y}}_{N+1} }{i\hbar} + 1 \Big) 
\cdot \Big( - \f{ \ov{\bs{y}}_{N+1} }{i\hbar} - 1 \Big)^{-N} \Big|^2
 \sul{k=1}{N+1} \Big| G(\bs{y}_N^{(k)} ) \,  \, \ex{\f{\pi}{2\hbar}\ov{\bs{y}}_{N}^{(k)} (2+N)} \,     \cdot 
\Big( \f{ \ov{\bs{y}}_{N+1}+i}{\hbar}  \Big)^{ \f{ N }{i\hbar} \ov{\bs{y}}_{N}^{(k)} } \Big|^2 \;. 
\nonumber
\enq

Inserting this expression into the integral, making the most of the symmetry of the integrand and finally
changing the variables from $y_{N+1}$ to $s=\ov{\bs{y}}_{N+1} - K$ recasts the norm in the form  
\beq
\mc{N}^2\big[ F_{K} \big]  \; = \;  (N!)^2  (2\pi\hbar)^{2N} \cdot 
\Int{ \R^{N}\times \R  }{}  |u(s)|^2 \cdot | G(\bs{y}_N) |^2
\msc{I}_K \big( s; \bs{y}_N \big) \cdot \dd \wt{\mu}(\bs{y}_N) \cdot \f{ \dd s }{2\pi \hbar }  
\enq
where 
\beq
\msc{I}_K \big( s; \bs{y}_N \big) \; = \; 
 \f{  \ex{\f{\pi}{\hbar}\ov{\bs{y}}_{N} (1+N)} }{  \Big( 1+ \f{ (K + s)^2 }{\hbar^2} \Big)^{N} } 
 \,     \cdot 
\Big( \f{ K+ s +i }{ K+ s -i }  \Big)^{ \f{ N }{i\hbar} \ov{\bs{y}}_{N} } 
\pl{a=1}{N} \Bigg\{ \f{ \Ga\Big( 1 + \f{ K+ s  }{i\hbar} ,  1 - \f{ K+ s  }{i\hbar} \Big)  }
	{\Ga\Big( \f{ s + K - y_a - \ov{\bs{y} }_N}{i\hbar} , \f{  y_a + \ov{\bs{y} }_N- s -K }{i\hbar}  \Big)  }  \Bigg\} \;. 
\enq
It is readily seen that 
\beq
 \msc{I}_K \big( s; \bs{y}_N \big)  \;\; \limit{ K }{ + \infty } \;\;   1 \qquad \e{uniformly} \; \e{in} \quad 
 (s; \bs{y}_N) 
\enq
belonging to compact subsets of $\R \times \R^{N}$. 
As a consequence, since the integration runs through the compact support of $u(s) G(\bs{y}_N)$, it follows that  
\beq
\lim_{K \tend +\infty} \mc{N}^2\big[ F_{K} \big] \; = \; (N!)^2  (2\pi\hbar)^{2N} \cdot 
||  G \, ||^2_{L^{2}_{\e{sym}}\big(\R^{N}, \dd \wt{\mu}(\bs{y}_{N})\big)}  \;. 
\label{ecriture N F K version fonction G seule}
\enq

\subsection*{The $K\tend +\infty$ limit of 
$\big| \big| \,  \mc{S}_N[F_{K}] \, \big| \big|_{ L^2_{\e{sym}\times -} \big( \R^N \times \R ,\,  \dd \mu(\bs{w}_{N}) \cdot \dd x \big) }$ }

 We first note that the kernel of $\mc{S}_N$ obviously contains partially antisymmetric functions. Hence,
 we can drop the symmetrization operator 
when inserting the function  \eqref{Definition fonction F necessaire pr reduction integrale} into the multiple integral 
\eqref{definition operateur integral S}. Further, when evaluating the $L^2$ norm of 
$ \mc{S}_N[F_{K}]$, the integrals in respect to $x$ can be taken due to the 
isometric character of the Fourier transform what allows one to recast the second representation for the norm as 
\bem
\mc{N}^2[F_{K}] \; = \; 
\Int{}{} \dd \mu(\bs{w}_{N})  \lim_{\a \tend 0^+} \Bigg\{ 
\Int{ \R^{N+1} }{} 
\pl{a=1}{N} \pl{b=1}{N+1} \Ga\Big( \f{ y_b - w_a +i\a}{ i \hbar}  \Big)
  \cdot \pl{a\not= b}{N+1} \Ga^{-1}\Big( \f{ y_b - y_a}{ i \hbar}  \Big)  
 \cdot F_K(\bs{y}_{N+1})  \cdot \f{ \dd^{N+1} y }{ 2 \pi \hbar }   \Bigg\}\\
\times  \lim_{ \a^{\prime} \tend 0^+}  \Bigg\{ 
\Int{ \R^{N} }{} \bigg\{  \pl{a=1}{N} \pl{b=1}{N+1} \Ga\Big( \f{  w_a- y_b^{\prime} +i\a^{\prime} }{ i \hbar}  \Big)
\cdot \pl{a\not= b}{N+1} \Ga^{-1}\Big( \f{ y_b^{\prime}  - y_a^{\prime}  }{ i \hbar}  \Big) 
\cdot F^*_K( \bs{y}_{N+1}^{\prime} ) \bigg\}_{\mid \ov{\bs{y}}_{N+1}=\ov{\bs{y}}^{\prime}_{N+1} } 
\hspace{-7mm}\cdot  \dd^N y^{\prime}  \Bigg\} \;. 
\label{ecriture formule des normes pr reduction integrale ac Fnelle S}
\end{multline}
Above, the symbol $\ov{\bs{y}}_{N+1}=\ov{\bs{y}}^{\prime}_{N+1}$ means that 
one should replace  $y_{N+1}^{\prime} = \ov{\bs{y}}_{N+1} - \ov{\bs{y}}^{\prime}_{N} $ in the part of 
the integrand depending on $\bs{y}_{N+1}^{\prime}$. 

Next, we implement the change of variables $y_{N+1} = s +K - \ov{\bs{y}}_{N}$
(which also imposes the relation  $y_{N+1}^{\prime} = s +K - \ov{\bs{y}}_{N}^{\prime}$), pull out the 
$\a \tend 0^+$ limit outside of the integrals and exchange the orders of integrations. 
Again, the justification for being able to do so basicall parallels the proof of lemma
\ref{Lemme well definiteness and L2 character of H-transform}. Thus, we solely sketch the main steps. 
One should first take the $\a \tend 0^+$ limit in each of the integrals, \textit{ie} represent each integrand
in the spirit of \eqref{ecriture HN comme succession Transfos Cauchy}, integrate
by parts in respect to the relevant variables so as to "regularize" the singular part of the 
integrand and explicitly ensure the convergence (in respect to the $\bs{w}_N$ variables)
 at $\infty$ (\textit{cf} \eqref{ecriture representation regularisee fonctionnelle HN}). 
At that stage it is already possible to invoke Fubbini's theorem 
so as to exchange the orders of integration. Further, one is then able to apply the dominated
convergence theorem so as to recast the integral as an $\a \tend 0^+$ limit. 
However, this time, the $\a \tend 0^+$ limit symbol is outside of all the integration symbols. 
It then solely remains to "undo" all of the integrations  by parts in respect to the
$\bs{y}_N$ and $\bs{y}^{\prime}_N$ variables. This last step is licit in that, for any $\a>0$,
the integral over $\bs{w}_N$ converges strongly. All in all, one obtains:
\beq
\mc{N}^2[F_K] \; = \; \lim_{\a \tend 0^+}
\Int{ \R^{N} }{} \hspace{-1mm} \dd^{N} y \, G(\bs{y}_N) \Int{ \R^{N} }{} \hspace{-1mm} \dd^{N} y^{\prime} \,  G^*(\bs{y}_N^{\prime})
\Int{ \R }{}  \f{ \dd s }{ 2\pi \hbar } |u( s )|^2
 \cdot \pl{a\not= b}{N} \Ga^{-1}\Big( \f{ y_b - y_a}{ i \hbar} , \f{ y_b^{\prime}  - y_a^{\prime}  }{ i \hbar} \Big) 
\cdot \mc{L}^{(\a)}_K\big(s, \bs{y}_{N}, \bs{y}^{\prime}_N \big)   \;, 
\label{forme norme FK ac lim alpha go to zero}
\enq
 where $\mc{L}_K^{(\a)}$ is given by the $N$-dimensional integral 
\beq
\mc{L}^{(\a)}_K\big(s, \bs{y}_{N}, \bs{y}^{\prime}_N \big) \; = \; \Int{ \R^N }{}
\pl{a,b=1}{N}  \bigg\{ \Ga\Big(  \f{ y_b - w_a +i\a}{ i \hbar} , \f{  w_a- y_b^{\prime} +i\a }{ i \hbar}  \Big)  \bigg\}
\mc{J}_K^{(\a)}\big( s ;   \bs{w}_N, \bs{y}_N, \bs{y}_N^{\prime} \big) \cdot  \dd \wt{\mu}(\bs{w}_{N})  \;. 
\label{definition de L cal alpha}
\enq
The measure $\dd \wt{\mu}(\bs{w}_N)$ has been already
introduced in \eqref{definition mesure mu tilde} whereas
\bem
\mc{J}_K^{(\a)}\big( s ;   \bs{w}_N, \bs{y}_N, \bs{y}_N^{\prime} \big) \;= \; 
\ex{\f{\pi}{2\hbar}(2+N)(\ov{\bs{y}}_N + \ov{\bs{y}}^{\prime}_N) } \ex{- \f{\pi }{\hbar } \ov{\bs{w}}_N }
\Ga^{N}\Big( 1+ \f{s+K}{i\hbar}  , 1 - \f{s+K}{i\hbar}  \Big)
\cdot 
\Big( - \f{s+K}{i\hbar} - 1 \Big)^{-N} \cdot \Big(  \f{s+K }{i\hbar} - 1 \Big)^{-N}  \\
\times \Big( \f{s+K+i}{\hbar}  \Big)^{ \f{ N}{i\hbar} \ov{\bs{y}}_{N} } 
 \Big( \f{ s+K -i}{\hbar}  \Big)^{ - \f{ N }{i\hbar} \ov{\bs{y}}_{N}^{\prime} }
\cdot \pl{a=1}{N} \Bigg\{ \Ga\bigg( \f{  w_a -s - K + \ov{\bs{y}}_N^{\prime}  + i\a }{ i \hbar}  , 
\f{   s + K - \ov{\bs{y}}_N -w_a + i\a }{ i \hbar}  \bigg)   \Bigg\} \\
\times \pl{a=1}{N} \Bigg\{ \Ga^{-1}\bigg( \f{ y_a^{\prime} + \ov{\bs{y}}_N^{\prime} - s - K  }{ i \hbar} , 
\f{ s+K - y_a^{\prime} - \ov{\bs{y}}_N^{\prime}   }{ i \hbar} ,
 \f{ y_a  + \ov{\bs{y}}_N - s-K   }{ i \hbar} , \f{   s + K - y_a  - \ov{\bs{y}}_N     }{ i \hbar} \bigg) \Bigg\} \;. 
\end{multline}

It is  a straightforward computation to check that, 
\beq
\mc{J}_K^{(0)}\big( s ;   \bs{w}_N, \bs{y}_N, \bs{y}_N^{\prime} \big)   \limit{ K }{ + \infty } 1
\qquad \e{uniformly} \; \e{in} \qquad (s, \bs{w}_N, \bs{y}_N ,\bs{y}_N^{\prime})
\enq
 belonging to compact subsets of $\R^{3N+1}$. 

This observation is however not enough so as to conclude.
Indeed, the integrand is expressed in a too singular way so as to be in position of invoking the dominated convergence theorem. 
Having this in mind, we implement several regularizing steps, allowing one, in the very end, to apply the latter theorem.

\vspace{2mm} 

We start by deforming the contours in \eqref{definition de L cal alpha} 
from $\R^N$ up to $(\R+i\eta)^N$ where $\eta > \a$ is taken small enough so that the only poles that are crossed 
correspond to those lying closest to the real axis.
Hence, in the course of doing so, one will cross, individually in each variable $w_a$, the poles at $w_a = y_b+i\a$, 
with\footnote{ Here, we agree upon the shorthand notation $y_{N+1} \; = \; s-\ov{\bs{y}}_N + K $} $b=1,\dots,N+1$.
 Because of the zeroes of the measure's density $\wt{\mu}(\bs{w}_N)$,
poles corresponding to two or more coinciding coordinates (\textit{ie} such that $w_a=w_b$) have zero residue. 
The evaluation of the  residues that result from such a handling can be further simplified by using that the integrand is symmetric in 
$\bs{w}_N$.  It is enough to choose the last $k$ coordiantes, $k\in \{ 1,\dots, N \}$,
of $\bs{w}_N$ as those which will correspond to residue evaluation, weight it with the combinatorial
factor $C^k_N$ and then sum up over all possible choices $k=0,\dots, N$. All in all, this yields  
\bem
\mc{L}^{(\a)}_K\big(s, \bs{y}_{N}, \bs{y}^{\prime}_N \big) \; = \; \sul{k=0}{N} \; C^k_N  \hspace{-3mm}
\Int{ (\R+i\eta)^{N-k} }{} \hspace{-4mm} \dd^{N-k} w  \cdot \pl{a=1}{N-k} \pl{b=1}{N} 
\bigg\{ \Ga\Big(  \f{ y_b - w_a + i\a}{ i \hbar} , \f{  w_a- y_b^{\prime} + i\a }{ i \hbar}  \Big)  \bigg\}
 \sul{p_1\not= \dots \not= p_k}{ N+1 }  (2i\pi)^k  \\
\e{Res}\Bigg(  
\wt{\mu}(\bs{w}_{N-k}, \bs{x}_k)  
\pl{b=1}{N} \pl{\ell=1}{k}  \bigg\{ \Ga\Big(  \f{ y_b - x_{\ell} +i\a}{ i \hbar} , 
\f{  x_{\ell}- y_b^{\prime} +i\a  }{ i \hbar}  \Big)  \bigg\}   \cdot 
\mc{J}_K^{(\a)}\big( s ;   (\bs{w}_{N-k},\bs{x}_k), \bs{y}_N, \bs{y}_N^{\prime} \big)  \cdot \dd^k x \, , \, x_a = y_{p_a} + i \a \Bigg) \;,  
\end{multline}

The above sum can be further simplified thanks to the effective symmetry of the function $\mc{L}^{(\a)}_K$. Indeed,
we are eventually interested in integrating it versus a symmetric function of $\bs{y}_N$ 
in \eqref{forme norme FK ac lim alpha go to zero}. It thus means that for our purpose,
 one can simplify further the expression for $\mc{L}_K^{(\a)}$ by using the permutation invariance
in respect to the components of $\bs{y}_N$. 

 Then $k=0$ term in the above sum is given by a $N$-dimensional integral over $\R+i\eta$ and it can be left as such. 
However, when $k \not= 0 $, one should prepare the corresponding expression; we partition the sums over the indices 
$p_1\not= \dots \not= p_k$ depending on whether the residue at $y_{N+1}+i\a$ has been computed or not. 
\begin{itemize}
\item Suppose that for any $a=1,\dots, k$ one has $p_a  \not= N+1$. There are, in total, $N!/(N-k)!$ such possible 
choices of the different configurations of the pairwise distinct integers $(p_1,\dots, p_k)$. Due 
to the aforementioned freedom of permuting the 
coordinates of $\bs{y}_N$, it is enough to weight by $\tf{N!}{(N-k)!}$ the contribution of the configuration 
$p_a = a$, with $a=1,\dots,k$, \textit{viz}. the one corresponding to $x_a = y_a+i\a$, $a=1,\dots, k$. 

\item Suppose that there is an $a\in \{1, \dots, k  \}$ such that $p_a=N+1$. Then, the other ones necessarily take values in $1,\dots, N$. 
In virtue of the symmetry of the integrand in the $x$'s and, again of the permutation freedom of the coordinates of $\bs{y}_N$,
it in enough to take into account the contribution 
of the configuration $p_a = a $ for $a=1,\dots, k-1$ and $p_k=N+1$, this weighted by the factor $k\cdot  N!/ (N-k+1)! $. 
\end{itemize}

Hence, one has 
\beq
\mc{L}^{(\a)}_K\big(s, \bs{y}_{N}, \bs{y}^{\prime}_N \big) \; \simeq \; 
\mc{L}^{(\a)}_{K;0}\big(s, \bs{y}_{N}, \bs{y}^{\prime}_N \big)  \; + \; 
\sul{k=1}{N}  \f{ N!C^k_N }{ (N-k)! }  \cdot 
\Big\{  \mc{L}^{(\a)}_{K;k}\big(s, \bs{y}_{N}, \bs{y}^{\prime}_N \big)   \;+ \; 
\f{k }{ N-k+1}   \msc{L}^{(\a)}_{K;k}\big(s, \bs{y}_{N}, \bs{y}^{\prime}_N \big) \Big\} \;. 
\label{ecriture decomposition L de alpha K apres deplacement ctr}
\enq
where $\simeq $ means that the \textit{lhs} and \textit{rhs} give the same result when integrated 
versus a symmetric function of $\bs{y}_N$. Also, in 
\eqref{ecriture decomposition L de alpha K apres deplacement ctr}, we have introduced
\beq
\mc{L}^{(\a)}_{K;0}\big(s, \bs{y}_{N}, \bs{y}^{\prime}_N \big) \; = \;
\Int{ (\R+i\eta)^{N} }{} \mc{V}_{0}\big( \bs{w}_{N} , \bs{y}_N+i\a \bs{e}_N, \bs{y}^{\prime}_N - i\a \bs{e}_N \big) 
\cdot \mc{J}_K^{(\a)}\big( s ;   \bs{w}_{N}, \bs{y}_N, \bs{y}_N^{\prime} \big)  \cdot  \dd \wt{\mu}(\bs{w}_N)   
\enq
in which we do remind the notation $\bs{e}_k = (1,\dots, 1) \in \R^k$ and have set
%
%
\beq
\mc{V}_{\ell}\big( \bs{w}_{p} , \bs{y}_N, \bs{y}^{\prime}_N  \big) \; = \; 
\ex{ \f{\pi}{\hbar}\ov{\bs{y}}_{\ell} } \cdot 
\pl{a=1}{p} \pl{b=\ell+1}{N}  
\bigg\{ \Ga\Big(  \f{ y_b - w_a   }{ i \hbar} , \f{  w_a- y_b^{\prime}  }{ i \hbar}  \Big)   \bigg\}  
\times \pl{a=1}{p} \pl{b=1}{\ell} \bigg\{ \Ga\Big( \f{  w_a- y_b^{\prime}  }{ i \hbar}  \Big) \cdot 
\Ga^{-1} \Big(  \f{  w_a- y_b  }{ i \hbar}  \Big)  \bigg\}   \;. 
\enq
 Further 
\bem
\mc{L}^{(\a)}_{K;k}\big(s, \bs{y}_{N}, \bs{y}^{\prime}_N \big) \; = \;  
\pl{b=1}{N} \pl{\ell=1}{k} \bigg\{ \Ga\Big(  \f{ y_{\ell} - y_b^{\prime} + 2i\a }{ i  \hbar } \Big)  \bigg\}
\pl{\ell=1}{k} \pl{ b=k+1  }{N} \bigg\{ \Ga\Big(  \f{ y_{b} - y_{\ell} }{ i  \hbar } \Big) \bigg\} \\
\times \Int{ (\R+i\eta)^{N-k} }{}   \mc{V}_{k}\big( \bs{w}_{N-k} , \bs{y}_N+i\a \bs{e}_N, \bs{y}^{\prime}_N - i\a \bs{e}_N \big)  \cdot 
  \mc{J}_{K}^{(\a)}\big( s ;  ( \bs{w}_{N-k}, \bs{y}_{k}+i\a\bs{e}_k), \bs{y}_N, \bs{y}_N^{\prime} \big)   
  \cdot \dd \wt{\mu}(\bs{w}_{N-k}) \;.  
\end{multline}

Note that, since $s$, $\bs{y}_N$ and $\bs{y}_N^{\prime}$ all belong to compact sets (and are thus bounded), 
for $K$ large enought, the $\a \tend 0^+$ limit of 
$\mc{J}_{K}^{(\a)}\big( s ;  ( \bs{w}_{N-k}, \bs{y}_{k} + i \a  \bs{e}_k), \bs{y}_N, \bs{y}_N^{\prime} \big)$ 
exists and defines a smooth function. Finally,
\bem
\msc{L}^{(\a)}_{K;k}\big(s, \bs{y}_{N}, \bs{y}^{\prime}_N \big) \; = \;
 \Ga\Big( \f{\ov{\bs{y}}^{\prime}_N - \ov{\bs{y}}_N   +2i\a}{i\hbar}\Big) \cdot
  \pl{b=1}{N}\pl{a=1}{k-1} \Ga\Big( \f{ y_a - y_b^{\prime} + 2i\a }{i\hbar} \Big) 
\cdot   \pl{b=k}{N}\pl{a=1}{k-1} \Ga\Big( \f{ y_b - y_a }{i\hbar} \Big)  \\
 \times   \Int{ (\R+i\eta)^{N-k}}{} \hspace{-3mm}
 \mc{V}_{k-1}\big( \bs{w}_{N-k} , \bs{y}_N+i\a \bs{e}_N, \bs{y}^{\prime}_N - i\a \bs{e}_N \big)  
\cdot \mc{J}^{(\a)}_{K;k}\big( s, \bs{w}_{N-k}, \bs{y}_N, \bs{y}^{\prime}_N \big)  \cdot \dd \wt{\mu}(\bs{w}_{N-k}) 
\end{multline}
where 
\bem
\mc{J}^{(\a)}_{K;k}\big( s, \bs{w}_{N-k}, \bs{y}_N, \bs{y}^{\prime}_N \big) \; = \; 
 \pl{b=1}{k-1} \bigg\{ \Ga^{-1}\Big( \f{  s + K - y_b - \ov{\bs{y}}_N }{i\hbar} \Big)
\cdot  \Ga\Big( \f{  s + K - y_b^{\prime} - \ov{\bs{y}}_N +2i\a }{i\hbar} \Big) \bigg\} \\
\times  \pl{b=k}{N} \bigg\{
 \Ga\Big( \f{ y_b + \ov{\bs{y}}_N- s - K  }{i\hbar}  \Big) \cdot \Ga\Big( \f{s + K - y_b^{\prime} - \ov{\bs{y}}_N +2i\a }{i\hbar}  \Big) \bigg\} \\
\times \pl{a=1}{N-k} \Ga^{-1}\Big( \f{w_a - s -K + \ov{\bs{y}}_N -i\a }{ i \hbar } , \f{ s + K - w_a -  \ov{\bs{y}}_N +i\a }{ i \hbar }  \Big) 
\cdot  \mc{J}^{(\a)}_{K}\big( s, (\bs{w}_{N-k}, \bs{y}_{k-1} + i\a \bs{e}_{k-1}) , \bs{y}_N, \bs{y}^{\prime}_N \big) \; . 
\end{multline}

\noindent The decomposition \eqref{ecriture decomposition L de alpha K apres deplacement ctr}, implies the 
associated decomposition for the norm:
\beq
\mc{N}^2\big[ F_K \big] \; = \; \mc{N}_{K;0}^2  \; + \; 
\sul{p=1}{N}  \f{ N!C^p_N }{ (N-p)! }  \cdot 
\bigg\{ \;   \mc{N}_{K;p}^2  \;+ \; 
\f{ p\cdot  \msc{N}_{K,p}^2 }{ N-p+1} \;   \bigg\} \;. 
\label{ecriture decomposition norme N et termes de diverses singularites}
\enq

We now study the $ K \tend +\infty$ of each of these terms separately.

\subsubsection*{ $\bullet$ Convergence of $\mc{N}_{K;0}$}

After some algebra, straightforwardly justifiable exchanges of orders of integration and a two-fold integration by parts,  $\mc{N}_{K;0}$ can be recast as 
\beq
\mc{N}_{K;0}^2 \; = \; N! \hspace{-5mm} \Int{ \substack{ \Re(w_1) < \dots < \Re(w_N)  \\ \Im(w_a) =  \eta }  }{} 
\hspace{-5mm}   \mc{G}[\mc{J}_K^{(0)}](\bs{w}_N) \cdot 
 \bigg( -i \hbar^{-1}   \ln \Big[ \f{ i w_N }{ \hbar} \cdot \f{ K }{K-w_N}  \Big]     \bigg)^{-2}  \cdot \dd \wt{ \mu }(\bs{w}_N) 
\enq
in which the operator  $\mc{G}$ is given by the below integral representation
\bem
\mc{G}[\mc{J}_K^{(0)}](\bs{w}_N) \; = \; \Int{ \R^{N} }{} \dd^{N} y \Int{ \R^{N} }{} \dd^{N} y^{\prime}  
\Int{ \R }{}  \f{ \dd s }{ 2\pi \hbar } |u( s )|^2   G^*(\bs{y}_N^{\prime})   \cdot
\Big( \f{i w_N}{\hbar} \cdot \f{ K }{K-w_N}  \Big)^{ - i \f{ y_N }{ \hbar} }  \cdot 
 \pl{a\not= b}{N}  \Ga^{-1}\Big( \f{ y_b^{\prime}  - y_a^{\prime}  }{ i \hbar} \Big)   \\
\hspace{-5mm} \times \f{ \Dp{}^2 }{ \Dp{}y_N^2 } \Bigg\{  \Big( \f{i w_N}{\hbar} \cdot \f{ K }{K-w_N} \Big)^{ i \f{ y_N }{  \hbar} }  G(\bs{y}_N) 
 \cdot \pl{a\not= b}{N} \Ga^{-1}\Big( \f{ y_b - y_a}{ i \hbar}  \Big)    \cdot 
\pl{a,b=1}{N} \bigg\{ \Ga\Big(  \f{ y_b - w_a    }{ i \hbar} , \f{  w_a- y_b^{\prime}  }{ i \hbar}  \Big)  \bigg\}
\cdot \mc{J}_K^{(0)}\big( s ;   \bs{w}_{N}, \bs{y}_N, \bs{y}_N^{\prime} \big) \Bigg\} \;. 
\nonumber
\end{multline}
It is readily seen that for $\bs{w}_N \in (\R+i\eta)^N$ and 
$(s, \bs{y}_N,\bs{y}_N^{\prime})$ belonging to a compact subset of $\R^{2N+1}$, one has 
\bem
\mc{J}_K^{(0)}\big( s ;   \bs{w}_{N}, \bs{y}_N, \bs{y}_N^{\prime} \big) \; = \; 
\mc{M}_K(\bs{w}_N) \cdot 
\pl{a=1}{N} \bigg\{ \Big( \f{K-w_a}{K} \Big)^{ \f{\ov{\bs{y}}_N^{\prime}-\ov{\bs{y}}_N }{i\hbar} } 
\cdot  \exp\Big\{ \f{2\pi}{\hbar} (s-\ov{\bs{y}}_N) \bs{1}_{\R^+}(\Re(w_a)-K)  \Big\}\bigg\}  \\
\cdot \bigg(  1+ \e{O}\Big( K^{-1} + \max_{a}\big| w_a-K\big|^{-1}   \Big) \bigg)
\label{ecriture DA fonction J cal K}
\end{multline}
with $\mc{M}_K(\bs{w}_N)$ being given by 
\beq
\mc{M}_K(\bs{w}_N) \; = \; \ex{ - \f{\pi}{\hbar} \ov{\bs{w}}_N }  \cdot 
\pl{a=1}{N} \bigg\{ \Ga\Big( \f{w_a  - K   }{i\hbar} ,\f{  K - w_a   }{i\hbar} \Big)   \cdot 
				\Ga^{-1} \Big( \f{K}{i\hbar} , -\f{K}{i\hbar}  \Big)	\bigg\} \;. 
\enq
It is readily seen that
\beq
\mc{M}_K(\bs{w}_N) \limit{K}{+\infty} 1 \qquad \e{pointwise} \; \e{in} \;  \bs{w}_N \in (\R+i\eta)^N \;. 
\enq

Thus, $ \mc{J}_K^{(0)}$ converges to 1, uniformly in $(s, \bs{y}_N,\bs{y}_N^{\prime})$
and pointwise in $\bs{w}_N \in (\R+i\eta)^N$ in the $K\tend +\infty$ limit. As a consequence
\beq
\lim_{K \tend +\infty} \mc{G}[\mc{J}_K^{(0)}](\bs{w}_N)  \; = \;  \mc{G}[1](\bs{w}_N)
\enq
pointwise in $\bs{w}_N$ \;. 

One needs however sharper bounds so as to apply the dominated convergence theorem to integrals involving $\mc{G}[\mc{J}_K](\bs{w}_N)$. 
Recall that given $\eps>0$ and $(x,y) \in \R^2$ such that $|x+iy|>\eps $ and $|x|< \tf{1}{\eps}$, one has the uniform bound 
\beq
\big| \Ga(x+iy) \big|  \leq C \cdot |x+iy|^{x-\f{1}{2}} \ex{ - \f{\pi}{2} |y| } 
\enq
for some constant $C>0$ depending on the choice of $\eps$. Thus, since $\Im(w_a) = \eta $, it follows that 
\beq
\Big|  \mc{M}_K(\bs{w}_N)  \Big|    
\; \leq \;  C^{\prime} \cdot \pl{a=1}{N} \f{ K }{ |w_a - K |  }    
\enq
for some constant $C^{\prime}>0$. 

This bound along with \eqref{ecriture DA fonction J cal K} and the remainder's differentiable uniformness in 
$(\bs{y}_N , \bs{y}_N^{\prime})$ implies that there exists a constant $\wt{C}>0$ such that uniformly in $K$ large, 
$\bs{w}_N \in (\R+i\eta)^N$ 
and $(s, \bs{y}_N, \bs{y}_N^{\prime}) $ belonging to a compact of $\R^{3N+1}$, one has 
\beq
\max_{\ell=0,1,2} \Big|  \f{ \Dp{}^{\ell} }{ \Dp{} y_N^{\ell} } \, 
\bigg\{ \Big( \f{ K }{K-w_N} \Big)^{ i \f{ y_N }{  \hbar} } \! \! 
\mc{J}_K^{(0)}\big( s ;   \bs{w}_{N}, \bs{y}_N, \bs{y}_N^{\prime} \big)  \bigg\} \Big|   \; \leq  \;
 \wt{C} \cdot \pl{a=1}{N} \bigg\{ \f{ K }{ |w_a - K |  } \bigg\} \;. 
\enq

Furthermore, one also has 
\beq
\pl{a,b=1}{N} \bigg\{ \Ga\Big(  \f{ y_b - w_a    }{ i \hbar} , \f{  w_a- y_b^{\prime}  }{ i \hbar}  \Big) \bigg\} \; = \; 
\pl{a=1}{N }  \Ga^N\Big(  \f{ - w_a    }{ i \hbar} , \f{  w_a  }{ i \hbar}  \Big) 
\cdot  \pl{a=1}{N}  \Big( \f{iw_a}{\hbar} \Big)^{ \f{ \ov{\bs{y}}_N }{ i \hbar } } \cdot 
\Big( \f{w_a}{i\hbar} \Big)^{ i \f{ \ov{\bs{y}}_N^{\prime} }{  \hbar } }
\cdot  \Big(  1 \; + \; \e{O}\big( \max_a |w_a|^{-1} \big) \Big) \;, 
\enq
with a uniformly differentiable remainder in respect to $(\bs{y}_N, \bs{y}_N^{\prime})$ belonging to 
compact subsets of $\R^{2N}$.

As a consequence, taking into account that the integration runs through the compact support of $u(s) G(\bs{y}_N) G^{*}(\bs{y}_N^{\prime})$, one gets that, for some other constant $\wt{C}$, given that $\Re(w_1)<\dots < \Re(w_N)$
\bem
\bigg|  \bigg( \ln \Big[ \f{ i w_N }{ \hbar} \cdot \f{ K }{K-w_N}  \Big]     \bigg)^{-2} \cdot 
\mc{G}\big[ \mc{J}_{K}^{(0)} \big] (\bs{w}_N) \cdot  \wt{\mu}(\bs{w}_N) \bigg|  \;   \leq  \; 
 \wt{C} \pl{a<b}{N} \Big( \f{ \Re(w_a) - \Re(w_b) }{ |w_a w_b| }  \Big)  \\
\times  \pl{a=1}{N-1}  \exp\bigg\{ \f{\pi}{\hbar} \Big[ \Re(w_a)(2a-N) - N | \Re(w_a) |  \Big] \bigg\} 
\cdot \pl{a=1}{N-1} \bigg\{ \f{   K (1+ \ln^2| w_a |) }{ |w_a(K-w_a) | } \bigg\} 
\cdot \f{ K }{ |w_N(K-w_N)| \ln^2|w_N|   } 
\; \leq \; g_K(\bs{w}_N)
\nonumber
\end{multline}
where the sequence $g_K$ reads
\beq
g_K(\bs{w}_N) \; = \; C^{\prime \prime} \pl{a=1}{N-1}  \Big\{  \ex{- \f{\pi}{\hbar} |\Re(w_a)| }  \Big\} 
\cdot  \f{ K }{ |w_N(K-w_N)| \ln^2|w_N|   }  
\enq
and $C^{\prime \prime}>0$ is some constant. 

It follows from lemma \ref{Lemme echangeability limite integrale} that the sequence $g_K$ satisfies 
\beq
\lim_{K \tend +\infty} \hspace{-1mm} \Int{ (\R+i\eta)^N }{} \hspace{-3mm} g_K(\bs{w}_N) \cdot  \dd^N w \; = \; 
		\Int{  (\R+i\eta)^N }{} \hspace{-3mm} \Big( \lim_{K \tend + \infty} g_K(\bs{w}_N) \Big) \cdot  \dd^N w  \;. 
\label{ecriture hypothese TCD pour gK}
\enq
Thus, by the dominated convergence theorem, 
\beq
\lim_{K \tend +\infty} \hspace{-1mm} \mc{N}_{K;0}^2 \; =   \hspace{-3mm}  \Int{ ( \R + i \eta )^N }{} 
 \hspace{-2mm} \dd \wt{ \mu }(\bs{w}_N)  \Int{ \R^{N} }{} \dd^{N} y \Int{ \R^{N} }{} \dd^{N} y^{\prime}  
  G^*(\bs{y}_N^{\prime})  G(\bs{y}_N)  \cdot
 \f{ \pl{a,b=1}{N} \bigg\{ \Ga\Big(  \f{ y_b - w_a    }{ i \hbar} , \f{  w_a- y_b^{\prime}  }{ i \hbar}  \Big)  \bigg\} }
 { \pl{a\not= b}{N} \bigg\{ \Ga\Big( \f{ y_b - y_a}{ i \hbar} , \f{ y_b^{\prime}  - y_a^{\prime}  }{ i \hbar} \Big) \bigg\}  }\;. 
\enq
%
%
%



\subsubsection*{$\bullet$ Convergence of $\mc{N}_{K;k}$ }

It follows by expanding the singular part of the prefactor in $\mc{L}^{(\a)}_{K;k}$ and then integrating the 
singular term by parts, that $\mc{N}_{K;k}$ can be recast as
\beq
\mc{N}_{K;k}^2 \; = \;\; \f{ N! }{N-k+1} \hspace{-7mm} \Int{ \substack{ \Re(w_1) < \dots < \Re(w_{N-k}) \\
									\Im(w_a)= \eta }		  }{  } 
\hspace{-5mm}  \mc{G}_k\big[ \mc{J}^{(0)}_K \big](\bs{w}_{N-k}) \cdot \dd \wt{\mu}(\bs{w}_{N-k}) 
\enq
where 
\bem
\mc{G}_k\big[ \mc{J}^{(0)}_K \big](\bs{w}_{N-k}) \; = \; \Int{\R^N}{} \dd^N y \Int{ \R^N }{} \dd^N y^{\prime}
\Int{\R}{} \dd s \cdot \pl{ a=1 }{ k } \Big\{ \ln | y_a - y^{\prime}_a |   +i\pi \bs{1}_{\R^+}( y_a^{\prime}  -  y_a )  \, \Big\}
 \\
\times  \pl{a=1}{k} \f{ \Dp{} }{\Dp{}y_a^{\prime}}  \, \cdot \, \Big\{  \wt{\mc{V}}_{k}(s, \bs{w}_{N-k}, \bs{y}_N, \bs{y}_N^{\prime} ) 
\cdot \mc{J}^{(0)}_K\big( s, (\bs{w}_{N-k},\bs{y}_k), \bs{y}_N, \bs{y}^{\prime}_N   ) \Big \} 
\end{multline}
with the function $\wt{\mc{V}}_{k}$ being given by  
\bem
\wt{\mc{V}}_{k}(s, \bs{w}_{N-k}, \bs{y}_N, \bs{y}_N^{\prime} ) \; = \; \f{ |u(s)|^2 }{ 2\pi \hbar } G(\bs{y}_N) G^{*}(\bs{y}_N^{\prime})
\pl{a=1}{k} \pl{b=k+1}{N} \bigg\{  \Ga\Big( \f{y_{b}- y_a}{i\hbar} \Big) \f{1}{y_a^{\prime}-y^{\prime}_{b}} \bigg\}
  \f{   \mc{V}_{k}(\bs{w}_{N-k}, \bs{y}_N, \bs{y}_N^{\prime} )   }
{ \pl{b>\ell}{k} \Big\{ \big(y_b \, - \, y_{\ell} \big) \cdot \big(y_{\ell}^{\prime}-y^{\prime}_{b} \big)  \Big\} }
  \\
 \times \pl{ \substack{ a\not = b } }{ N } 
 \bigg\{ \Ga^{-1}\Big( \f{y_b^{\prime} - y_a^{\prime} }{i\hbar} , \f{y_b - y_a }{i\hbar} \Big)\bigg\}
 \cdot  \pl{b=1}{N} \pl{a=1}{k}\bigg\{ i\hbar \Ga\Big( \f{y_a - y_b^{\prime} }{i\hbar}  + 1 \Big)   \bigg\}  \;.
\end{multline}

 \noindent The analysis of the $K\tend  +\infty$ limit slightly differs depending on whether $k=N$ or $1\leq k \leq N-1$. 

\vspace{2mm} 
Suppose that $k=N$, then it is readily seen from the uniform differentiability of the $K\tend +\infty$  remainder
to $\mc{J}^{(0)}_K\big( s, \bs{y}_N, \bs{y}_N, \bs{y}^{\prime}_N   ) $ that 
\beq
\max_{ A \subset \intn{1}{N}}  \bigg| \pl{a \in A }{}  \f{ \Dp{} }{ \Dp{}y_a^{ \prime} }  \cdot 
\Big[ \mc{J}^{(0)}_K\big( s, \bs{y}_N, \bs{y}_N, \bs{y}^{\prime}_N   )  - 1\Big]  \bigg| \; = 0
\enq
and thus, 
\beq
\mc{G}_N\big[ \mc{J}^{(0)}_K \big] \limit{K}{+\infty} \mc{G}_N\big[ 1 \big]   \qquad ie \qquad 
\lim_{K\tend +\infty} \mc{N}_{K;N} \; = \; N! \cdot  \mc{G}_N\big[ 1 \big]  \; . 
\enq

Now assume that $1\leq k \leq N-1$. Then, we change the variables $y_N^{\prime}= v^{\prime} - \ov{\bs{y}}_{N-1}^{\prime}$ and integrate
 by parts the oscillatory asymptotic behaviour in $\bs{w}_{N-k}$. Agreeing upon 
\beq
\bs{v}_N^{\prime} \; = \; (\bs{y}_{N-1}^{\prime}, v^{\prime} - \ov{\bs{y}}_{N-1}^{\prime}) 
\enq
we recast $\mc{G}_k$ as
\bem
\mc{G}_k\big[ \mc{J}^{(0)}_K \big](\bs{w}_{N-k}) \; = \;
 \bigg[  \f{ -i }{ \hbar } \ln \bigg(  \f{ i w_{N-k} \cdot K }{ \hbar (K - w_{N-k}) }  \bigg)     \bigg]^{ -2  }
\Int{\R^N}{} \dd^N y \hspace{-2mm} \Int{ \R^{N-1} \times \R }{} \hspace{-3mm} \dd^{N-1}\! y^{\prime}  \cdot \dd v^{\prime}
\Int{\R}{} \dd s \cdot \pl{ a=1 }{ k } \Big\{ \ln | y_a - y^{\prime}_a |   + i\pi \bs{1}_{\R^+}(y_a^{\prime} - y_a)  \Big\}
 \\
\times  \bigg(  \f{ i w_{N-k}\cdot K }{ \hbar (K - w_{N-k}) } \bigg)^{ i\f{y_N}{\hbar} } \cdot 
  \f{ \Dp{}^{2} }{ \Dp{}y_N^{2} }   \pl{a=1}{k} \f{ \Dp{} }{\Dp{}y_a^{\prime}}  \, \cdot \, 
\Bigg\{ \bigg(   \f{ i w_{N-k} \cdot K }{ \hbar (K - w_{N-k}) }  \bigg)^{ \f{y_N}{i\hbar} } 
 \wt{\mc{V}}_{k}(s, \bs{w}_{N-k}, \bs{y}_N, \bs{v}_N^{\prime} ) 
\mc{J}^{(0)}_K\big( s, (\bs{w}_{N-k},\bs{y}_k), \bs{y}_N, \bs{v}^{\prime}_N   )  \Bigg\} \;. 
\nonumber
\end{multline}

It then follows from the large $w_a$ representation 
\beq
\mc{V}_{k}( \bs{w}_{N-k}, \bs{y}_N, \bs{v}_N^{\prime} )  \; = \; 
\ex{\f{\pi}{\hbar} \ov{\bs{y}}_k} \pl{a=1}{N-k} \bigg\{ \Ga\Big( \f{-w_a}{i\hbar} , \f{w_a}{i\hbar} \Big) \bigg\}^{N-k}
\pl{a=1}{N-k} \bigg\{   \Big(  \f{ - w_a }{ i\hbar }  \Big)^{ \f{ \ov{\bs{y}}_N -\ov{\bs{y}}_k }{i \hbar} }  
 \cdot \Big(  \f{  w_a }{ i\hbar }  \Big)^{   \f{  \ov{\bs{y}}_k - v^{\prime}  }{i\hbar} }   \bigg\}  
 \cdot \Big(  1+ \e{O}\big( \max_a |w_a|^{-1} \big) \Big)
\label{ecriture comportement dominant en w de Vk}
\enq
that 
\bem
\max_{A \subset \intn{1}{k}} \max_{\ell \in \intn{0}{2} } 
\bigg|  \f{ \Dp{}^{\ell} }{ \Dp{}y_N^{\ell} }  \pl{a \in A }{}  \f{ \Dp{} }{ \Dp{}y_a^{\prime} }  \cdot 
\bigg\{ \Big(  \f{ i w_{N-k} }{ \hbar }  \Big)^{  i \f{y_N}{  \hbar} } 
 \wt{\mc{V}}_k(s, \bs{w}_{N-k}, \bs{y}_N, \bs{v}_N^{\prime} ) \bigg\} \bigg| \\
\; \leq \; C \pl{a=1}{N-k-1} \! \!  \Big\{ 1+ \big| \ln |w_a| \big|^{2+k} \Big\}  \cdot
 \pl{a=1}{N-k} \Big| \Ga\Big( \f{-w_a}{i\hbar} , \f{w_a}{i\hbar} \Big) \Big|^{N-k} \;. 
\label{ecriture majorant V tilde k}
\end{multline}

Likewise, it follows from \eqref{ecriture DA fonction J cal K} specialised to the case 
where $\bs{w}_N = \big( \bs{w}_{N-k}, \bs{y}_k \big)$ that 
\beq
\max_{A \subset \intn{1}{k}} \max_{\ell \in \intn{0}{2} } 
\bigg|  \f{ \Dp{}^{\ell} }{ \Dp{}y_N^{\ell} }  \pl{a \in A }{}  \f{ \Dp{} }{ \Dp{}y_a^{\prime} }  \cdot 
\bigg\{ \Big( \f{ K }{K-w_{N-k} } \Big)^{i\f{y_N}{\hbar}} \cdot  
 \mc{J}^{(0)}_K\big( s, (\bs{w}_{N-k},\bs{y}_k), \bs{y}_N, \bs{v}^{\prime}_N   ) \bigg\}  \bigg| 
\; \leq \; \wt{C} \cdot \pl{a=1}{N-k} \bigg\{ \f{ K }{ | w_a - K | } \bigg\}  \;.
\nonumber  
\enq

Thus, by repeating the previously discussed schemes of majorations, 
we arrive to 
\bem
\bigg| 	\mc{G}_k\big[ \mc{J}^{(0)}_K \big](\bs{w}_{N-k})  \cdot \wt{\mu}(\bs{w}_{N-k}) \bigg| \; \leq \; 
 \wt{C} \cdot 
\pl{a=1}{N-k-1} \bigg\{ K \cdot \f{ 1+ |\ln w_a |^{2+k} }{ |w_a(K-w_a)| } \cdot \ex{ - 2\f{\pi}{\hbar} |\Re(w_a) | }   \bigg\} 
 \cdot 
		\f{  K \cdot \big( |\ln w_{N-k} | \big)^{-2}  }{ |w_{N-k}(K-w_{N-k})| }  \\
\; \leq \; 	\wt{C}^{\prime} \cdot 	\pl{a=1}{N-k-1} \Big\{  \ex{ - \f{\pi}{\hbar} |\Re(w_a) | }   \Big\}
\times \bigg\{ \f{  K \cdot \big( |\ln w_{N-k} | \big)^{-2} }{ |w_{N-k}(K-w_{N-k})| }    \bigg\}   \;. 
\end{multline}

The \textit{rhs} of the last inequality does already fulfil the hypothesis on the dominant
function in Lebesgue's dominated convergence theorem, so that 
\beq
\lim_{K\tend +\infty} \mc{N}_{K;\, k}^2 \; = \;  \f{N!}{(N-k+1)!}  \hspace{-3mm} \Int{    (\R + i\eta)^{N-k} }	{  } 
\hspace{-3mm}  \mc{G}_k\big[  1  \big](\bs{w}_N) \cdot \dd \wt{\mu}(\bs{w}_{N-k}) \;. 
\enq
%
%
%



\subsubsection*{$\bullet$ Convergence of $\msc{N}_{K;k}$ }

Upon repeating the aforediscussed manipulations and carrying out the change of variables
\beq
y_N  \; = \;  v - \ov{\bs{y}}_{N-1} \qquad \e{and} \qquad  y_N^{\prime} \; = \;  v^{\prime} - \ov{\bs{y}}_{N-1}^{\prime}  
\enq
it is readily seen that $\msc{N}_{K;k}$ can be recast as 
\beq
\msc{N}_{K;\,k}^{2} \; = \;\;  N! \hspace{-5mm} \Int{ \substack{ \Re(w_1) < \dots < \Re(w_{N-k}) \\
									\Im(w_a)=  \eta }		  }{  } 
\hspace{-5mm}  \msc{G}_k\big[ \mc{J}^{(0)}_{K;k} \big](\bs{w}_N) \cdot \dd \wt{\mu}(\bs{w}_{N-k}) 
\enq
where 
\bem
\msc{G}_k\big[ \mc{J}^{(0)}_{K;k} \big](\bs{w}_{N-k}) \; = \; i\hbar \hspace{-3mm}\Int{ \R^{N-1}\times \R }{} \hspace{-3mm} \dd^{N-1} y \cdot \dd v 
 \hspace{-2mm} \Int{ \R^{N-1} \times \R }{} \hspace{-3mm} \dd^{N-1} y^{\prime} \cdot \dd v^{\prime}
\Int{\R}{} \dd s 
\; \;  \pl{ a=1 }{ k-1 } \Big\{ \ln | y_a - y^{\prime}_a |  + i\pi \bs{1}_{\R^+}(y_a^{\prime} - y_a)  \, \Big\}
 \\
\times \Big\{ \ln | v - v^{\prime} |  +i\pi \bs{1}_{\R^+}(v^{\prime} - v)  \, \Big\}  \cdot 
 \f{ \Dp{} }{\Dp{}v^{\prime}} \pl{a=1}{k-1} \f{ \Dp{} }{\Dp{}y_a^{\prime}}  \, \cdot \, 
 \bigg\{ \Ga\Big( \f{v-v^{\prime}}{i\hbar} + 1  \Big) \wt{\mc{V}}_{k-1}(s, \bs{w}_{N-k}, \bs{v}_N, \bs{v}_N^{\prime} ) 
\cdot \mc{J}^{(0)}_{K;k}\big( s, \bs{w}_{N-k}, \bs{v}_N, \bs{v}^{\prime}_N   ) \bigg\} \;.
\nonumber
\end{multline}
On this occasion, we do remind that 
\beq
\bs{v}_N \; = \; \big( \bs{y}_{N-1}, v-\ov{\bs{y}}_{N-1} \big) \quad \e{and} \quad 
\bs{v}_N^{\prime} \; = \; \big( \bs{y}_{N-1}^{\prime}, v^{\prime} - \ov{\bs{y}}_{N-1}^{\prime} \big)
\enq
The function $\mc{J}^{(0)}_{K;k}$ exhibits the large $K$ behaviour 
\bem
\mc{J}^{(0)}_{K;k}\big( s, \bs{w}_{N-k}, \bs{v}_N, \bs{v}^{\prime}_N   )  \; = \;
\exp\Big\{   \f{\pi}{\hbar}(N-k+1)(v-s) \, - \, \f{\pi}{2\hbar}(v+v^{\prime}-2\ov{\bs{y}}_{k-1})   \Big\}
 \\
\times \Big(  \f{ K }{ \hbar } \Big)^{ \f{v-v^{\prime} }{i\hbar} } \cdot
  \Ga \bigg( - \f{ K }{ i \hbar } , \f{ K }{ i \hbar } \bigg) \cdot \ex{- \f{\pi}{\hbar} \ov{\bs{w}}_{N-k} }
\pl{a=1}{N-k} \bigg\{   \Big( \f{K-w_a}{K} \Big)^{ \f{v^{\prime}-v}{i\hbar} }  \bigg\} 
\cdot \Bigg(  1 \;  + \; \e{O}\Big( K^{-1} + \max_{a}|w_a - K |^{-1} \Big)   \bigg)
\end{multline}
Furthermore, the behaviour of $\wt{\mc{V}}_{k-1}$ is given by 
\eqref{ecriture comportement dominant en w de Vk}-\eqref{ecriture majorant V tilde k}, with the sole difference 
that one should replace in these expressions the variable $\ov{\bs{y}}_N$ by $v$. 

\vspace{2mm}
\noindent These informations along with the uniform differentiability of the remainders lead to the bound
\bem
\Bigg| \wt{\mu}(\bs{w}_{N-k}) \f{ \Dp{} }{\Dp{}v^{\prime}} \pl{a=1}{k-1} \f{ \Dp{} }{\Dp{}y_a^{\prime}}  \, \cdot \, 
 \bigg\{ \Ga\Big( \f{v-v^{\prime}}{i\hbar} + 1  \Big) \wt{\mc{V}}_{k-1}(s, \bs{w}_{N-k}, \bs{v}_N, \bs{v}_N^{\prime} ) 
\cdot \mc{J}^{(0)}_{K;k}\big( s, \bs{w}_{N-k}, \bs{v}_N, \bs{v}^{\prime}_N   ) \bigg\} \Bigg|  \\
\; \leq \; C  \cdot   \Ga \big( - \f{ K }{ i \hbar } , \f{ K }{ i \hbar } \Big) 
\cdot \pl{a<b}{N-k} \Big| \f{w_a - w_b}{w_a w_b} \Big|\; \cdot \; 
\pl{a=1}{N-k} \Big\{ |w_a|^{-2} \big( 1+|\ln w_a| \big) \cdot  \big( 1+|\ln [\tf{(K-w_a)}{K}] | \big)   \Big\} \\
\times \pl{a=1}{N-k}\bigg\{  \ex{-\f{\pi}{\hbar}\big( (N-k+1)|\Re(w_a)| - (2a-N+k+1) \Re(w_a)\big) } \bigg\} 
\; \leq \; C^{\prime}  \cdot   \Ga \big( - \f{ K }{ i \hbar } , \f{ K }{ i \hbar } \Big) 
 \pl{a=1}{N-k} \f{1}{ |w_a|^{ \f{3}{2} } } \;, 
\end{multline}
for some constants $C,C^{\prime}>0$. 

Taking into account that the $\dd^N\nu \cdot \dd^N\nu^{\prime}$ integration runs through the compact support of $u(s) G(\bs{y}_N) G^{*}(\bs{y}_N^{\prime})$,
the above bound allow us to assert, in virtue of the dominated convergence theorem, that, independently of $k\in \intn{1}{N}$, 
\beq
\lim_{K\tend +\infty} \msc{N}_{K;k} \; = \; 0 \;. 
\enq

In order to conclude, it is enough to carry out  backwards, once that the $K\tend +\infty$ limits of interest have been taken, 
the chain of contour deformation that originally led to \eqref{ecriture decomposition norme N et termes de diverses singularites}. The
reverse chain of transformations leads to a representation for $\lim_{K\tend +\infty} \mc{N}^2[F_K]$ that has 
the desired form. Indeed, all of our manipulations simply amount to having set, from the
very beginning, $\mc{J}^{(0)}_K =1$ in \eqref{definition de L cal alpha}.  
We do stress that the role of the function $\mc{J}^{(\a)}_K$ in generating pole contributions in the process
of deforming the contours was passive, apart from generating poles at $w_a = s+K-\ov{y}_N$. The latter 
have been shown to yield vanishing contributions in the $K\tend + \infty$ limit. Hence, setting $\mc{J}^{(0)}_K =1$  is not an
obstruction to taking the reverse chain of transformations. 
Then, as soon as one sets $\mc{J}^{(0)}_K =1$ in \eqref{definition de L cal alpha},
one can readily permute the orders of integrations in \eqref{forme norme FK ac lim alpha go to zero}
and reconstruct the product of two $\msc{H}_N$ transforms leading to 
\beq
\lim_{K\tend +\infty} \mc{N}^2\big[ F_K \big] \; = \; (2\pi \hbar)^{2N} \cdot (N!)^2 \cdot 
\big| \big| \msc{H}_N[G] \big| \big|^2_{L^2_{\e{sym}}\big( \R^N , \dd \wt{\mu}(\bs{w}_N) \big)} \;. 
\label{ecriture limite N F K version 2 avec lien HN transfo}
\enq
A comparison of \eqref{ecriture limite N F K version 2 avec lien HN transfo} and \eqref{ecriture N F K version fonction G seule}
yields the claim. \qed

\subsection{The integral transform $\msc{J}_N$}

We now use the isometricity of the $\msc{H}_N$-transform so as to establish the one of a transform $\msc{J}_N$.
The latter is already directly of interest to the problem. The results of the present section imply the
the isometricity of $\mc{V}_N$. 

\begin{prop}
\label{Prop charactere isometrique J}

Let $R \in \msc{C}^{\infty}_{ \e{c}; \e{sym}\times- }(\R^{N-1}\times \R)$, then the integral transform
\beq
\msc{J}_N\big[ R \big](\bs{w}_N) \; = \;   \lim_{\a \tend 0^+} 
\Int{\R^{N-1}}{} \hspace{-2mm} \f{\dd^{N-1} y }{ (2\pi \hbar)^{N-1} }
\Int{\R}{} \hspace{-1mm}  \dd x \cdot \f{  \ex{ \f{i}{\hbar} \ov{\bs{w}}_{N}x} \cdot   R(\bs{y}_{N-1},x) }
{ \sqrt{N} \cdot (N-1)! }
\cdot \pl{a\not= b}{N-1} \Ga^{-1}\Big( \f{y_a-y_b}{i\hbar} \Big)
\cdot \pl{a=1}{N} \pl{b=1}{N-1} \Ga\Big(  \f{y_b-w_a + i\a }{ i \hbar }  \Big)
\enq
belongs to $\mc{S}( \R^{N} ) \cap  L^{2}_{\e{sym}}\big(\R^N, \dd \mu(\bs{y}_N)\big)$ and satisfies to 
the identity 
\beq
\big| \big| \msc{J}_N\big[ R \big] \, \big| \big|_{L^{2}_{\e{sym}}\big(\R^N, \dd \mu(\bs{w}_N)\big)}
\; = \; ||\,  R \, ||_{L^{2}_{\e{sym}\times-}\big(\R^{N-1}\times \R, \dd \mu(\bs{y}_{N-1})\otimes \dd x\big)}
\label{charactere isometrique application JN}
\enq

\end{prop}

The isometric identity \eqref{charactere isometrique application JN} allows one to raise 
$\msc{J}_N$ into an isometric operator 
\beq
\msc{J}_N \; : \;  L^{2}_{\e{sym}\times-}\big(\R^{N-1}\times \R, \dd \mu(\bs{y}_{N-1})\otimes \dd x\big) \; \tend \; 
 L^{2}_{\e{sym}}\big(\R^N, \dd \mu(\bs{w}_N)\big) \;. 
\enq

\Proof 

Given an $R \in \msc{C}^{\infty}_{\e{c};\e{sym}\times -}(\R^{N-1}\times \R) $, the 
$\mc{S}(\R^{N-1}\times \R)\cap  L^{2}_{\e{sym}}\big(\R^N, \dd \mu(\bs{y}_N)\big)$ nature of the transform
is established with the help of technique discussed in the course of the analysis of the $\msc{H}_N$ transform. 
Accordingly, we shall not reproduce these arguments once again. 

Let  $R \in \msc{C}^{\infty}_{\e{c};\e{sym}\times -}(\R^{N-1}\times \R) $ and $G_K $ be defined as 
\beq
G_K(\bs{y}_N) \; = \;   \e{Sym} \bigg\{  R\big( \bs{y}_{N-1}, \ln(-y_{N}/K) \, \big) 
\f{ \bs{1}_{\R^-}(y_N) }{ y_N }  \ex{-\f{\pi}{\hbar} \ov{\bs{y}}_{N-1}}
\Ga^{N-1}\Big(-\f{y_N}{i\hbar} \Big) \cdot \Ga^{-1}\Big(\f{y_N}{i\hbar} \Big)   \bigg\}
\enq
where $\bs{1}_{A}$ stands for the indicator function of the set $A$. 
It is then readily seen that $G_K \in \msc{C}^{\infty}_{\e{c};\e{sym}}(\R^N)$. It follows from 
the proposition \ref{Proposition charactere isometrique de la transfo H} that 
\beq
\wt{\mc{N}}[ G_K \,] \; = \; || \, G_K \,  ||_{ L^{2}_{\e{sym}}\big(\R^N, \dd \wt{\mu}(\bs{y}_N)\big) }
 \; = \;  \big| \big| \, \msc{H}_N \big[ G_K \big] \, \big| \big|_{ L^{2}_{\e{sym}}\big(\R^N, \dd \wt{\mu}(\bs{w}_N)\big) }\;. 
\enq
Just as in the proof of that proposition, we use this equality so as to estimate the $K \tend +\infty$ limit 
of $\mc{N}[ G_K \,]$ in two different ways. These estimations will then allow us to extract the isometricity
relation \eqref{charactere isometrique application JN}. 

\vspace{3mm}

First, we focus on the $L^2$-norm of $G_K$. Then, in virtue of the symmetry of the function $R$, we have the decomposition 
\beq
G_K(\bs{y}_N) \; = \;  \f{ 1 }{N}\sul{p=1}{N} R\big( \bs{y}_{N-1}^{(p)},   \ln(-y_{p}/K) \, \big) 
\f{ \bs{1}_{\R^-}(y_p) }{ y_p } \cdot  \ex{ - \f{\pi}{\hbar} \ov{\bs{y}}_{N-1}^{(p)} }
\cdot \Ga^{N-1}\Big(-\f{y_p}{i\hbar} \Big) \cdot \Ga^{-1}\Big(\f{y_p}{i\hbar} \Big) 
\enq
where
\beq
\bs{y}_{N-1}^{(p)} \; = \; \big( y_1,\dots, y_{p-1}, y_{p+1}, \dots, y_N \big) \;.
\enq
The "off-diagonal" products in $ | G_K(\bs{y}_N) |^2 $ will involve 
$R\big( \bs{y}_{N-1}^{(p)},  \ln(-y_{p}/K) \, \big) \cdot \Big( R\big( \bs{y}_{N-1}^{(j)},  \ln(-y_{j}/K) \, \big) \Big)^*$
with $p\not= j$. However, since $R$ is compactly supported, it follows that there exists compacts 
$L \Subset \R^{N-1}$ and $J\Subset \R$ such that $R(\bs{z}_{N-1},x)=0$ if either 
$\bs{z}_{N-1} \not \in L$ or $x \not\in J$. Yet, $ \ln(-y_{p}/K) \in J$ imples that 
$y_p \in K \ex{-J}$. Observe that, for $K$ large enough, since $0\not \in \ex{-J}$, 
should $y_j, y_p \in K \ex{-J}$ with $j\not= p$, then necessarily $\bs{y}_{N-1}^{(p)} \not \in L$.
As a consequence, then, $R\big( \bs{y}_{N-1}^{(p)}, \ln(-y_{p}/K) \, \big)=0$ and 
the corresponding term does not contribute to the norm. All in all, this means 
\beq
\wt{\mc{N}}^2[G_K\,] \; = \;  \f{1}{N}\Int{ \R^{N-1} \times \R }{} 
\big| R(\bs{y}_{N-1},x)\big|^2 \cdot 
\mc{T}_K^{(1)}(\bs{y}_{N-1}, x ) \cdot \dd\mu(\bs{y}_{N-1}) \otimes \f{ \dd x }{ (2\pi \hbar)^2 }
\enq
in which
\beq
\mc{T}_K^{(1)}( \bs{y}_{N-1}, x) \; = \;\f{  2\pi \hbar  }{   K  \ex{x} }
\cdot \f{  \ex{ -\f{\pi}{\hbar} (K \ex{x} + \ov{\bs{y}}_{N-1}) }  }
{ \Ga\Big(-K\f{\ex{x}}{i\hbar} \, , K\f{\ex{x}}{i\hbar}\Big)  }
\cdot \pl{a=1}{N-1} \Bigg\{ \f{  \Ga\Big(-K\tf{\ex{x}}{i\hbar} , K\tf{\ex{x}}{i\hbar}\Big)    }
{ \Ga\Big(-\f{K\ex{x}+y_a}{i\hbar}\,  ,  \f{ K\ex{x}+y_a }{i\hbar}\Big) }
 \Bigg\} \;. 
\enq
It is readily seen that, uniformly in $(\bs{y}_{N-1},x)$ belonging to compact subsets of $\R$, 
\beq
\lim_{K \tend +\infty} \mc{T}_K^{(1)}(\bs{y}_{N-1},x)  \; = \; 1 \;. 
\enq
Thus, by the dominated convergence theorem, 
\beq
\lim_{K\tend +\infty} \wt{\mc{N}}[G_K  ] \; = \; \f{1}{\sqrt{N} \cdot (2\pi \hbar) } \cdot 
\norm{ R }_{L^{2}_{\e{sym}\times -} \big(\R^{N-1}\times \R, \, \dd \mu(\bs{y}_{N-1})\otimes \dd x\big)} \;. 
\label{ecriture version 1 de lim N tile G K}
\enq

We now estimate the same limit while using the second representation for $\wt{\mc{N}}[G_K\,]$. 
Straightforward calculations based on the previously introduced ideas then lead to 
\bem
\wt{\mc{N}}^2[G_K\,] \; = \;  \lim_{\a\tend 0^+} \Int{\R^N}{} \dd\mu(\bs{w}_N) 
\hspace{-3mm}\Int{\R^{N-1}\times \R }{} \hspace{-2mm} \f{ \dd^{N-1} y \otimes \dd x }{ N! (2\pi \hbar)^N }  \; R(\bs{y}_{N-1},x)  
\ex{i\f{x}{\hbar} \ov{\bs{w}}_N}
\pl{a\not= b}{N-1} \Ga^{-1}\Big( \f{y_a-y_b}{i\hbar} \Big)
\pl{a=1}{N} \pl{b=1}{N-1} \Ga\Big(  \f{y_b-w_a + i\a }{ i \hbar }  \Big) \\
\times \Int{\R^{N-1}\times \R }{} \hspace{-4mm}\f{ \dd^{N-1} y^{\prime} \otimes \dd x^{\prime} }{ N! (2\pi \hbar)^N }
\Bigg\{  R(\bs{y}_{N-1}^{\prime},x^{\prime})  \ex{i\f{\pi}{\hbar}x^{\prime} \ov{\bs{w}}_N} 
 \pl{a\not= b}{N-1} \Ga^{-1}\Big( \f{y_a^{\prime}-y_b^{\prime}}{i\hbar} \Big)
\pl{a=1}{N} \pl{b=1}{N-1} \Ga\Big(  \f{y_b^{\prime}-w_a + i\a }{ i \hbar }  \Big) \Bigg\}^*
\mc{T}_{K;\a}^{(2)}\Big( x ,  x^{\prime} ,\bs{y}_{N-1}, \bs{y}_{N-1}^{\prime} \Big)
\end{multline}
where, this time, we have set 
\bem
\mc{T}_{K; \a}^{(2)}\big( x, x^{\prime},\bs{y}_{N-1}, \bs{y}_{N-1}^{\prime} \mid \bs{w}_N\big) \; = \;
\ex{ \f{\pi}{\hbar}\ov{\bs{w}}_N } \cdot \ex{ -i\f{\pi}{\hbar}(x-x^{\prime})\ov{\bs{w}}_N  }
\cdot 
\pl{a=1}{N} \Ga \left( \ba{c} i\tf{ (K\ex{x}+w_a - i \a ) }{\hbar} \, , -i\tf{(K\ex{x^{\prime}}+w_a + i \a) }{\hbar}    \\ 
  i \tf{ K \ex{x} }{\hbar} \, , \;  -i \tf{ K \ex{x^{\prime}} }{\hbar} \ea \right) \\
\times \ex{-\f{\pi}{\hbar} (\ov{\bs{y}}_{N-1} + \ov{\bs{y}}^{\prime}_{N-1}) }  \cdot 
\pl{a=1}{N-1} \Ga\left(\ba{c c c c} 
i \tf{ K \ex{x} }{\hbar} \, , &  -i \tf{ K \ex{x} }{\hbar} \, , &
i \tf{ K \ex{x^{\prime}} }{\hbar} \, , & -i \tf{ K \ex{x^{\prime}} }{\hbar} \\
 i\tf{(K\ex{x}+y_a)}{\hbar} \, , & -i\tf{ (K\ex{x}+y_a) }{\hbar} \, , &
i\tf{(K\ex{x^{\prime}}+y_a^{\prime}) }{\hbar} \, , & i\tf{(K\ex{x^{\prime}}+y_a^{\prime}) }{\hbar}   \ea\right) \;. 
\end{multline}

A straightforward computation shows that uniformly in $(\bs{y}_{N-1},x), (\bs{y}_{N-1}^{\prime},x^{\prime})$
and $\bs{w}_N$ belonging to compact subsets of $\R^N$, one has that 
\beq
\lim_{K \tend +\infty} \mc{T}_{K;0}^{(2)}\big( x, x^{\prime},\bs{y}_{N-1}, \bs{y}_{N-1}^{\prime} \mid \bs{w}_N \big) \; = \; 1 \;. 
\enq
It then remains to repeat the handlings outlined in the course of the proof of proposition 
\ref{Proposition charactere isometrique de la transfo H}
so as to show that this type of convergence is, in fact, enought so as to 
take the $K\tend + \infty$ limit under the integral sign. Since these are basically the same, we do not reproduce them here again. 
One thus gets 
\beq
\lim_{K \tend + \infty}  \wt{\mc{N}}[\, G_K\,] \; = \; 
\f{1}{\sqrt{N}\cdot (2\pi \hbar) } \norm{ \msc{J}_N\big[ R \big] }_{L^{2}_{\e{sym}} \big(\R^N, \dd \mu(\bs{w}_N)\big)} \;. 
\label{ecriture version JN de lim N tile G K}
\enq
Equations \eqref{ecriture version 1 de lim N tile G K} and \eqref{ecriture version JN de lim N tile G K}
put together lead to the claimed identity. \qed

\subsection{ Isometric character of the transform $\mc{V}_N$}

\begin{theorem}

The transform $\mc{V}_N$ defined through \eqref{definition transformation VN} is such that 
given  any
\beq
F\in \msc{C}^{\infty}_{\e{c}}\big( \R^N \big) \; ,  \qquad 
\mc{V}_N\big[ F\big]  \in \mc{S}(\R^N )\cap L^{2}_{\e{sym}}\big(\R^N, \dd \mu(\bs{y}_N) \big) \;. 
\enq
Furthermore,  one has
\beq
\big|\big|  \mc{V}_N\big[ F\big]     \big|\big|_{L^{2}_{\e{sym}}\big( \R^N, \dd \mu(\bs{y}_N) \big) } \; = \; 
 || F ||_{L^{2}\big(\R^N, \dd^N x\big)} \;. 
\enq

\end{theorem}

As a consequence, $  \mc{V}_N$ extends to an isometric operator
\beq
  \mc{V}_N \; :  \;  L^{2}\big(\R^N, \dd^N x\big)  \tend   
  L^{2}_{\e{sym}}\big(\R^N, \dd \mu(\bs{y}_N) \big) \;. 
\enq
This property, written formally, amounts to the so-called completeness of the system $\{ \vp_{\bs{y}_N}\big( \bs{x}_N \big) \}$, 
\textit{cf} \eqref{ecriture relation completude par rapport espace SoV}. 

\Proof 

It follows from the recurrence relation \eqref{definition fct Whittaker par Mellin-Barnes} 
satisfied by the functions $\vp_{\bs{y}_N}\big( \bs{x}_N \big)$  that
 $\mc{V}_N[F]$  can be recast as  
\bem
\mc{V}_N[ F ](\bs{y}_N) \; = \; \hbar^{\f{ \ov{\bs{y}}_N }{i \hbar}(N-1)  } \cdot 
 \lim_{\a \tend 0^+} \hspace{-2mm} \Int{ \R^{N-1}\times \R }{} \hspace{-3mm}
\f{  \dd\mu( \bs{w}_{N-1} ) \otimes \dd x }{ \sqrt{N} \cdot (N-1)!  } \cdot \ex{ \f{i}{\hbar} x \ov{\bs{y}}_{N-1} }
\cdot \pl{a= 1}{ N-1 }  \pl{b= 1}{ N } \Ga\Big( \f{ y_b - w_a+i\a }{ i\hbar }  \Big) 
\wt{\mc{V}}_{N-1}\big[ F(*,x) \big](x,\bs{w}_{N-1}) \\
\; = \; \hbar^{-\f{i}{\hbar}(N-1) \ov{\bs{y}}_N }  \cdot \f{1}{\sqrt{N} }
\ov{\msc{J}}_N \big[ \wt{\mc{V}}_{N-1} \big[ F(*,x) \big](x,\bs{w}_{N-1})  \big] (\bs{y}_N) 
\; ,
\end{multline}
where $*$ indicates the couple of variables of the function $F$ on which the transform 
$\wt{\mc{V}}_{N-1} $ acts and the $\ov{\msc{J}}_N$ transform is understood to act 
on the variables $(\bs{w}_{N-1}, x )$. The latter is defined as  
\beq
\ov{\msc{J}}_N\big[ R \big](\bs{w}_N) \; = \;   \lim_{\a \tend 0^+} 
\Int{\R^{N-1}}{} \hspace{-2mm} \f{\dd^{N-1} y }{ (2\pi \hbar)^{N-1} }
\Int{\R}{} \hspace{-1mm}  \dd x \cdot \f{ - \ex{ \f{i}{\hbar} \ov{\bs{w}}_{N}x} \cdot   R(\bs{y}_{N-1},x) }
{ \sqrt{N} \cdot (N-1)! }
\cdot \pl{a\not= b}{N-1} \Ga^{-1}\Big( \f{y_a-y_b}{i\hbar} \Big)
\cdot \pl{a=1}{N} \pl{b=1}{N-1} \Ga\Big(  \f{w_a -y_b-+ i\a }{ i \hbar }  \Big)
\enq
Finally, the $\wt{\mc{V}}_{N-1} $-transform is expressed, 
given any $G \in \msc{C}^{\infty}_{\e{c}}(\R^{N-1})$, as
\beq
\wt{\mc{V}}_{N-1}[ G ] (x,\bs{w}_{N-1})  \; = \; \hbar^{\f{i}{\hbar}N \ov{\bs{w}}_{N-1} } 
 \ex{ -i\f{x}{\hbar}\ov{\bs{w}}_{N-1} } \cdot  \mc{V}_{N-1}[ G ]( \bs{w}_{N-1} )  \;. 
\enq

Since the isometric nature of the $\ov{\msc{J}}_N$-transform is equivalent to the 
one of the $\msc{J}_N$-transform, proposition \ref{Prop charactere isometrique J} ensures that 
\beq
\big|\big|  \mc{V}_N\big[ F\big]     \big|\big|_{L^{2}_{\e{sym}}\big( \R^N, \dd \mu(\bs{y}_N) \big) } \; = \;  
 \big|\big|  \mc{V}_{N-1}\big[ F(*,x) \big]  
   \big|\big|_{L^{2}_{\e{sym}\times -}\big( \R^{N-1}\times \R, \dd \mu(\bs{y}_{N-1})\otimes \dd x \big) }  \;. 
\enq
As a consequence, a straightforward induction leads to the claim. \qed



\section*{Conclusion}

In this paper we have developed a technique allowing one to prove the unitarity of the SoV transform
in the case of integrable models with a infinite dimensional representation attached to each of its sites. 
Although we have developed the method on the example of the Toda chain, there do not seem to appear
any obstruction to applying it to more complex models such as the lattice discretization of the 
sinh-Gordon model \cite{BytskoTeschnerSinhGordonFunctionalBA}. 
The original contribution of this paper to the  right invertibility of the transform, \textit{ie}
$\mc{U}^{\dagger}_{N} \cdot  \mc{U}_N \; = \; \e{id}_{L^{2}_{\e{sym}}\big( \R^N , \dd \mu(\bs{y}_N) \big) }$ 
consisted in bringing in several elements of rigour to the scheme invented in \cite{DerkachovKorchemskyManashovXXXSoVandQopNewConstEigenfctsBOp}
and applied to the case of the Toda chain in \cite{SilantyevScalarProductFormulaTodaChain}. 
However, the part relative to the left invertibility of the map 
$\mc{U}_N \cdot  \mc{U}^{\dagger}_N \; = \; \e{id}_{L^{2}\big( \R^N , \dd^N x \big) }$ 
was entirely based on brand new ideas. In the case of the Toda chain, our approach provides an alternative in respect to 
a purely group theoretic handling of the issue
\cite{GoodmannWallachQuantumTodaI,
KostantIdentificationOfEigenfunctionsOpenTodaAndWhittakerVectors,
SemenovTianShanskyQuantOpenTodaLatticesProofOrthogonalityFormulaForWhittVectrs,WallachRealReductiveGroupsII}. 
On the one hand, our approach is much simpler as solely based on a direct calculation. 
On the other hand, our proof does not rely, at any stage,
on the group theoretical interpretation of the Toda chain but solely on objects naturally arising in the framework
of the quantum inverse scattering method approach to the quantum separation of variables. Hence, it will 
most probably work as well for other quantum integrable models through the separation of variables 
where the interpretation of the transform's kernel in terms of a suitable Whittaker function does not exist. 
In a forthcoming publication, we plan to study the implementation of our method for proving the unitarity 
of the SoV transform to other quantum integrable models solvable by the quantum separation of variables.

\section*{Acknowledgements}

The author would like to thank O. Babelon, E. van der Ban, 
M. Semenov-Tian-Shansky, E. K. Sklyanin, J. Teschner and N. R. Wallach
for stimulating discussions or expert informations relative to various topics treated in this paper. 
The authors also thanks D. An from 
kindly providing a copy of his doctoral thesis. 
The author is supported by CNRS. The author acknowledges support from the 
the Burgundy region PARI 2013 FABER grant "Structures et asymptotiques d'int\'{e}grales multiples".
The present work has been carried out, in part, within the financing of the grant PEPS-PTI "Asymptotique d'int\'{e}grales multiples". 
This research has been initiated when the author was supported by the EU Marie-Curie Excellence Grant MEXT-CT-2006-042695 and DESY.
The author is indebted to the Laboratoire de Physique Th\'{e}orique d'Annecy-Le-Vieux 
and Katedra Metod Matematycznych dla Fizyki for their warm hospitality
and providing excellent working conditions during my visits 
when part of the research has been carried out.




\appendix

\section{Proof of proposition \ref{Proposition charactere de transfo UN}}
\label{Appendix Proof of explicit behavior function varphi}

The Mellin-Barnes multiple integral representation for $\vp_{\bs{y}_N}(\bs{x}_N)$ can be recast as 
\bem
\vp_{\bs{y}_N}(\bs{x}_N) \; = \; \ex{ \f{i}{\hbar} \ov{\bs{y}}_{N} x_{N} }   \pl{s=1}{N-1} \Int{ (\R-i\a_s)^{N-s} }{  } \hspace{-3mm} 
 \dd^{N-s} w^{(s)} 
\pl{s=1}{N-1}  \bigg\{ \ex{ \f{i}{\hbar} \ov{\bs{w}}_{N-s}^{(s)} \big( x_{N-s} - x_{N-s+1} \big) } \bigg\}
\cdot  \pl{b>a}{N}\big(  y_a - y_b  \big) \\
\times \;  \pl{s=1}{N-1} \Bigg\{  
 \f{ \pl{a \not= b}{N-s} \big( w_a^{(s)}-w_b^{(s)} \big) }{ \pl{b=1}{N-s+1}\pl{a=1}{N-s}  \big( w_b^{(s-1)} \, - \,   w_a^{(s)} \big) } 
\Bigg\} 
\; \cdot \; \mc{W}_N \big( \big\{ \bs{w}_{N-s}^{(s)} \big\}_0^{N-1}  \big) \;,
\end{multline}
where $\a_{N-1} > \dots > \a_1 >0$. We also remind that $\bs{w}_N^{(0)}=\bs{y}_N$ and  agree upon
\beq
\mc{W}_N \Big( \big\{ \bs{w}_{N-s}^{(s)} \big\}_0^{N-1}  \Big) \; = \; \pl{b>a}{N} \bigg\{  \f{ 1 }{  y_a - y_b  }  \bigg\}
\cdot \pl{s=1}{N-1} \Bigg\{ \f{ \varpi\big( \bs{w}_{N-s}^{(s)} \mid \bs{w}_{N-s+1}^{(s-1)} \big) }{ (N-s)! (2\pi \hbar)^{N-s} } 
 \cdot 
 \f{ \pl{b=1}{N-s+1}\pl{a=1}{N-s}  \big( w_b^{(s-1)} \, - \,   w_a^{(s)} \big) }{ \pl{a \not= b}{N-s} \big( w_a^{(s)}-w_b^{(s)} \big) }
\Bigg\}
  \;. 
\enq
It is a direct consequence of the identity \eqref{ecriture identite vers decomposition produits transfo Cauchy simples}
that
\beq
\pl{b>a}{N}\big(  y_a - y_b  \big) \; \cdot \;  \pl{s=1}{N-1} \Bigg\{  
\f{ \pl{a \not= b}{N-s} \big( w_a^{(s)}-w_b^{(s)} \big) }
				{ \pl{b=1}{N-s+1}\pl{a=1}{N-s}  \big( w_b^{(s-1)} \, - \,   w_a^{(s)} \big) }  \Bigg\} \; = \; 
\sul{ \substack{ \sg_s \in \mf{S}_{N+1-s}  \\ s=1,\dots, N-1  }   }{}	  \pl{s=1}{N-1} (-1)^{\sg_s}   \cdot
		\pl{s=1}{N-1}\pl{a=1}{N-s} \f{1}{  \big(  w_a^{(s)} \, - \, w_{\sg_s(a)}^{(s-1)} \big)	} \;. 
\enq
 Define the sequence of permutations $\tau_{s}\in \mf{S}_{N+1-s}$ as
\beq
\tau_{N-1} \; = \; \sg_{N-1} \qquad \e{and} \qquad 
 \left\{ \ba{cc}  \tau_{s}(N+1-s) \; = \; \sg_{s}(N+1-s) &\\
				\tau_{s}(a) \; = \; \sg_{s}\circ \tau_{s+1}(a) & a=1,\dots, N-s \ea \right. \quad \e{for} \quad 
s=1,\dots, N-2	\;, 		
\enq
so that 
\beq
\pl{a=1}{N-s} \f{1}{  \big(  w_a^{(s)} \, - \, w_{\sg_s(a)}^{(s-1)} \big) } \; = \; 
\pl{a=1}{N-s} \f{1}{  \big(  w_{\tau_s(a)}^{(s)} \, - \, w_{\tau_{s-1}(a)}^{(s-1)} \big) }\;. 
\enq
Since, for $s=1,\dots, N-2$ 
\beq
(-1)^{\tau_s} \; = \; (-1)^{\sg_{s} } (-1)^{\tau_{s+1}}  \qquad \e{it} \; \e{follows} \; \e{that} \qquad
\pl{s=1}{N-1} (-1)^{\sg_s} \; = \; (-1)^{\tau_1} \;. 
\enq
Then, the symmetry in each vector $\bs{w}_{N-s}^{(s)}$, $s=1,  \dots, N-1$ taken singly of 
$\mc{W}_N \big( \big\{ \bs{w}_{N-s}^{(s)} \big\}_1^{N-1}  \big)$ and its anti-symmetry in 
$\bs{w}_{N}^{(0)}=\bs{y}_N$ implies that 
\beq
\vp_{\bs{y}_N}(\bs{x}_N) \; = \; \pl{s=2}{N-1} s! \cdot 
\sul{ \tau_1 \in \mf{S}_N }{} J\big( \bs{y}_{N;\tau_1} \big) \qquad \e{with} \qquad 
\bs{y}_{N;\tau_1} \; = \; \big( y_{\tau_1(1)}, \dots, y_{\tau_1(N)} \big)
\enq
where
\beq
J\big( \bs{w}_{N}^{(0)} \big) \; = \; \ex{ \f{i}{\hbar} \ov{\bs{w}}_{N}^{(0)}\cdot  x_{N} }   
\pl{s=1}{N-1} \Int{ (\R-i\a_s)^{N-s} }{  } \hspace{-3mm} 
 \dd^{N-s} w^{(s)} 
\pl{s=1}{N-1}  \bigg\{ \ex{ \f{i}{\hbar}  r_s \cdot \ov{\bs{w}}_{N-s}^{(s)}  } \bigg\}
\cdot \;   \f{  \mc{W}_N \big( \big\{ \bs{w}_{N-s}^{(s)} \big\}_0^{N-1}  \big)   }
				{   \pl{s=1}{N-1} \pl{a=1}{N-s} \big( w_a^{(s)}-w_a^{(s-1)} \big)   }  \; \;. 
\enq
Above, we agree upon 
\beq
r_s \; = \; x_{N-s} \, - \, x_{N-s+1} \;. 
\enq
In order to obtain an explicit formula allowing one to bound the function $\vp_{\bs{y}_N}(\bs{x}_N)$, one should
move the contours of integration for variables associated with $r_{s}>0$ from the lower half-plan to the upper half-plane. 
The matter is that, in doing so, one will cross poles which will generate new type of exponents, say 
containing the combinations $r_{a} + \dots + r_{b}$ with $a>b$. This last factor can be positive or negative. In the latter
case, one should then move the integration in respect to the associated variables also to the upper half-plane. 
In the former case, there is nothing else to do. In fact, the most optimal way of
expressing  the result of contour shifting is in terms of a sum over all sequences $R_{a s}$, with $a=1, \dots, N-s$ and $s=1, \dots N-1$
that can be built according to the below algorithm. This algorithm is well defined provided that 
all the partial sums do not vanish, \textit{ie}
\beq
r_{a} + \dots + r_{b} \not= 0 \qquad \e{for}  \; \e{any} \qquad a\geq b \;.  
\enq
The case when some of the partial sums vanish is readily obtain by taking appropriate limits. 

\noindent The algorithm starts at $N-1$
\begin{itemize}
\item If $\left\{  \ba{ccc} r_{N-1} \; \leq \; 0 & \e{then} \; \e{set}  \; & R_{1,N-1} = r_{N-1} \\   
							r_{N-1} \; > \;  0 & \e{then} \; \e{pick}  \; & R_{1,N-1} \in \{ 0 \, ,  r_{N-1} \} \ea	\right. \;  .$ 
\end{itemize}
The quantities $R_{a,N-2}$, $a=1,2$ are built as
\begin{itemize}
\item suppose  $R_{1,N-1}=0$,  then if 	$\left\{  \ba{ccc} 
	 r_{N-1}+ r_{N-2} \; \leq \; 0 & \e{then} \; \e{set}  \; & R_{1,N-2} = r_{N-1}+ r_{N-2}  \vspace{2mm} \\   
					r_{N-1}+ r_{N-2} \; > \; 0 & \e{then} \; \e{pick}  \; & R_{1,N-2} \in \{ 0 \, ,  r_{N-1}+ r_{N-2} \} \ea	\right.	$ \; , 	

\item if 		$R_{1,N-1} \not= 0$,  should
$\left\{  \ba{ccc} r_{N-2} \; \leq \; 0 & \e{then} \; \e{set}  \; & R_{1,N-2} = r_{N-2} \vspace{2mm}  \\   
							r_{N-2} \; > \;  0 & \e{then} \; \e{pick}  \; & R_{1,N-2} \in \{ 0 \, ,  r_{N-2} \} \ea	\right. $ \;  , 

\item finally, 	if $\left\{  \ba{ccc} r_{N-2} \; \leq \; 0 & \e{then} \; \e{set}  \; & R_{2,N-2} = r_{N-2} \vspace{2mm} \\   
							r_{N-2} \; > \;  0 & \e{then} \; \e{pick}  \; & R_{2,N-2} \in \{ 0,  r_{N-2} \} \ea	\right. \;  .$ 	
\end{itemize}
 Assume having chosen $\{ R_{a,s^{\prime} } \}$	with $1 \leq a \leq N-s^{\prime}$ and $s+1\,  \leq \, s^{\prime} \, \leq \, N-1$. 		
Then define
\beq
\left\{   \ba{l}  k_{a,s} \; = \; \min \big\{ k \geq s+1 \; : \; R_{a,k} \; = \; 0  \big\} \quad  \e{if} \quad  R_{a, s+1} =0  
\vspace{3mm} \\
 k_{a,s} \; = \; s \quad \e{otherwise}  \ea \right. \;. 
\enq
Note that, necessarily, for $k_{a,s} \geq s+1$ one has $r_{k_{a,s}} + \dots + r_{s+1} \; > \; 0$. 	
\begin{itemize}
\item If $\left\{  \ba{ccc} r_{k_{a,s}} + \dots + r_{s} \; \leq \; 0 & \e{then} \; \e{set}  \; 
&   R_{a,s} = r_{k_{a,s}} + \dots + r_{s} \vspace{2mm} \\   
							r_{k_{a,s}} + \dots + r_{s}  \; > \;  0 & \e{then} \; \e{pick}  \; & 
							R_{a,s} \in \{ 0,  r_{k_{a,s}} + \dots + r_{s} \} \ea	\right. \;  .$ 	
\end{itemize}

One continues in this way up to $s=0$ when there is a unique choice possible for the sequence $R_{a,0}$.
For $a=1,\dots,N-1$
\begin{itemize}
\item if $\left\{ \ba{ccc}  k_{a,0} = 0  & \e{then} &  R_{a,0} = 0 \\  
							k_{a,0} \not= 0  & \e{then} &  R_{a,0} = r_{k_{a,s}} + \dots + r_{1}   \ea  \right. $  
							\; \qquad and  \qquad  $R_{N,0}=0$ \;. 
\end{itemize}

We denote by $\mc{R}$ the set of all possible sequences $R_{a,s}$ that can be obtained by application of the above algorithm,
\beq
\mc{R} \; = \; \bigg\{  \{R_{a,s} \} \; ,\, \e{with} \; \; 
 \Big| \ba{c}  a=1,\dots, N-s \; \\  s=0,\dots,N-1  \ea  \; \;  \e{such} \; \e{that}  \; \; 
R_{a,s} \; \; \e{built} \; \e{by} \; \e{algorithm} \Big\} \;. 
\enq

A given sequence $\{ R_{a,s} \} $ defines uniquely which residues have been computed in the course 
of moving the contours of integrations. Indeed, set
\beq
\Om_{a} \; = \; \big\{ s \;  \geq \; 1  \; \; : \; \; R_{a,s} \, = \,  0 \big\} \quad \e{and} \quad
\Om_{a}^{\pm} \; = \; \big\{ s \;  \geq \; 1  \; \; : \; \; \pm R_{a,s} \,  > \,  0 \big\} \;. 
\enq
The set $\Om_a$ will have $\ell_a$ neighbouring  components in the sense that 
\beq
\Om_{a} \; = \; \Big\{ b_{a,1},b_{a,1}+1,\dots, c_{a,1} \Big\} \cup \dots \cup \Big\{ b_{a,\ell_a},\dots, c_{a,\ell_a} \Big\}
\enq
where
\beq
1 \; \leq \; b_{a,1} \; \leq \; c_{a,1} \; \leq \; b_{a,2}-2 \; \dots \; \leq \;  b_{a,p} \;  \leq \; 
c_{a,p} \; \leq \; b_{a,p+1}-2 \; \leq \dots \; \leq \; c_{a,\ell_a} \; \leq \; N-a \;.  
\enq
Furthermore, since there is at least one integer in between each of the "connected parts", one has
$\ell_a \leq \big[ \tf{(N-a)}{2}  \big]$. 
Then, the result of contour deformation is to integrate solely over the variables $w_a^{(s)}$ with $s \in \Om^{\pm}_a$, $a=1, \dots, N-1$.
The variables belonging to $ \Om_a$ should be reduced according to 
\beq
w_a^{(b_{a,p})} \; = \; \dots\; = \; w_a^{(c_{a,p}+1)}  \qquad \e{with} \qquad p=1,\dots, \ell_a \;. 
\label{ecriture reduction chain variables poles}
\enq
It is the reduction \eqref{ecriture reduction chain variables poles} that corresponds to computing 
the poles. Thus $J(\bs{w}_N^{(0)})$ is recast as 
\bem
J\big( \bs{w}_{N}^{(0)} \big) \; = \; \ex{ \f{i}{\hbar} \ov{\bs{w}}_{N}^{(0)}\cdot  x_{N} }   \cdot \sul{ \{R_{a,s} \} \in \mc{R} }{}
\pl{a=1}{N-1}  \Big\{  \ex{ \f{i}{\hbar}R_{a,0} w_a^{(0)} }  \Big\}  \\
\times \pl{a=1}{N-1} \Bigg\{  \pl{s \in \Om_a^+ }{} \Int{ \R+i\eta_s }{  } \hspace{-3mm} 
 \f{ \dd w^{(s)} }{ 2i\pi }    \ex{ \f{i}{\hbar}R_{a,s} w_a^{(s)} }    
\cdot \pl{s \in \Om_a^- }{} \Int{ \R-i\a_s }{  } \hspace{-3mm} 
 \f{ \dd w^{(s)} }{ 2i\pi }    \ex{ \f{i}{\hbar}R_{a,s} w_a^{(s)} }       \Bigg\}
 \cdot  \wt{\mc{W}}_{N }^{(\e{red})} \big(  \{  R_{a,s} \} \, ; \,  \big\{ \bs{w}_{N-s}^{(s)} \big\}_0^{N-1}  \big) \;. 
\end{multline}
where $\eta_1>\dots > \eta_{N-1}>0$ and $\a_{N-1}> \dots > \a_1 >0$, 
\beq
\wt{\mc{W}}_{N }^{(\e{red})} \big(  \{  R_{a,s} \} \, ; \,  \big\{ \bs{w}_{N-s}^{(s)} \big\}_0^{N-1}  \big)  \; = \; 
 \f{   \mc{W}_{N }^{(\e{red})} \big(  \{  R_{a,s} \} \, ; \,  \big\{ \bs{w}_{N-s}^{(s)} \big\}_0^{N-1}  \big)  }
{   \pl{a=1}{N-1} \Big\{ 
 \pl{ \substack{ s \in \Om_a^+ \cup \Om_a^- \\ \not\in B_a}  }{}  \big( w_a^{(s)}  - w_a^{(s-1)} \big)  
  \cdot  \pl{p=1}{\ell_a} \big( w_a^{(c_{a,p}+1)} \; - \; w_a^{(b_{a,p}-1)} \big) \Big\} } 
\; . 
\enq
Note that, above, we agree upon 
\beq
B_a \; = \; \big\{ b_{a,1},\dots, b_{a,\ell_a}    \big\} \;, 
\enq
and $\mc{W}_{ N }^{(\e{red})} $ is obtained from $\mc{W}_{N} $ by implementing the 
reduction \eqref{ecriture reduction chain variables poles}. 

\vspace{2mm}

It follows from the previous handlings that the integral transform $\mc{U}_N[F](\bs{x}_N)$ can be recast as 
\bem
\mc{U}_N[F](\bs{x}_N) \; = \; \f{ \pl{s=1}{N} s! }{ \sqrt{N!} } \Int{ \R^N }{} \dd^N y 
F\big( \bs{y}_N \big) \cdot \ex{ \f{i}{\hbar} \ov{\bs{y}}_{N}\cdot  r_{N} }   \cdot \sul{ \{R_{a,s} \} \in \mc{R} }{}
\pl{a=1}{N-1}  \Big\{  \ex{ \f{i}{\hbar}R_{a,0} y_a }  \Big\}  \\
\hspace{-5mm} \times \pl{a=1}{N-1} \Bigg\{  \pl{s \in \Om_a^+ }{} \Int{ \R+i\eta_s }{  } \hspace{-2mm} 
 \f{ \dd w^{(s)} }{ 2i\pi }  \,   \ex{ \f{i}{\hbar}R_{a,s} w_a^{(s)} }    
\cdot \pl{s \in \Om_a^- }{} \Int{ \R-i\a_s }{  } \hspace{-2mm} 
 \f{ \dd w^{(s)} }{ 2i\pi }  \,   \ex{ \f{i}{\hbar}R_{a,s} w_a^{(s)} }       \Bigg\}_{\mid \bs{w}_N^{(0)}=\bs{y}_N }  
\hspace{-3mm} \mu(\bs{y}_N) \cdot \wt{\mc{W}}_{N }^{(\e{red})} \big(  \{  R_{a,s} \} \, ; \,  \big\{ \bs{w}_{N-s}^{(s)} \big\}_0^{N-1} 
   \big)_{\mid \bs{w}_N^{(0)}=\bs{y}_N } \;. 
\nonumber
\end{multline}
Above, we agree upon the identification $r_N \equiv x_N$. 
We do stress that the singularities of $\wt{\mc{W}}_{N }^{(\e{red})} $ at $y_a=y_b$, $a \not= b$, are compensated by the 
zeroes of the measure's density $\mu(\bs{y}_N)$. The integrand is thus a smooth, compactly supported function of 
$\bs{y}_N$.

We now build on the above representation so as to ensure the Schwartz class of $\mc{U}_N[F]$. 
By the very construction of the sequence $\{ R_{a,s} \}$, for every fixed $a$,
the numbers $R_{a,s}$ with $0\leq s \leq N-a$ cannot all be zero. 
Defining
\beq
s_a \; = \; \min \Big\{ s \; : \;  R_{a,s} \not= 0 \Big\} \quad \e{one} \; \e{has} \quad 
R_{a,s_a} \; = \; \left\{ \ba{cc}  r_{N-a} + \dots + r_{s_a}  & \e{if} \;\;  s_a \; \geq \; 1\\ 
							r_{N-a} + \dots + r_{1}  & \e{if} \;\;  s_a \; = \; 0	 \ea \right. 
\enq

One can then integrate by parts in respect to all the variables 
$y_a$ such that $R_{a,0} \not= 0$ and bound the exponentially decreasing factors 
by a power law for all the variables $y_a$ such that $s_{a} \geq 1$. 
This readily shows that there exists a $k,F$-dependent constant $C_{k,F}$ such that 
\beq
\Big|  \mc{U}_N[F](\bs{x}_N)  \Big| \; \leq \; 
C_{k,F} \cdot \sul{ \{ R_{ab} \} \in \mc{R} }{}  
\pl{a=0}{N-1}  \bigg(  \f{   1  }{  1 \, + \, |r_{N-a} \, + \,  \dots \, + \, r_{s_a} |  }  \bigg)^k \;. 
\enq
This condition is readily translated into one in respect to the "position" variables $\bs{x}_N$, thus 
ensuring the Schwartz class of the integral transform $\mc{U}_N[F](\bs{x}_N)$. \qed





\section{From Mellin-Barnes to Gauss-Givental}
\label{Appendix Section from MB 2 GG}

\begin{lemme}
\label{Proposition calcul relation echange operateur L cal}

Let $\mc{L}^{(N-1)}_{y} : L^{\infty}(\R^{N-1}) \tend L^{\infty}(\R^{N-1})$ be an integral operator with a kernel
\beq
\mc{L}^{(N-1)}_{y}\big( \bs{x}_{N-1} \mid \bs{\tau}_{N-1} \big) \; = \; 
\ex{\f{i y }{2\hbar} (x_1 - \tau_{N-1}) } \pl{n=1}{N-1}V_{y;-}(x_n-\tau_n) \pl{n=1}{N-2}V_{y;+}(x_{n+1}-\tau_n) \;. 
\enq
Then, for $\Im(y) > \Im(w)$ the function $\Big( \mc{L}^{(N-1)}_{y}\cdot \La_w^{(N-1)}  \Big) \big( \bs{x}_{N-1} \mid \bs{z}_{N-2} \big)$
is well-defined and one has the relation
\beq
\mc{L}^{(N-1)}_{y}\cdot \La_w^{(N-1)} \; = \; \hbar^{\f{i}{\hbar}(w-y)} \Ga\Big( \f{y-w}{i\hbar} \Big)
\times  \La_w^{(N-1)} \cdot \mc{L}^{(N-2)}_{y} \;. 
\label{ecriture relation echange op L cal et La big}
\enq

\end{lemme}

The proof of this lemma goes similarly to the one of lemma \ref{Lemme exhange operateur Lambda bar Lambda}, 
so we do not reproduce it here. The operators $\mc{L}_y$ along with the above lemma have been introduced for the
first time in \cite{GerasimovLebedevOblezinBAxtOpMixedRepsForTodaWhittakerAndMore}.

\begin{lemme}
\label{Lemme nouvelle integrale type many gamma functions}

Let $s \in \R$. Let   $y_1, \cdots, y_n$ and  $x_1, \dots, x_{n-1}$ be two sets of generic variables 
 in $\Cx$ and $\msc{C}$ a contour 
that circumvents all the points $y_a + i \hbar n_a $, with $n_a \in \mathbb{N}$, $a=1,\dots, n$
from below whereas it circumvents the points $x_a - i \hbar n_a $, with $n_a \in \mathbb{N}$, $a=1,\dots, n-1$
from above. Then one has the integral identity
\beq
\Int{\msc{C}^n}{}  \ex{\f{s \ov{\bs{w}}_n }{ i\hbar } } 
\f{ \pl{a=1}{n} \bigg\{ \pl{b=1}{n} \Ga\Big( \f{y_b-w_a}{i\hbar}\Big) \pl{b=1}{n-1} \Ga\Big( \f{w_a-x_b}{i\hbar}\Big)  \bigg\} }
{  \pl{a\not =b }{n} \Ga\Big( \f{w_a-w_b}{i\hbar}\Big)  } \cdot \f{ \dd^n w }{ (2\pi \hbar)^n} 
\; = \; n! \ex{\f{s \ov{\bs{y}}_n }{ i\hbar } }  \ex{- \ex{s}} \cdot
\pl{a=1}{n} \pl{b=1}{n-1} \Ga\Big( \f{y_a-x_b}{i\hbar} \Big) \;. 
\enq

\end{lemme}

\Proof 

The starting point is given by the multi-dimensional integral computed by Gustafsen \cite{GustafsonqBetaIntegralsForTodaLikeCase}. 
Given any 
two generic sets of points $y_1, \cdots, y_{n+1}$ and $x_1, \dots, x_{n+1}$  and  a contour $\msc{C}$
that circumvents all the points $y_a + i \hbar n_a $, with $n_a \in \mathbb{N}$, $a=1,\dots, n+1$
from below whereas it circumvents the points $x_a - i \hbar n_a $, with $n_a \in \mathbb{N}$, $a=1,\dots, n+1$
from above, one has  
\beq
\Int{\msc{C}^n}{} 
\f{ \pl{a=1}{n} \bigg\{ \pl{b=1}{n+1} \Ga\Big( \f{y_b-w_a}{i\hbar}, \f{w_a-x_b}{i\hbar}\Big)  \bigg\} }
{  \pl{a\not =b }{n} \Ga\Big( \f{w_a-w_b}{i\hbar}\Big)  } \cdot \f{ \dd^n w }{ (2\pi \hbar)^n} 
\; = \; n! \cdot \f{  \pl{a,b=1}{n+1} \Ga\Big( \f{y_a-x_b}{i\hbar} \Big)  }{ \Ga\Big(  \sul{a=1}{n+1} \f{y_a-x_a}{i\hbar}  \Big) }\;. 
\label{ecriture integrale Gustafsen}
\enq
For the time being, we assume that the parameters $\{x_a\}_1^{n+1}$ and $\{y_a\}_1^{n+1}$ are such that one can choose
$\msc{C}$ lying sufficiently close to $\R$, \textit{ie} $\eta=\sup_{w \in \msc{C}} |\Im(w)|$ is small, and 
that $\msc{C}$ avoids $0$. We then set $y_{n+1}=-K\ex{-s}$ with $(K,s) \in \R^+\times \R$ and $x_{n+1}=K$ and divide both sides of \eqref{ecriture integrale Gustafsen}
by 
\beq
\Ga^{n} \Big(  - \f{K}{i\hbar} , -\f{ K\ex{-s} }{ i\hbar } \Big) \;, 
\enq
what leads to 
\beq
\Int{\msc{C}^n}{}   
\mc{I}_K\big( \{w_a\}_1^n ; \{y_a\}_1^n; \{x_a\}_1^n\mid s \big)  \cdot \f{ \dd^n w }{ (2\pi \hbar)^n} 
\; = \; n! \cdot  \pl{a,b=1}{n} \Ga\Big( \f{y_a-x_b}{i\hbar} \Big)  \cdot  u_K\big( \{y_a\}_1^n; \{x_a\}_1^n\mid s \big) \;, 
\enq
in which 
\beq
u_K( \{y_a\}_1^n; \{x_a\}_1^n\mid s ) \; = \; \f{ \Ga\Big( -K \f{1+\ex{-s}}{i\hbar} \Big)    }
{ \Ga\Big( -K \f{1+\ex{-s}}{i\hbar}  \; + \; \sul{a=1}{n} \f{y_a-x_a}{i\hbar}  \Big)   }
\pl{k=1}{n}  \f{ \Ga\Big( - \f{K\ex{-s} +x_k }{i\hbar} ,   \f{y_k-K }{i\hbar}  \Big)  }
{  \Ga \Big(  - \f{K}{i\hbar} , -\f{ K\ex{-s} }{ i\hbar } \Big) }
\enq
whereas the integrand reads 
\beq
\mc{I}_K\big( \{w_a\}_1^n ; \{y_a\}_1^n; \{x_a\}_1^n\mid s \big) \; =  \; 
 \pl{k=1}{n}  \f{ \Ga\Big( - \f{K\ex{-s} + w_k }{i\hbar} ,   \f{w_k-K }{i\hbar}  \Big)  }
{  \Ga \Big(  - \f{K}{i\hbar} , -\f{ K\ex{-s} }{ i\hbar } \Big) }
\cdot \f{ \pl{a=1}{n} \bigg\{ \pl{b=1}{n} \Ga\Big( \f{y_b-w_a}{i\hbar}, \f{w_a-x_b}{i\hbar}\Big)  \bigg\} }
{  \pl{a\not =b }{n} \Ga\Big( \f{w_a-w_b}{i\hbar}\Big)  } \; . 
\enq
It is readily seen that pointwise in $\{y_a\}_1^n; \{x_a\}_1^n$ and in  $s$
\beq
u_K( \{y_a\}_1^n; \{x_a\}_1^n\mid s ) \quad \limit{K}{+\infty} \quad 
\ex{ \f{s \ov{\bs{x}}_n }{ i\hbar } } \cdot \big(1+\ex{-s}  \big)^{ \f{ \ov{\bs{x}}_n - \ov{\bs{y}}_n }{ i\hbar } }  \; . 
\enq
It is likewise readily seen that, pointwise in $\{w_a\}_1^n ; \{y_a\}_1^n; \{x_a\}_1^n$ and in  $s$,
\beq
\mc{I}_K\big( \{w_a\}_1^n ; \{y_a\}_1^n; \{x_a\}_1^n\mid s \big) \quad  \limit{K}{+\infty} \quad  
 \ex{\f{s \ov{\bs{w}}_n }{ i\hbar } } \cdot \f{ \pl{a=1}{n} \bigg\{ \pl{b=1}{n} \Ga\Big( \f{y_b-w_a}{i\hbar}, \f{w_a-x_b}{i\hbar}\Big)  \bigg\} }
{  \pl{a\not =b }{n} \Ga\Big( \f{w_a-w_b}{i\hbar}\Big)  }
\enq
and that, furthermore, 
\beq
\big| \mc{I}_K\big( \{w_a\}_1^n ; \{y_a\}_1^n; \{x_a\}_1^n\mid s \big) \big|  \; \leq \; 
\wt{C} \cdot \pl{a=1}{n} \Big\{  |w_a|^{2 (\eps+\eta) \f{n}{\hbar} }  \ex{- \f{3\pi}{4\hbar}|\Re(w_a) | }   \Big\}  \cdot
\pl{a=1}{n} f_K(w_a) \;,
\enq
with $\eps = \e{max}_{t \in S} |\Im(t)|$, $S=\{ x_1, \dots, x_n, y_1, \dots, y_n\}_1^n$, $\eta= \max_{w \in \msc{C}}|\Im(w)|$ and 
\beq
f_K(w) \; = \; \f{  K \cdot  \ex{- \f{\pi}{4\hbar}|\Re(w) | }  }{|w| \sqrt{ |K-w|\cdot |w+K\ex{-s}| }  }\cdot 
  \Big|  \f{ K-w }{ w+K\ex{-s}  }  \Big|^{\f{\Im(w)}{\hbar} }   \cdot
\exp\bigg\{ \f{\pi}{2\hbar} \Big[ K(1+ \ex{-s}) \, - \,  |\Re(K-w)| \, - \, |\Re(K\ex{-s}+w)|   \Big]  \bigg\} \;. 
\enq
Elementary analysis then shows that, uniformly in $K$,  $w \in \msc{C}\mapsto f_K(w)$ is bounded\footnote{we do recall that $0 \not \in \msc{C}$
and that $\msc{C}$ also avoids $K$ and $-K\ex{-s}$}. As a consequence, 
we are thus in position to apply the dominated convergence theorem, leading to 

\beq
\Int{\msc{C}^n}{}   
\ex{\f{s \ov{\bs{w}}_n }{ i\hbar } } \cdot \f{ \pl{a=1}{n} \bigg\{ \pl{b=1}{n} \Ga\Big( \f{y_b-w_a}{i\hbar}, \f{w_a-x_b}{i\hbar}\Big)  \bigg\} }
{  \pl{a\not =b }{n} \Ga\Big( \f{w_a-w_b}{i\hbar}\Big)  }\cdot \f{ \dd^n w }{ (2 \pi \hbar)^n} 
\; = \; n! \cdot  \pl{a,b=1}{n} \Ga\Big( \f{y_a-x_b}{i\hbar} \Big) \cdot \ex{i  \f{s}{\hbar} \ov{\bs{y}}_n } 
\cdot \big( 1+\ex{s} \big)^{ \f{\ov{\bs{x}}_n- \ov{\bs{y}}_n }{i\hbar} }\;.
\label{ecriture identite intermediaire mellinBarnes integrals}
\enq

It then follows from an analytic continuation that, in fact, the formula holds for $|\Im(s)|< \pi$ and for all 
sets of points $\{x_a\}_1^n$ and $\{y_a\}_1^n$ that can be separated by a curve $\msc{C}$ in the sense of the statement
of the present lemma. 

In the newly obtained identity, we substitute
\beq
x_n = K \qquad s=v-\ln \Big( \f{-K}{i\hbar}  \Big) \qquad \e{with} \qquad v \in \R
\enq
and again consider sets of points $\{x_a\}_1^{n-1}$, $\{y_a\}_1^n$ such that one can
take the contour $\msc{C}$ lying sufficiently close to $\R$ and avoiding $0$.  
Then, we divide both sides of \eqref{ecriture identite intermediaire mellinBarnes integrals}
by $\Ga^{n}\Big( -\tf{iK}{\hbar} \Big)$. This yields the representation
\beq
\Int{\msc{C}^n}{}   
\mc{J}_K\big( \{w_a\}_1^n ; \{y_a\}_1^n; \{x_a\}_1^{n-1}\mid v \big)  \cdot \f{ \dd^n w }{ (2\pi \hbar)^n} 
\; = \; n! \cdot  \pl{a=1}{n} \pl{b=1}{n-1}  \Ga\Big( \f{y_a-x_b}{i\hbar} \Big)  
\cdot  \wt{u}_K\big( \{y_a\}_1^n; \{x_a\}_1^{n-1}\mid v \big) \;, 
\enq
in which 
\beq
\wt{u}_K\big( \{y_a\}_1^n; \{x_a\}_1^{n-1}\mid v \big) \; = \; 
\Big( 1 - \f{i\hbar}{K} \ex{v} \Big)^{ \f{K}{i\hbar}   + \f{ \ov{\bs{x}}_{n-1} - \ov{\bs{y}}_{n} }{i\hbar} }
\cdot \ex{ \f{ v \ov{\bs{y}}_n }{ i\hbar } } \times
\pl{a=1}{n} \bigg\{   \Ga\Big(  \f{y_a-K }{i\hbar}  \Big)  \cdot
 \Ga^{-1} \Big(  - \f{K}{i\hbar}   \Big)   \cdot \Big( - \f{K}{i\hbar} \Big)^{-\f{y_a}{i\hbar} } \bigg\}
\enq
whereas the integrand reads 
\beq
\mc{J}_K\big( \{w_a\}_1^n ; \{y_a\}_1^n; \{x_a\}_1^{n-1}\mid s \big) \; =  \; \ex{ \f{v \ov{\bs{w}}_n}{i\hbar} }
 \pl{a=1}{n}  \bigg\{  \Big( - \f{K}{i\hbar} \Big)^{-\f{w_a}{i\hbar} }  \cdot 
					\Ga\Big(  \f{w_a-K }{i\hbar}  \Big)  \cdot  \Ga^{-1} \Big(  - \f{K}{i\hbar} \Big) \bigg\}
\cdot \f{ \pl{a=1}{n} \bigg\{ \pl{b=1}{n} \Ga\Big( \f{y_b-w_a}{i\hbar}\Big)
		 \cdot \pl{b=1}{n-1} \Ga\Big( \f{w_a-x_b}{i\hbar}\Big)  \bigg\} }
{  \pl{a\not =b }{n} \Ga\Big( \f{w_a-w_b}{i\hbar}\Big)  } \; . 
\enq
It is readily seen that pointwise in $\{y_a\}_1^n; \{x_a\}_1^{n-1}$ and in $\nu$
\beq
\wt{u}_K\big( \{y_a\}_1^n; \{x_a\}_1^{n-1}\mid v \big) \quad \limit{K}{+\infty} \quad 
\ex{ \f{v \ov{\bs{y}}_n }{ i\hbar } } \cdot \exp\Big\{ -\ex{v} \Big\} \;. 
\enq
It is likewise readily seen that, pointwise in $\{w_a\}_1^n ; \{y_a\}_1^n; \{x_a\}_1^{n-1}$ and in  $s$,
\beq
\mc{J}_K\big( \{w_a\}_1^n ; \{y_a\}_1^n; \{x_a\}_1^{n-1}\mid s \big) \quad  \limit{K}{+\infty} \quad  
 \ex{ \f{v \ov{\bs{w}}_n}{i\hbar} } \cdot \f{ \pl{a=1}{n} \bigg\{ \pl{b=1}{n} \Ga\Big( \f{y_b-w_a}{i\hbar}\Big)
		 \cdot \pl{b=1}{n-1} \Ga\Big( \f{w_a-x_b}{i\hbar}\Big)  \bigg\} }
{  \pl{a\not =b }{n} \Ga\Big( \f{w_a-w_b}{i\hbar}\Big)  } \; ,
\enq
and that, furthermore, 
\beq
\big| \mc{J}_K\big( \{w_a\}_1^n ; \{y_a\}_1^n; \{x_a\}_1^n\mid s \big) \big|  \; \leq \; 
\wt{C} \cdot \Big| \ex{ \f{v \ov{\bs{w}}_n}{i\hbar}} \Big| \cdot 
\pl{a=1}{n} \Big\{  |w_a|^{(\eps+\eta) \f{2n}{\hbar} }  \ex{- \f{\pi}{4\hbar}|\Re(w_a) | }   \Big\}  \cdot
\pl{a=1}{n} \wt{f}_K(w_a) \;,
\enq
with
\beq
\wt{f}_K(w) \; = \; \f{  \sqrt{K} }{ \sqrt{ |K-w|\cdot |w| }  }\cdot 
 \Big\{ \ex{- \f{\pi}{4\hbar}|\Re(w) | }   \Big|  \f{ K-w }{ K }  \Big|^{\f{\Im(w)}{\hbar} }  \Big\} 
\cdot  \exp\Big\{ \f{\pi}{2\hbar} \big[ K- \Re(w) - |K- \Re(w)|  \big] \Big\} \;. 
\enq
We are, again, in position to apply the dominated convergence theorem, thus leading to the claim in the case 
of $s \in \R$ and $x_a , y_a$ lying sufficiently far from $\R$. The general result then follows by 
analytic continuation in $(\bs{x}_{n-1},\bs{y}_n)$ since the result already holds on an open subset of $\Cx^{n-1}\times \Cx^{n}$. 

 \qed

\begin{prop}
 \label{Proposition relation de recurrence fcts vp}
The unique solution $\vp_{\bs{y}_{N}}\big( \bs{x}_{N} \big)$  \eqref{definition fonction vp apres resolution MellinBarnes}
to the Mellin-Barnes induction 
\eqref{definition fct Whittaker par Mellin-Barnes}  satisfies to the 
induction
\beq
\vp_{\bs{y}_{N}}\big( \bs{x}_{N} \big) \; = \; \Int{ \R^{N-1} }{} 
\La^{(N)}_{y_n}\big(\bs{x}_{N} \mid \bs{\tau}_{N-1} \big) \vp_{\bs{y}_{N-1}}\big( \bs{\tau}_{N-1} \big) \cdot \dd^{N-1} \tau \;. 
\label{ecriture explicite relation recurrence type Gauss-Givental}
\enq
\end{prop}

The multiple integral \eqref{ecriture explicite relation recurrence type Gauss-Givental} is well defined since
$\vp_{\bs{y}_N} \in  L^{\infty}(\R^{N})$ and, for fixed $\bs{x}_{N} \in \R^N$, 
the function 
\beq
\tau_{\bs{N-1}} \;  \mapsto  \; \La^{(N)}_{y_n}\big(\bs{x}_{N} \mid \bs{\tau}_{N-1} \big) \in L^{1}(\R^{N-1}) \;. 
\enq
The recurrence relation \eqref{ecriture explicite relation recurrence type Gauss-Givental} 
provides a connection between the Mellin-Barnes and Gauss-Givental 
representation for $\vp_{\bs{y}_{N+1}}\big(\bs{x}_{N+1}\big)$ and, in fact, shows that the 
Gauss-Givental representation, seen as an encased integral, is well defined. 
Proposition \ref{Proposition relation de recurrence fcts vp} has been first derived in 
\cite{GerasimovLebedevOblezinBAxtOpMixedRepsForTodaWhittakerAndMore}. 
However, the proof given in \cite{GerasimovLebedevOblezinBAxtOpMixedRepsForTodaWhittakerAndMore} 
utilizes the completeness and orthogonality of the system of functions 
$\vp_{\bs{y}_N}$. Here, we provide a different proof of this induction that 
does not build on the completeness and orthogonality of the system $\vp_{\bs{y}_N}(\bs{x}_N)$.

\Proof 

The statement holds for $N=0$ since 
\beq
\vp_{y_1}(x_1) \; = \; \La_{y_1}^{(1)}\big( x_1 \mid -  \big)  \; . 
\enq
Now assume that it holds up to some $N$. A straightforward induction based on the exchange relation 
\eqref{ecriture relation echange op L cal et La big} shows that given $y_{N+1} \in \R$ and 
$\bs{w}_N \in (\R-i\a)^N$, $\a >0$ one has 
\beq
\mc{L}_{y_{N+1} }^{(N)} \cdot \vp_{\bs{w}_{N}}(\bs{x}_{N}) \; = \; 
\pl{a=1}{N} \bigg\{ \hbar^{ \f{i}{\hbar}(w_a-y_{N+1} ) }  \cdot \Ga\Big( \f{y_{N+1}-w_a}{ i \hbar } \Big)   \bigg\}
\cdot \vp_{\bs{w}_{N}}(\bs{x}_{N}) \;. 
\enq
Thus
\begin{itemize}
\item inserting this relation into the Mellin-Barnes recurrence relation \eqref{definition fct Whittaker par Mellin-Barnes}, 
\item exchanging the order of $\tau_{N}$ and $\bs{w}_N$ integrations,
\item applying the Mellin-Barnes induction a second time to $\vp_{\bs{w}_N}(\bs{\tau}_{N})$,
\item exchanging the order of $\ga_{N-1}$ and $\bs{w}_N$ integrations, 
\end{itemize}
leads us to the representation
\bem
\vp_{\bs{y}_{N+1}}(\bs{x}_{N+1}) \; = \; \Int{ \R^{N} }{} \dd^N \tau \hspace{-3mm}   \Int{ (\R-2i\a)^{N-1} }{  } \hspace{-3mm}  
\dd^{N-1}\ga
\f{ \mc{L}^{(N)}_{y_{N+1}}\big(\bs{x}_N \mid \bs{\tau}_N \big) }{ (N-1)! (2\pi \hbar)^{N-1} } \cdot 
\f{  \ex{ \f{i}{\hbar}  ( \ov{\bs{y}}_{N+1} x_{N+1} - \ov{\bs{\ga}}_{N-1} \tau_{N} ) } }
{ \pl{ a \not= b }{ N-1 } \Ga \Big( \f{ \ga_a-\ga_b }{ i\hbar } \Big)  }
\cdot \hbar^{ \f{i}{\hbar}N  ( \ov{\bs{\ga}}_{N-1} -\ov{\bs{y}}_{N} ) } \cdot \vp_{\bs{\ga}_{N-1}}\big( \bs{\tau}_{N-1} \big)
\\
\times \Int{ (\R-i\a)^{N} }{  } \hspace{-3mm} 
\ex{ \f{i}{\hbar} \ov{\bs{w}}_{N+1} (\tau_N-x_{N+1} ) }  \hbar^{ \f{i}{\hbar} \ov{\bs{w}}_N }
\pl{a=1}{N} \bigg\{ \pl{b=1}{N} \Ga\Big( \f{y_{b}-w_a}{ i \hbar } \Big) \cdot  
		\pl{b=1}{N-1} \Ga\Big( \f{w_a-\ga_{b} }{ i \hbar } \Big)    \bigg\} 
\pl{a \not= b }{N}  \Ga^{-1}\Big( \f{w_{b}-w_a}{ i \hbar } \Big) 
 \f{ \dd^{N} w  }{ N! (2\pi \hbar)^{N} } \;. 
\label{ecriture formule intermerdiare pour preuve GG 2 MB}
\end{multline}
Note that we were able to exchange the orders of integration twice since, in each case, the 
integrand can be bounded by a strictly positive function that is readily seen to be integrable
for at least one ordering of the integration variables. By Fubbini's theorem, 
this is already enough.  
The integral arising in the last line of \eqref{ecriture formule intermerdiare pour preuve GG 2 MB} 
can be computed thanks to the results of 
lemma \ref{Lemme nouvelle integrale type many gamma functions} leading to  
\bem
\vp_{\bs{y}_{N+1}}(\bs{x}_{N+1}) \; = \; \Int{ \R^{N} }{} \dd^N \tau \cdot 
\f{ \mc{L}^{(N-1)}_{y_{N+1}}\big(\bs{x}_N \mid \bs{\tau}_N \big) }{ (N-1)! (2\pi \hbar)^{N-1} } \cdot 
 \exp\Big\{  -\f{1}{\hbar} \ex{ x_{N+1}- \tau_N } \Big\}
\\
\times  \Int{ (\R-i\a)^{N-1} }{  } \hspace{-3mm}  \dd^{N-1}\ga
\cdot \f{ \ex{ \f{i}{\hbar}  ( \ov{\bs{y}}_{N}  - \ov{\bs{\ga}}_{N-1} )\tau_{N}  } }
{ \pl{ a \not= b }{ N-1 } \Ga \Big( \f{ \ga_a-\ga_b }{ i\hbar } \Big)  }
\pl{a=1}{N} \pl{b=1}{N-1} \bigg\{ \hbar^{ \f{i}{\hbar}(\ga_b-y_a) }  \cdot  \Ga\Big(  \f{y_a-\ga_b }{ i\hbar } \Big)  \bigg\} 
\cdot  \vp_{\bs{\ga}_{N-1}}\big( \bs{\tau}_{N-1} \big) \;. 
\end{multline}
Hence, using that 
\beq
\La^{(N+1)}_{y_{N+1}}\big( \bs{x}_{N+1} \mid \bs{\tau}_{N} \big) \; = \; 
\mc{L}^{(N)}_{y_{N+1}}\big( \bs{x}_N \mid \bs{\tau}_{N} \big) \exp\Big\{  -\f{1}{\hbar} \ex{ x_{N+1}- \tau_N } \Big\}
\cdot \ex{ \f{i}{\hbar} y_{N+1} x_{N+1} }
\enq
we obtain the claim. \qed

\section{An auxiliary lemma}

\begin{lemme}
\label{Lemme echangeability limite integrale}

Let $\eta >0$. The sequence 
\beq
f_K(s) \; = \; \f{   K }{  \sqrt{s^2 + \eta^2} \cdot \ln^2 \big[ s^2 + \eta^2 \big] \cdot \sqrt{ (s-K)^2 + \eta^2}   } 
\enq
satisfies
\beq
\lim_{K\tend +\infty} \Int{ \R }{} f_K(s) \cdot \dd s \; = \; \Int{ \R }{} \Big[ \lim_{K\tend +\infty} f_K(s) \Big] \cdot \dd s
\enq

\end{lemme}

\Proof 

One has 
\beq
\Int{ \R }{}  f_K(s) \cdot \dd s  \; = \; \hspace{-5mm} \Int{ \R \setminus \intff{(1-\eps)K}{ (1+\eps)K }  }{} \hspace{-7mm} f_K(s) \cdot \dd s 
\; \;  + \; \Int{ (1-\eps)K  }{ (1+\eps)K}  f_K(s) \cdot \dd s  \;. 
\enq
Since
\beq
\big|  f_K(s) \bs{1}_{\R \setminus \intff{(1-\eps)K}{ (1+\eps)K }}(s) \big| \; \leq \; 
  \f{  C_{\eps}    }{  \sqrt{s^2 + \eta^2} \cdot \ln^2 \big[ s^2 + \eta^2 \big]   }  \; \in \; L^1(\R, \dd s)
\enq
for some $\eps$-dependent constant $C_{\eps}>0$, 
one can apply the dominated convergence theorem to the first integral leading to
\beq
\lim_{K\tend +\infty} \hspace{-5mm} \Int{ \R \setminus \intff{(1-\eps)K}{ (1+\eps)K }  }{} \hspace{-5mm} f_K(s) \cdot \dd s 
\; = \; \Int{ \R }{}  \f{ \dd s  }{ \sqrt{s^2 + \eta^2} \cdot \ln^2 \big[ s^2 + \eta^2 \big]   } \;. 
\enq
It thus remains to establish that the second integral goes to zero. 
It is readily seen that 
\beq
\Int{ (1-\eps)K  }{ (1+\eps)K}  f_K(s) \cdot \dd s  \; = \; \Int{0}{\eps K }  
h_K(s) \cdot \Big[  \f{1}{\sqrt{s^2 + \eta^2}}  -  \f{1}{s+i\eta }\Big]  \cdot \dd s  
\; + \; \Int{0}{\eps K }   \f{ h_K(s) }{s+i\eta }   \cdot \dd s 
\enq
where 
\beq
h_K(s) \; = \; \sul{ \upsilon = \pm }{}  \; 
 \f{  K \cdot \big[ (s+ \upsilon K)^2 + \eta^2 \big]^{-\f{1}{2}}  }{ \ln^2 \big[ (s+ \upsilon K)^2 + \eta^2 \big] } \;. 
\enq
Since, uniformly in $K$,  $h_K$ is bounded  $ \intff{0}{\eps K}$, one can apply the dominated convergence theorem to the first 
integral so as to get that it goes to zero with $K$. Finally, 
\beq
\Int{0}{\eps K }   \f{ h_K(s) }{s+i\eta }   \cdot \dd s  \; = \; 
h_K( \eps K ) \ln (\eps K + i \eta) \; - \; h_K(0) \ln(i\eta)  \; - \; 
\Int{0}{\eps K } h_K^{\prime}(s) \ln (s + i \eta) \cdot \dd s \;. 
\enq
A straightforward calculation leads to 
\beq
\big| \big| h_K^{\prime} \big| \big|_{L^{\infty}(\intff{0}{\eps K})} \; \leq \;  \f{ C_{\eps} }{ K \ln^2(K)}  \;, 
\enq
hence allowing one to conclude. \qed.

\end{document}